\documentclass[pdftex,twocolumn,epjc3]{svjour3}          

\RequirePackage[T1]{fontenc}

\smartqed  

\RequirePackage{graphicx}
\RequirePackage{mathptmx}      
\RequirePackage{flushend}
\RequirePackage[numbers,sort&compress]{natbib}
\RequirePackage[colorlinks,citecolor=blue,urlcolor=blue,linkcolor=blue]{hyperref}
\RequirePackage{amsfonts}
\RequirePackage{bbm}
\RequirePackage{amsmath}
\RequirePackage{verbatim}
\RequirePackage{booktabs}

\journalname{Eur. Phys. J. C}

\begin{document}

\title{The Practical Pomeron for High Energy Proton Collimation}


\author{R.B. Appleby\thanksref{addr1}
        \and
        R.J. Barlow\thanksref{addr2} 
        \and
        J.G Molson\thanksref{addr3}
        \and
        M. Serluca\thanksref{addr4}
        \and
        A. Toader\thanksref{addr2}
}

\thankstext{e1}{e-mail: robert.appleby@manchester.ac.uk}

\institute{The Cockcroft Institute and the University of Manchester, Oxford Road, Manchester, M13 9PL, UK.\label{addr1}
          \and
          The University of Huddersfield, Huddersfield, West Yorkshire HD1 3DH, United Kingdom
\label{addr2}
       \and
         LAL, Univ. Paris-Sud, CNRS/IN2P3, Universit\'{e} Paris-Saclay, Orsay, France.
\label{addr3}
       \and
         CERN, Geneva, 1203, Switzerland
\label{addr4}
          }

\date{Received: date / Accepted: date}

\maketitle

\begin{abstract}
We present a model which describes proton scattering data from ISR to Tevatron energies, and which
can be applied to collimation n high energy accelerators, such as the LHC and FCC.  Collimators remove beam halo particles, so that they do not impinge on vulnerable regions of the machine, such as the superconducting magnets and the experimental areas.
In simulating the effect of the collimator jaws it is crucial to model the scattering of protons at small momentum transfer~$t$,
 as these protons can subsequently survive several turns of the ring before being lost.  At high energies these soft processes  are well described by Pomeron exchange models.
We study the behaviour of elastic and single-diffractive dissociation cross sections over a wide 
range of energy, and show that the model can be used as a global description of the wide variety of high energy elastic and diffractive data presently available. 
In particular it models low mass diffraction dissociation, where a rich resonance structure is present, and thus predicts the differential and integrated cross sections in the kinematical range appropriate to the LHC.    
We incorporate the physics of this model into the beam tracking code MERLIN and use it to simulate the resulting loss maps of the beam halo lost in the collimators in the LHC.
\end{abstract}

\section{\label{sec:part1}Introduction and motivation}

The world's highest energy particle accelerator, the  Large Hadron Collider (LHC), contains two high-energy proton beams travelling in opposite directions, guided around the accelerator ring by  superconducting (SC) magnets. 
Its nominal stored beam energy of 360~MJ is orders of magnitude greater than previous accelerators, such as the Tevatron.
This high energy stored beam passes close to SC magnets with a quench limit of about 15~mW~cm$^{-3}$~\citep*{LHCDesign}.
 A powerful cleaning system is vital to the machine protection in order to operate below the quench limit, with a highly efficient collimation system necessary in order to remove any stray halo protons.
The halo is generated by various effects~\citep*{LHCDesign} and it is characterised as an off-momentum halo (in which particle energies deviate from the reference) and a betatron halo (in which particles have large transverse amplitudes).
Although the collimation system is adequate for the current configuration of the LHC, 
for the future High-Luminosity (Hi-Lumi) machine~\citep*{hilumibook} upgrade the physics of the scattering of protons in the collimators must be accurately
 simulated, to avoid any quench of the SC magnets and to protect the vulnerable parts of the machine such as the detectors.

The tracking of protons around the ring and inside the collimator material is based on complex simulations where many different physics effects are involved. Here we focus on the scattering. 
Protons interact with both electrons and nuclei in the collimator material. The former
give multiple Coulomb scattering, leading to angular deflections and energy loss that modify the beam particle's momentum. The latter can be divided into elastic ($pp \to pp$), Single-Diffractive (SD) ($p p \to p X$ or $p p \to X p$), double diffractive ($p p \to X Y$) and inelastic scatters. Note that we ignore nuclear effects and  consider a nucleus as a collection of protons and neutrons, and interactions with neutrons are treated similarly to those with protons. To study the beam halo we do not consider inelastic scatters, double-diffractive scatters, or SD interactions $p p \to X p $, in which the beam proton breaks up: 
for such events all the energy is lost locally, within a few metres. With elastic and single-diffractive scattering ($p p \to p X$) the emerging protons are only slightly affected and may survive several turns before being lost. 
The elastic scattering contributes to the betatron halo creation, and SD to the off-momentum halo.

The LHC ring is divided into 8 regions. For the nominal layout, as described in the design report~\citep*{LHCDesign}, there are two collimation regions. In the third Interaction Region (IR3), the removal of off-momentum halo particles, known as momentum cleaning, takes place in a dispersive region. In IR7, particles with large transverse amplitude are removed; this is known as betatron cleaning. There is also an accelerating region in IR4, and a beam dump region in IR6. The remaining four regions are dedicated to the detector insertions:
there are two at low $\beta^{*}$ in IR1 (ATLAS) and IR5 (CMS), and two at  high $\beta^{*}$ in IR2 (ALICE) and IR8 (LHCb), 
where $\beta^{*}$ is the betatron function of the magnetic lattice at the interaction point. 
In each collimation region there is a cleaning hierarchy,  and the primary collimators (TCP) in IR7 have the tightest apertures of the machine. In addition, tertiary collimators (TCT) are installed at both sides of the detectors to protect them.
 


In table~\ref{tab:ft_energy}, the equivalent centre-of-momentum energy is given for various LHC proton energies on a `fixed target' proton in the collimator.
It varies from 29~GeV at injection ($E_{beam} = 450$~GeV) to 115~GeV for the nominal beam energy, 176~GeV for the LHC energy upgrade and 306 GeV for the FCC-hh. 

\begin{table}[h]   
\caption{\label{tab:ft_energy}%
The relevant beam energies required for protons impinging on a collimator.}
\begin{tabular}{ccc}
\hline
\textrm{State}&
\textrm{$E_{beam}$ [GeV]}&
\textrm{Fixed target $\sqrt{s}$ [GeV]}\\
\hline
LHC Injection & 450 & 29\\ 
LHC 2011 Collision & 3500 & 81\\ 
LHC 2012 Collision & 4000 & 84\\ 
LHC Nominal collision & 7000 & 115 \\
FCC-hh & 50000 & 306 \\ \hline
\end{tabular}
\end{table}

Experimental data for $pp$ and $ p\bar{p}$ reactions exist for many energies from different experiments and accelerators, principally the  Intersecting Storage Rings (ISR)  at $\sqrt{s}= $23-63~GeV
and the Tevatron at 2~TeV.  There are also data from the SP$\bar{\mathrm{P}}$S. With plentiful data both above and below the range required, our model parameters are obtained by interpolation, rather than extrapolation.
 
In this paper we create a model within the Pomeron and Reggeon exchange framework of Donnachie and Landshoff~\cite{D-L1,QCDDonnachie}. The model is an elegant description of the strong interaction at high energies, and describes the experimental data for total, elastic and SD scattering with minimal assumptions. The fit uses a small number of parameters to describe data for 21 energies
and 11 experiments, aiming to achieve the best possible fit. 
We use an extension of the model which we fit to most of the available elastic and SD data,  in order to obtain a parametrisation which covers the required  
proton-target kinematical range at LHC energies.  

The extended model, which we simply call the DL model, is implemented into the beam tracking library MERLIN~\cite{molsonicap12,serlucaIpac1,serlucaIpac2,MerlinWEB}, which is then used to simulate the loss maps for the nominal LHC. 

We use this model to simulate the LHC loss maps, demonstrating the cleaning performance of the collimation system.
This performance determines whether the accelerator can safely run at higher intensity, or whether additional shielding or collimators will be required.  Realistic simulations of particle loss maps are fundamental to our ability to predict eventual quenching locations, for the nominal LHC and possible upgraded collimation systems, new materials and advanced collimation concepts such as hollow electron lenses~\cite{Shiltsev} and crystal collimation~\cite{Biryukov200523}.  

The layout of this paper is as follows. In section~\ref{sec:part2} we introduce the kinematics and discuss the requirements for the simulation of proton scattering within the collimator materials. In section~\ref{sec:part3} we model the elastic scattering, performing a fit which 
achieves a good description of the available data. Then in section~\ref{sec:part4} we describe the single diffractive model and obtain a fit for the double differential cross section for low and high missing mass regions, producing a good description of a wide range of data.
We illustrate the fitting procedure and present the results at LHC energies and the prediction of the total SD cross section as a function of the centre-of-mass energy $s$. We show that it is possible to use the DL fit approach for elastic and SD scattering to cover the required range of kinematical variables for the LHC.    
In section~\ref{sec:part5} we introduce the MERLIN code and the implementation of the model. The resulting loss maps for the nominal LHC at 7~TeV are presented with a detailed examination of the betatron cleaning region and the losses in the dispersion regions.


The data sources for elastic and single diffraction dissociation at different energies are reported in~\ref{sec:app_datalist} and~\ref{sec:app_datalist2}, along with references and the fit of the model to data.

\section{\label{sec:part2}Proton scattering and beam dynamics}
\subsection{Particle beam dynamics and dispersion}

The horizontal transverse motion of a particle in an accelerator is given by the Courant-Snyder parameterization of the solution to Hill's equation \cite{accphys}
\begin{equation}
\label{eq:twiss}
x\left(s\right) = \sqrt{2J \beta_x\left(s\right)} \sin(\mu_x\left(s\right)+\mu_{x0}) + D_x \frac{\Delta p}{p}.
\end{equation}
Here $s$ is the longitudinal position along the accelerator lattice, $J$ is the particle action, $\beta_x$ the betatron function of the accelerator magnetic lattice, and $\mu_x$ the betatron phase.  $p$ is the reference momentum for the lattice and $\Delta p$ is the deviation of the particle from this reference momentum.
$D_x$ is the dispersion function, describing the motion of particles with such a deviation.
Protons that have lost momentum in diffractive interactions may have large transverse displacements from the reference orbit in regions where $|D_x|$ is large.
 
\begin{figure}
\includegraphics[width=0.5\textwidth]{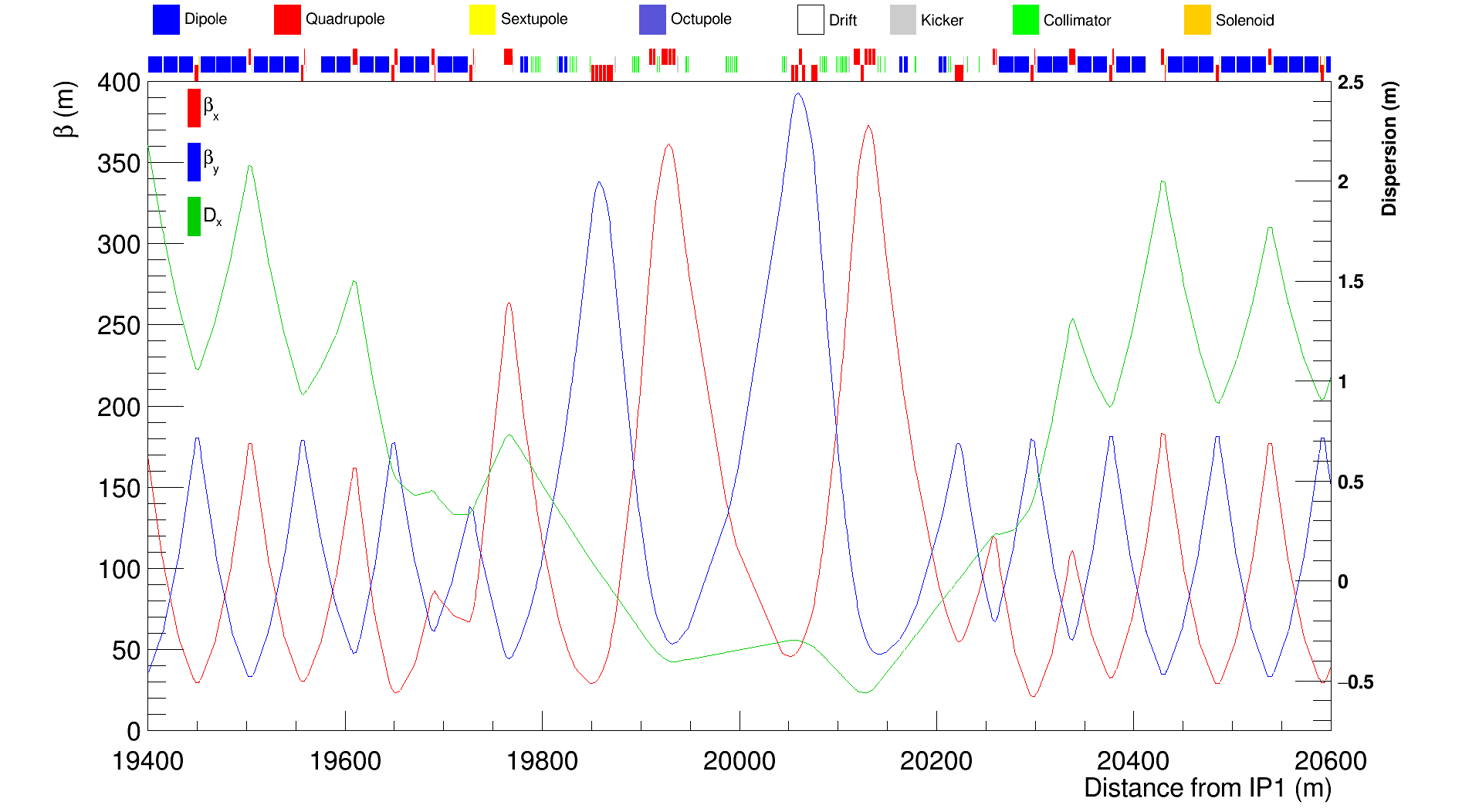}
\caption{\label{fig:IR7_optics}Horizontal dispersion (green line) and horizontal and vertical $\beta$ functions (red and blue lines) for beam 1 around the IR7 region in the LHC as generated by MERLIN using the V6.503 optics layout. The horizontal axis represents the distance from ATLAS interaction point.}
\end{figure}

Figure~\ref{fig:IR7_optics} shows the $\beta$-functions and the horizontal dispersion in the betatron collimation region IR7. 
There, the dispersion is small but the $\beta$-functions are large, so the collimators placed here remove protons in the betatron halo but not the energy halo.

\begin{figure}
\includegraphics[width=0.5\textwidth]{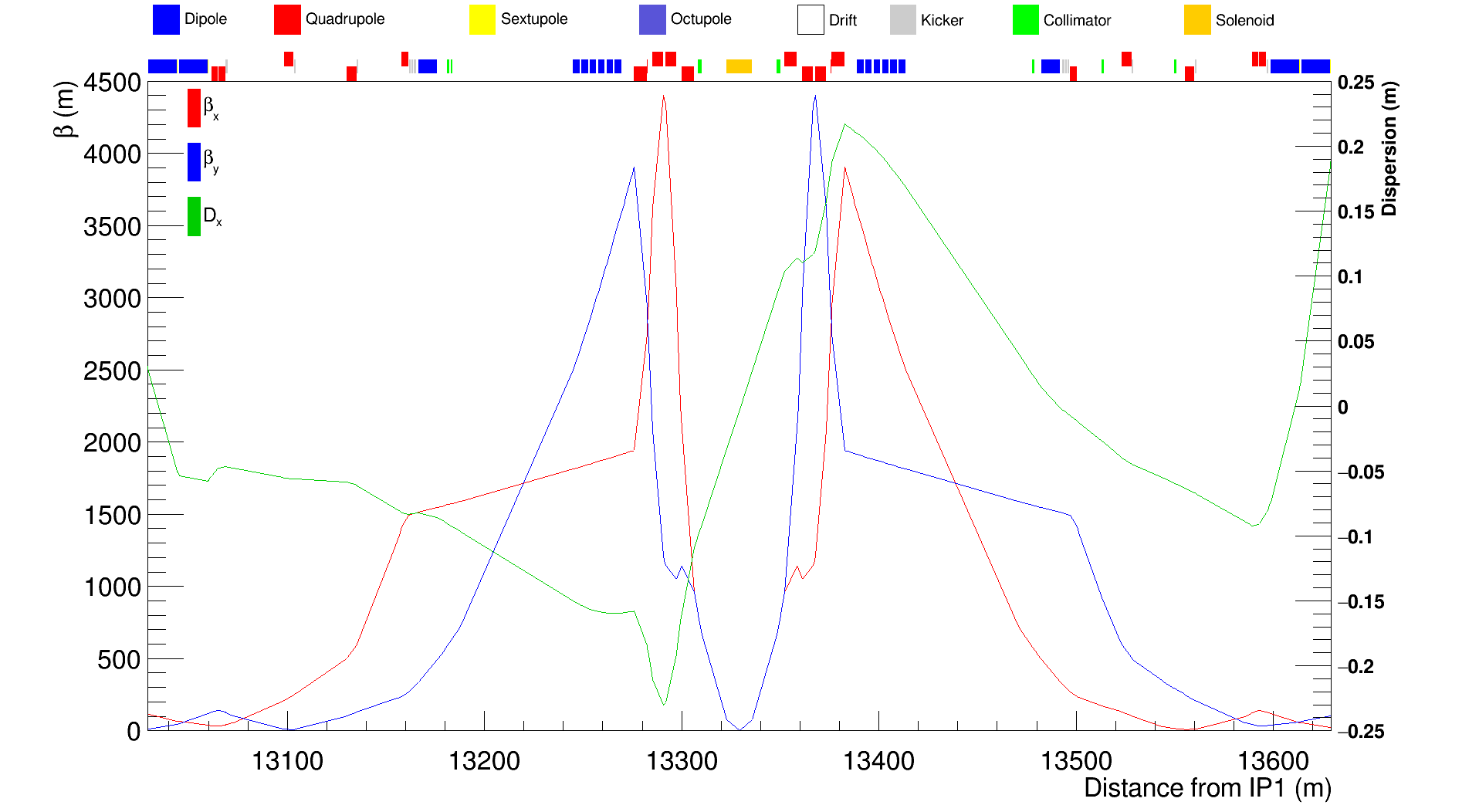}
\caption{\label{fig:IR5_optics}Horizontal dispersion (green line) and $\beta$ functions (red and blue lines) for beam 1 around the CMS detector (IR5) in the LHC as generated by MERLIN using the V6.503 optics layout.}
\end{figure}

Figure~\ref{fig:IR5_optics} shows the $\beta$-functions and horizontal dispersion in the IR5 region, where  the CMS detector is located. The magnetic elements are shown above the plot, including the quadrupole triplets on both sides of the detector that squeeze the beam at the interaction point. For the nominal LHC, the value of  $\beta$ at the interaction point is 55~cm  in IR1 (ATLAS) and  IR5 (CMS) and 10~m in IR2 (ALICE) and  IR8 (LHCb).

\subsection{Kinematics and the relevant range of $t$ and $M_X$}
If a proton with mass $M_p$ and  4-momentum $p_i^{\mu} = (E_i,\vec{p_i})$ in the lab frame interacts with a stationary proton $P^{\mu} = (M_p,0)$, and  scatters to $p_f^{\mu} = (E_f, \vec{p_f})$, the invariants $s$ and $t$ are defined as
\begin{equation}
\label{eq:energy}
s = (P^{\mu}+p_i^{\mu})^2 =2 M_{p}^2 + 2 M_{p}E_{i},
\end{equation}
\begin{equation}
\label{eq:t1}
t=(p_i^\mu - p_f^\mu)^2 =  (E_i - E_f)^2 - (\vec{p}_i -\vec{p}_f)^2.
\end{equation}
$\sqrt{s}$ is the energy in the centre of momentum system. The expression for 
 $t$ can usefully be rewritten in terms of the proton scattering angle~$\theta$,
\begin{equation}
\label{eq:t2}
t =  2M_p^2-2E_iE_f+2p_i p_f \cos\theta.
\end{equation}
The invariant mass  of the diffracted proton-target, $M_X$ also called the  missing mass, is given by
energy-momentum conservation

\begin{equation}
\label{eq:MX}
M_X^2=(p_i^\mu+P^\mu-p_f^\mu)^2=(E_i-E_f+M_P)^2 - (\vec p_i-\vec p_f)^2.
\end{equation}
It is convenient to define the dimensionless variable
\begin{equation}
\label{eq:xi}
\xi={M_X^2 \over s}.
\end{equation}

The energy loss by the scattered proton is given by combining equations \ref{eq:t2} and \ref{eq:MX}
\begin{equation}
\Delta E = E_f-E_i= \frac{t +M_p^2 - M_X^2}{2 M_p}.
\end{equation}
(For elastic scattering $M_X$ is equal to $M_p$ and this simplifies further.)
Thus the quantities relevant for the simulation: $\Delta E$  the energy change, and $\theta$, the direction of the outgoing particle (apart from a random azimuthal angle) 
are determined by the quantities $t$ and $M_X$ or $\xi$, and it is the distributions for these two quantities that are predicted by the model.

If the scattering angle is significantly larger than the beam divergence, the scattered proton will be lost immediately, or in the nearby downstream region of the machine.
Thus our model of elastic scattering must be accurate at small $|t|$/small $\theta$ but need not model large $|t|$/large $\theta$, where `large' and `small' 
refer to comparison of the scattering angle with  the angular beam divergence at that location.
 Table~\ref{tab:colloptics} shows the LHC V6.503 machine optics, characterised by the Twiss parameters $\alpha$ and $\beta$, at three typical collimators in IR7 (defined earlier) and, assuming a normalised nominal beam emittance of 3.75~mm.mrad, shows the
angular divergences corresponding to $20$ times their nominal values at the collimator locations, and the corresponding  $|t|$ values. 

\begin{table*}[tH]
\caption{\label{tab:colloptics} A list of collimators in IR7 with their Twiss parameters, and assuming a normalised emittance of $3.75~\mu$m.rad, the value of $|t|$ values corresponding to 20 times the nominal beam divergence at collimator location.}
\begin{center}
\begin{tabular}{lcccccccccc}
\hline
Collimator	& $\beta_x$ &  $\beta_y$ &  $\alpha_x$ &  $\alpha_y$ & $\sigma_{\mathrm{max}}' (\mu{\rm rad})$ at 7 TeV & $\sigma_{\mathrm{max}}'(\mu{\rm rad})$ at 3.5 TeV & 
$|t|_{\mathrm{max}}$ 7 TeV & $|t|_{\mathrm{max}}$ 3.5 TeV\\ \hline
D6L7.B1&  158& 78 &2.1 & -1.1& 4.2 & 5.9 & 0.17 & 0.34\\
C6L7.B1&  150& 83 & 2.0& -1.2& 4.2 & 5.9 & 0.17& 0.34\\
A6L7.B1&  129&  97& 1.9& -1.3& 4.2 & 5.9 & 0.17& 0.34\\
EXTREME&  100& 100 &3 &3 & 7.1 & 10.0 &0.49 &0.98 \\ \hline
\end{tabular}
\end{center}
\end{table*}

The table shows that, for the collimation optics in IR7, modelling elastic scattering events up to $|t|=0.34~\mathrm{GeV}^{2}$ is sufficient to correctly 
model scattering events which could change the collimator-induced loss map beyond the immediately vicinity of the scattering collimator.
The table also includes an extreme case of collimation optics (arbitrarily chosen) which shows that for larger values of $\alpha$ at collimator locations we need to model an approximately double $|t|$ range of elastic events. 
The elastic fits in this paper are valid over the range of available data, and extend to $t$ = - 14.2~GeV$^2$, which is more than ample for our simulations.

Detailed modelling of scattering at very small $|t|$ is also unnecessary as
very small angle scatters do not lead to beam loss.
To investigate this  we have used MERLIN to perform a full phase space aperture scan in both planes, injecting a beam filling one plane of phase space, i.e. 
a grid in $x$ and $x^\prime$, with the other coordinates matched to the optical lattice at each collimator.
These particles were then tracked for 100 turns, with particles removed if they touch any aperture restrictions, and the surviving particles' initial angles at the collimator jaws recorded.
The smallest possible angle for which a particle is lost gives the minimum $t$ value required.
For the collimator jaw around the experimental regions, which have the minimum aperture available for scattering, an appropriate minimum value of $|t|$ is 0.0001~GeV$^2$. 

The ranges of $M_X$ and $|t|$ required for modelling diffractive scattering
depend on the beam's angular divergence and its intrinsic energy spread. 
For the former, referring to table~\ref{tab:colloptics},  we take a conservative value of $\sigma'=15 \,\,\mu$rad 
to cover all possible current and future cases including deviation from the specified normalised emittance value. 
For the latter, the  LHC beam energy spread $\sigma_e$ is the nominal 1.1$\cdot$10$^{-4}$ at 7 TeV,
 and has been  measured to be 1.36$\pm$0.04 $\cdot$ 10$^{-4}$ at 3.5 TeV.
The dependence of the scattering angle on $\xi$ is weak, and a $|t|$ limit of 4 GeV$^{2}$ 
corresponds to 20$\sigma'$ over all relevant $\xi$. 
 For a 3.5 TeV beam energy, $\xi$ = 0.12 corresponds to $M_X$ = 28 GeV and, even at our maximum $|t|$ of 4 GeV$^2$,  this gives an energy deviation 420 GeV, which is
  856$\sigma_e$, and also 41$\sigma'$. At 7 TeV it corresponds to 1109$\sigma_e$, The conclusion for all energies is that a kinematical range of $\xi$ up to 0.12 and $|t|$  up to 4 GeV$^2$ is sufficient and conservative for the single diffractive fit. The fits are not sensitive to the minimum values, and we take the fits down to threshold for $\xi$ and down to $t$ = - 0.0001 GeV$^2$.

\section{\label{sec:part3}Elastic Proton scattering and the Pomeron}

The differential cross-section $d\sigma/dt$ of elastic $p p$ and $p \bar{p}$ scattering is described by Coulomb scattering at very small $|t|$ and nuclear scattering for larger $|t|$. Early measurements at the ISR~\cite{Kwak1975233,Bohm1974491} with energies between 23 GeV and 63 GeV revealed at low $|t|$ an approximately exponential behaviour, $e^{-B |t|}$, where $B$ is known as the slope parameter. This is followed by a diffractive minimum at around $|t| \simeq$ 1.4 GeV$^2$, and subsequently a broad peak. The energy dependence of $d\sigma/dt$ shows a shrinkage of the elastic peak, i.e. an increase in $B$, with increasing $\sqrt{s}$~\cite{Amaldi1978367}.

The DL nuclear model includes Regge ($\rho,\, \omega\,\mathrm{ and }\, a_2, f_2$ trajectories) and Pomeron exchange \cite{D-L4}, including multiple Regge and Pomeron exchanges \cite{D-L2,D-L1}.
At large $t$ triple gluon exchange is also present \cite{D-L3,D-L5}. Recently, in the light of the LHC data from the TOTEM experiment, a hard Pomeron term has also been added \cite{D-L-arxiv}.

We extend the DL nuclear model to take into account the low $t$ Coulomb peak in order to simulate elastic scattering in the energy ranges given in table~\ref{tab:ft_energy}. The DL model has been fitted to all elastic data to obtain the fit parameters of the model. The fitting procedure is different from the one originally used in~\cite{D-L-arxiv} where the normalisations were kept constant. 
Further details of our approach can be found in~\citep*{JamesThesis}.

In this section we describe the elastic model general formulation and the fitting procedures. We then present the fitted differential cross section and the total elastic cross section.





\subsection{The general formulation}

The method used to calculate the elastic differential cross section is well established~\cite{RevModPhys.57.563}.
It is given by
\begin{equation}
\label{eq:simple1}
\frac{d\sigma}{dt} = \pi \left| f_c + f_n \right|^2,
\end{equation}
where $f_c$ and $f_n$ are the Coulomb and nuclear amplitudes.
In general, there is a phase difference between the Coulomb and nuclear amplitudes, $e^{i\alpha\phi\left(t\right)}$, such that
\begin{equation}
\label{eq:simple}
\frac{d\sigma}{dt} = \pi \left| f_c e^{i\alpha\phi\left(t\right)} + f_n \right|^2.
\end{equation}

To find the Coulomb phase $\phi$, we use a fit to the cross section slope at $t=0$, and the calculation by Cahn~\cite{Cahn},
\begin{equation}
\phi = \mp\left(\gamma + \ln\left[\frac{B}{2}\right]\right).
\end{equation}
The upper sign refers to $pp$ scattering, the lower to $p\bar{p}$, $\gamma$ is Euler's constant and $B$ is given by
\begin{equation}
B = 8.1 + 1.2 \log\sqrt{s}.
\end{equation}

\subsubsection{Photon exchange}
The Coulomb amplitude $f_{c}$ is given by~\citep*{Amos1985689}
\begin{equation}
f_{c} = \mp 2 \alpha_{em} \frac{F\left(t\right)^{2}}{t.},
\end{equation}
where $F\left(t\right)$ is the proton electromagnetic form factor, given by equation 3.17 in~\cite{QCDDonnachie},
\begin{equation}
F(t)= \frac{4M^2_p-2.79t}{4M^2_p-t}\frac{1}{(1-t/0.71)^2}.
\label{eq:proton_ff}
\end{equation}
However it is well approximated by the simpler formula~\cite{QCDDonnachie},
\begin{equation}
F\left(t\right)^2 = 0.27 e^{8.38t} + 0.56e^{3.78t} + 0.18 e^{1.36t}.
\end{equation}

\subsubsection{Hadronic exchange}

In the DL model, 4 Regge trajectories are used, the hard Pomeron, the soft Pomeron, the $f_2$ and $a_2$ trajectory, and  the $\omega$ and $\rho$ trajectory. In the following these are labelled by $0$ to $3$ respectively. 
This gives, for the nuclear amplitude,
\begin{equation}
f_n  = A_{ggg}\left(s,t\right) + \sum\limits_{i=0}^3 A_i\left(s,t\right).
\end{equation}
where we have included a triple-gluon exchange amplitude. Its form varies like $1/t^4$ at large $t$, is exponential at small $t$ and the expressions are forced to match at an intermediate $t=t_0$, such that~\cite{D-L4}
\begin{eqnarray}
A_{ggg}(t)&=&\frac{ \sqrt{16 \pi}\sqrt{0.09}}{t^4}  (|t| > |t_0| )\nonumber \\
&= &\frac{\sqrt{16 \pi}\sqrt{0.09} }{t_0^4}\exp\left( 4-4t/t_0\right)  ( |t| < |t_0|)
\end{eqnarray}
The amplitude is purely real and energy independent. $t_0$ is used as a free parameter in the fitting procedure. 

The exchange amplitudes $A_i$ are
\begin{equation}
A_i\left(s,t\right) = Y_i\left(2\nu \alpha_{i}^\prime\right)^{\alpha_{i}\left(t\right)} e^{\frac{i\pi}{2}\alpha_{i}\left(t\right)}F^{2}(t),
\end{equation}
with
\begin{equation}
2\nu = \left(\frac{s-u}{2}\right) \qquad  Y_i = - X_i(i=0,1,2) \qquad Y_3 = iX_3
\end{equation}
where $s$ and $u$ are the Mandelstam variables  and $F(t)$ is the form factor of the proton. The $X_i$ are real and positive; the factor $i$ multiplying $X_3$ ensures the correct signature factor~\cite{D-L1} for negative $C$-parity exchange. The amplitude for $p\bar{p}$ scattering is the same except that $Y_3$ has the opposite sign.

The form of each Regge trajectory is
\begin{equation}
\alpha_i\left(t\right) = 1 + \epsilon_{i} + \alpha_{i}^{\prime} t
\end{equation}
where $\alpha_{i}^{\prime}$ is the slope, and $1+ \epsilon_i$ is the intercept at $t = 0$. 

We extend this single scattering model to include double Pomeron exchange~\cite{D-L-arxiv,JamesThesis}, which is necessary to account for the observed dip in the elastic differential cross section. The appropriate term\footnote{The full set of equations can be obtained from the authors.} is included our computed amplitude; it involves no new parameters apart from parameterising higher order scattering terms
not included through scaling the double scattering amplitude by a factor $\lambda$.


\subsubsection{The full model }
Combining the contributions to equation~(\ref{eq:simple}) gives the differential cross section for elastic scattering as
\begin{eqnarray}
\label{eq:full}
\frac{d\sigma}{dt}=\pi \left[A_c\left(s,t\right)\right]^2 + \frac{1}{4\pi}\left(\mathbb{R}e\left[A_n\left(s,t\right)\right]^2 
+ \mathbb{I}m\left[A_n\left(s,t\right)\right]^2\right) &+& \nonumber \\
\left(\rho+\alpha_{em}\phi\right)A_c\left(s,t\right)~\mathbb{I}m\left[A_n\left(s,t\right)\right],
\end{eqnarray}
where the optical theorem has been used for the cross term and we define the ratio of the real and imaginary components of the nuclear term as $\rho$.
The real and imaginary terms are the sum of all the corresponding terms from Regge exchange, and both single and double Pomeron exchange.
We have assumed that $\alpha\phi$ is small, so that the exponential term could be expanded.
For $p\bar{p}$ scattering, the sign of the terms must be inverted for the $C=-1$ $\omega$ and $\rho$ trajectory, and the triple gluon term.

\subsection{The elastic model fit}
Using this model, we fit all suitable available elastic data. Since the electromagnetic cross section diverges as~$t \longrightarrow 0$, a minimum $t$ value must be defined otherwise the integrated cross section will be infinite.


Using MINUIT within ROOT~\citep{ROOTcite}, a global fit is performed over the data shown in table~\ref{tab:elastic_data_sources}, where $\sqrt{s}~>~23$~GeV. At and below this value, the fit quality of the model starts to degrade.

In the fitting, both the Regge trajectories in the model are ``effective'' trajectories and are initialized with values taken from a Chew-Frautschi plot.
The Pomeron trajectories, $X_i$ factors, and both $\lambda$ and $t_0$  are taken as free parameters, making a
total of 14 parameters. However the Regge trajectory slopes are fixed to control the stability of the fit.  
Full systematic uncertainties are taken into account and correlations between experimental data sets are included.
The fit is performed over all data, and over the full $t$ range available, yielding a $\chi^2/NDF = 4.00$.
This overall figure covers a considerable variation: for many datasets the fit quality is acceptable ($\chi^2/NDF \sim 1$) but there are some features of some datasets where the model and the data systematically disagree in terms of the statistical errors, which are, particularly at low $t$ in the peak, sometimes very small.

The resulting fit parameters and uncertainties given by MINUIT are given in table~\ref{tab:fit_parms}. We note the two Pomeron intercepts are both soft ($\epsilon_0$ and $\epsilon_1$) - in essence we started off with a hard and a soft Pomeron and the fit to the data pushed them together, leaving a single soft Pomeron~\cite{sandyplb1,sandyplb2} with a complicated
$t$ dependence. 


\begin{table}[h]
\begin{center}
\caption{\label{tab:fit_parms}%
The fitted parameters for the elastic scattering model.
}
\begin{tabular}{ccc}
\hline
\textrm{Parameter}&
\textrm{Value}&
\textrm{Fit uncertainty}\\
\hline
$X_0$ & 228 & 12\\
$X_1$ & 194 & 2 \\
$X_2$ & 519  & 24\\
$X_3$ & 10.8 & 3.3\\
\hline
$\epsilon_0$ & 0.1062 & 0.0007\\
$\epsilon_1$ & 0.0972 & 0.0002\\
$\epsilon_2$ & -0.511 &  0.007\\
$\epsilon_3$ & -0.3 &  0.05\\
\hline
$\alpha_0^{\prime}$ & 0.045 & 0.003\\
$\alpha_1^{\prime}$ & 0.28 & 0.001\\
$\alpha_2^{\prime}$ & 0.82 & Fixed\\
$\alpha_3^{\prime}$ & 0.90 & Fixed\\
\hline
$\lambda$ & 0.5212 & 0.0006\\
$t_0$ & 5.03 & 0.01\\
\hline
\end{tabular}
\end{center}
\end{table}

\subsection{Total and differential elastic cross section}

The total elastic cross section is usually quoted as the contribution from the nuclear term only,
\begin{equation}
\sigma_{el} = \pi \int \left|f_{n} \left(t\right)\right|^2 dt.
\end{equation}
Our values, calculated by integration over the differential cross section according to this convention, for
the total elastic $pp$ cross sections at LHC energies  are listed in table~\ref{tab:elasticTot}.

\begin{table}[h]   
\caption{\label{tab:elasticTot} The integrated elastic proton-proton cross section obtained from integration over the differential cross section at LHC energies.}
\begin{center}
\begin{tabular}{ccc}
\hline
\textrm{Energy [GeV]}&
\textrm{$\sqrt{s}$ [GeV]}&
\textrm{$\sigma_{el}$ [mb]}\\ \hline
450 &  29.0 & 6.8\\
3500 & 81.0 & 8.1\\
4000 & 83.9 & 8.2\\
7000 & 114.6 & 8.8\\
\hline
\end{tabular}
\end{center}
\end{table}

\begin{figure}
\includegraphics[width=0.5\textwidth]{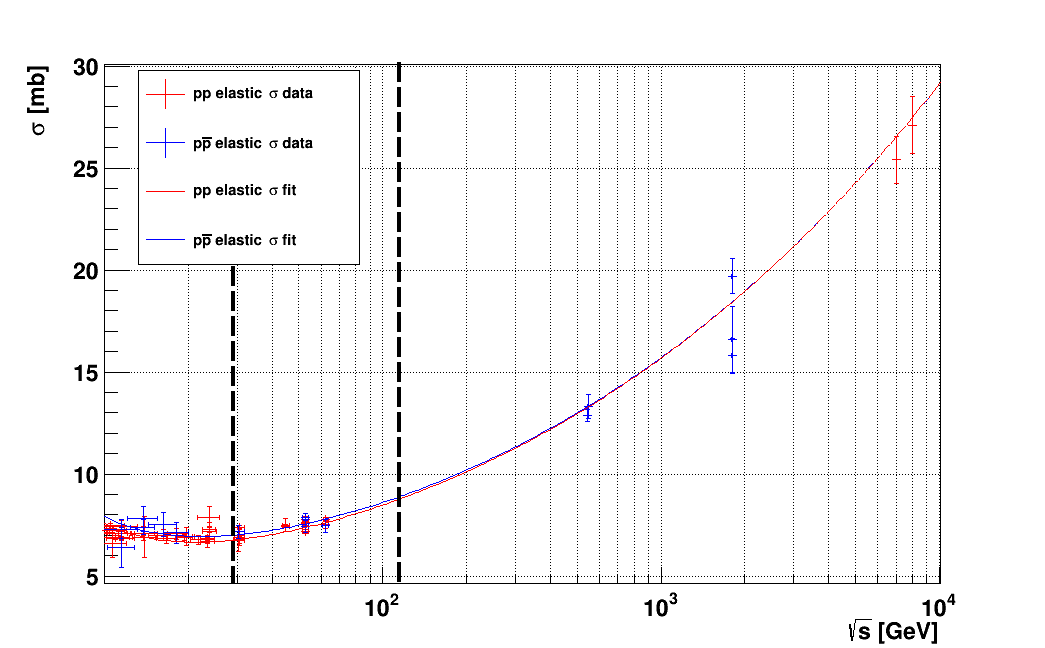}
\caption{\label{fig:elastictot_full_pp}The total elastic cross section fit against experimental $pp$ and $p\bar{p}$ data. The black dashed vertical lines at  29 and 114.6 GeV ($\sqrt{s}$), that is a beam energy of 450 and 7000~GeV,  show the energy range of interest for the LHC collimation system.}
\end{figure}

In Figure~\ref{fig:elastictot_full_pp} we show a subset of results for the total elastic cross section fit against experimental $pp$ and $p\bar{p}$ data. The black dashed vertical lines at 450 and 7000 GeV show the energy range of interest for the LHC collimation system. 

\begin{figure}
\includegraphics[width=0.5\textwidth]{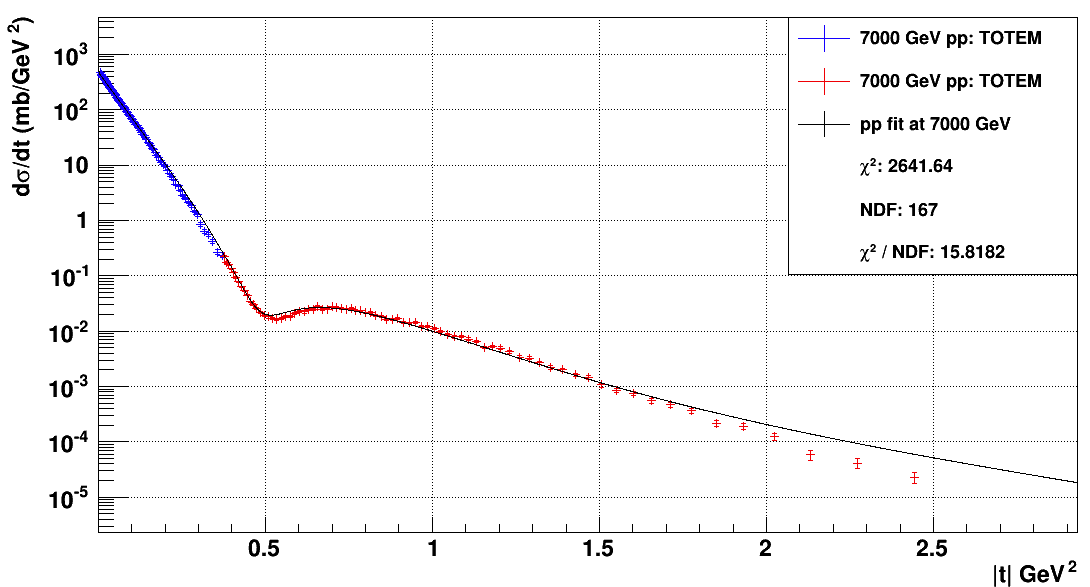}
\caption{\label{fig:elastic_full_pp7000}The $pp$ elastic scattering model model fit (fitted over all data) shown
for $\sqrt{s}=$7000 GeV,  over the full published $t$ range.}
\end{figure}
\begin{figure}
\includegraphics[width=0.5\textwidth]{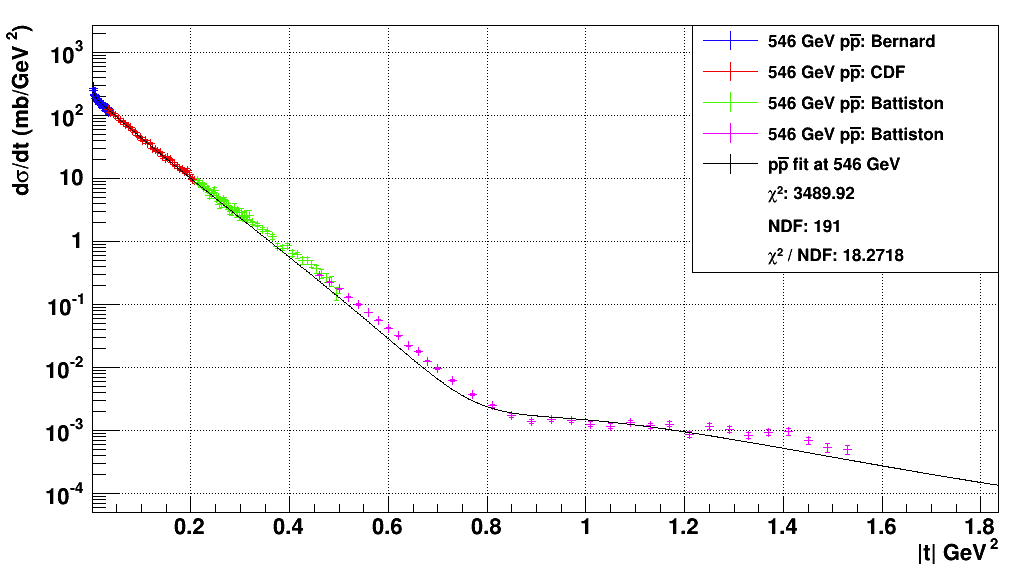}
\caption{\label{fig:elastic_full_pp23}The $p\bar{p}$ elastic scattering model model fit (fitted over all data) shown
for $\sqrt{s}=$546 GeV,  over the full published $t$ range.}
\end{figure}
\begin{figure}
\includegraphics[width=0.5\textwidth]{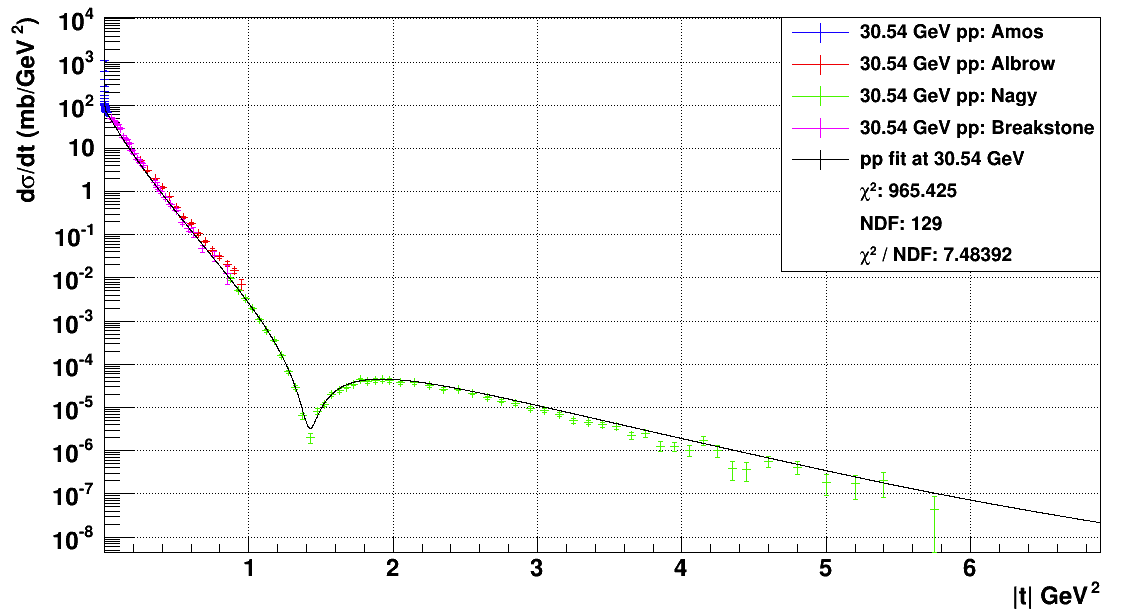}
\caption{\label{fig:elastic_full_pp3054}The $pp$ elastic scattering model model fit (fitted over all data) shown
for $\sqrt{s}=$30.54 GeV,  over the full published $t$ range.}
\end{figure}

Figure~\ref{fig:elastic_full_pp7000} shows the fit for $pp$ elastic scattering at $\sqrt{s} = 7000$~GeV using TOTEM data from the LHC, showing the fit performance at very high energy. Figure~\ref{fig:elastic_full_pp23} shows the fit for $p\bar{p}$ elastic scattering from $\sqrt{s} = 546$~GeV, showing the fit performance in an energy range just above
the range for proton interactions in collimators. Finally,  figure~\ref{fig:elastic_full_pp3054} shows the fit for $pp$ elastic scattering from $\sqrt{s} = 30.54$~GeV, showing the accurate description of the elastic dip for this kinematic region.

The remaining plots for  $pp$ and  $p\bar{p}$ elastic scattering are presented in~\ref{sec:app_datalist}.

All figures show the combined systematic and statistical errors. 
Each plot shows the differential $d\sigma/dt$ distribution, with each experimental data set in a different colour. The black line is the fitted function with the parameters given in table~\ref{tab:fit_parms}. 
The normalisations given in~\ref{sec:app_datalist} have been applied to the data. 

\section{\label{sec:part4}Single diffraction dissociation}
Single diffraction dissociation in $p p$ interactions is the process 
\begin{equation}
p p \to p X,
\end{equation}
in which one proton breaks up into a system $X$  while the other scatters elastically. 
Diffractive kinematics are described by $s$, ~$t$ and $M_X$. In fixed-target SD events at LHC energies $M_X$ can vary from $M_p + m_{\pi}$ to more than 50~GeV. 

\begin{figure*}
\includegraphics[width=0.7\textwidth]{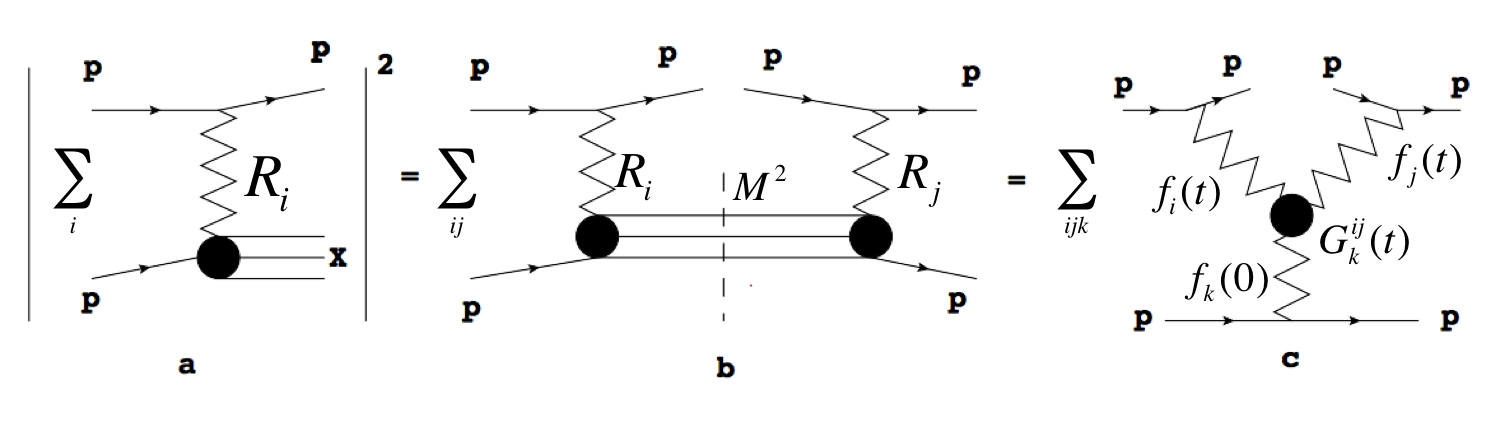}
\caption{\label{fig:3RR}The triple-Regge description of high-mass diffractive dissociation. a) The squared amplitude summed over all possible system X in the large triple-Regge limit ($s \gg M_X^2 \gg |t|$). b) The discontinuity across $M_X^2$ of the scattering amplitude. c) The total cross section $d^2\sigma / dtd\xi$ as the sum of triple-Reggeon contributions.}
\end{figure*}

The simplest description of high energy process is given in the diagram of figure \ref{fig:3RR}(a) in which a Reggeon or a Pomeron is 
exchanged between the elastically-scattered proton and the system $X$.  In the limit $s \gg M_X^2 \gg |t|$ and $M_X^2$ not too small the process may be 
described by the triple-Regge model \cite{Gribov,D-L1,QCDDonnachie,DL_2003,LKMR09} as illustrated in Figure~\ref{fig:3RR}(c) and discussed in section~\ref{sec:part4-1}. 
For small values of the missing mass $M_X$, around a few GeV,  the system $X$ is dominated by baryon resonances and requires a different treatment. A simple model~\cite{sandyup}, based on duality arguments,
allows us to extend the fit to low mass where existing data are scarce. This is discussed in section \ref{sec:part4-2}.

The advent of the LHC has renewed interest in diffraction dissociation \cite{Kaidalov,Ostapchenko,Gotsman,Ryskin}. The associated models go beyond
the simple triple-Regge model, principally by the inclusion of absorptive corrections, and they are successful in describing the total single diffraction cross section and, in some cases, the double differential cross section $d^2\sigma/dtd\xi$ at small $t$. In one sense our approach is less ambitious in that we use the standard triple-Regge model without modification. In another sense, however, it is much more ambitious as we attempt, successfully, to describe all existing single diffractive dissociation data in $p p$ interactions. 

\subsection{\label{sec:part4-1} High mass: triple-Regge formalism}

The triple-Regge description, shown in figure~\ref{fig:3RR}, describes the $pp$ SD cross section in the region of high $\xi$ as the sum of contributions from triple-Regge exchanges~\cite{D-L1} (and applied to the LHC in~\cite{LKMR09}). In figure \ref{fig:3RR}(c) each of the upper exchanges carry momentum transfer~$t$ while the lower one carries zero momentum transfer; the $f_i(t)$, $i=1,2,3$
are the couplings of the exchanges to the relevant hadrons and $G^{12}_3(t)$ is the triple-Reggeon vertex. In addition to the Pomeron, Reggeised $f_2$, $a_2$, $\omega$, $\rho$ exchanges are allowed so in principle we require a whole series of terms, given by
\begin{equation}
 \label{eq.4}
  \begin{tabular}{cccccccc}
    $\mathbb{P P}$ &$\mathbb{P P}$& $\mathbb{R R}$ & $\mathbb{R R}$ & $\mathbb{R P}$ &$\mathbb{P R}$ & $\mathbb{R P}$ & $\mathbb{P R}$\\
   $\mathbb{P}$    &$\mathbb{R}$ & $\mathbb{P }$ & $\mathbb{R}$ &  $\mathbb{P}$  &$\mathbb{P}$ & $\mathbb{R}$ & $\mathbb{R}$ 
 \end{tabular}.
\end{equation}
  
Here $ \mathbb{P}$ refers to the Pomeron and $ \mathbb{R}$ to any of $f_2$, $a_2$, $\omega$ and $\rho$ trajectories. A term $(^{1~2}_{~3})$ contributes to $d^2\sigma/dtd\xi$ as
\begin{equation}
\resizebox{.95\hsize}{!}{$f_1(t)f_2(t)f_3(0)G^{12}_3(t)e^{i(\phi(\alpha_1(t))-\phi(\alpha_2(t)))}
\xi^{1-\alpha_1(t)-\alpha_2(t)}\Big(\frac{M_X^2}{s_0}\Big)^{\alpha_3(0)-1}$}.
\label{eq:contrib}
\end{equation}

Here the $\alpha_i(t)$, $i=1,2,3$ are the relevant Pomeron and Reggeon trajectories, $s_0$ is a scale factor that we take to be 1 GeV$^2$ and $\phi(\alpha(t))$ is the Reggeon phase. In practice only the first four terms 
of the series are required \cite{DL_2003}, those being $(^{\mathbb{P}~\mathbb{P}}_{~\mathbb{P}})$, $(^{\mathbb{P}~\mathbb{P}}_{~\mathbb{R}})$, $(^{\mathbb{R}~\mathbb{R}}_{~\mathbb{P}})$ and $(^{\mathbb{R}~\mathbb{R}}_{~\mathbb{R}})$. Further it is sufficient to consider the $f_2$,  $a_2$, $\omega$ and $\rho$ trajectories to be degenerate and use a single 
generic Reggeon exchange.  The differential cross section may be written as

\begin{eqnarray}
\frac{\partial^2 \sigma}{\partial t \partial \xi} &=& g_{\mathbbm{PPP}}(t)s^{\alpha_{\mathbbm{P}}(0)-1}\xi^{\alpha_{\mathbbm{P}}(0)-2\alpha_{\mathbbm{P}}(t)} + \nonumber\\
 & & g_{\mathbbm{PPR}}(t)s^{\alpha_{\mathbbm{R}}(0)-1}\xi^{\alpha_{\mathbbm{R}}(0)-2\alpha_{\mathbbm{P}}(t)} + \nonumber \\
 & & g_{\mathbbm{RRP}}(t)s^{\alpha_{\mathbbm{P}}(0)-1}\xi^{\alpha_{\mathbbm{P}}(0)-2\alpha_{\mathbbm{R}}(t)} + \nonumber \\
 & & g_{\mathbbm{RRR}}(t)s^{\alpha_{\mathbbm{R}}(0)-1}\xi^{\alpha_{\mathbbm{R}}(0)-2\alpha_{\mathbbm{R}}(t)}, \nonumber \\
\label{eqn:triple_regge_model}
\end{eqnarray}
with
\begin{equation}
g_{iik}(t) = f_i(t)^{2} f_k(0)G^{ii}_{k}(t),
\end{equation}
where $i$ and $k$ denote $\mathbbm{P}$ or $\mathbbm{R}$ as appropriate.

The Pomeron trajectory~\cite{DL_2003} is given by
\begin{eqnarray}
 \alpha_{\mathbbm{P}}(t) &=& 1+ \epsilon_{\mathbbm{P}}+ \alpha_{\mathbbm{P}}'t \\
  & =  &1 + 0.08 + 0.25t, 
\label{eqn:pom_trajectories}
\end{eqnarray}
and  
\begin{eqnarray}
 \alpha_{\mathbbm{R}}(t) & = & 1+ \epsilon_{\mathbbm{R}}+ \alpha_{\mathbbm{R}}'t \\
& = &  1 - 0.45 + 0.93t,  \\
\label{eqn:preg_trajectories}
\end{eqnarray}
is an ``effective'' reggeon trajectory, which is a reasonable average of the $f_2$, $a_2$, $\omega$, $\rho$ trajectories.

For the fitting procedure various parametrisations have been tested for the function $g_{iik}(t)$, 
\begin{eqnarray}
g_{iik}(t) & = & A_{i} e^{B_{k} t} \label{eq:expct1} \\
  g_{iik}(t) & = & A_{i} e^{B_{i}t} + C_{k} \label{eq:expct2} \\
    g_{iik}(t) & = & A_{i} e^{B_{i} t + C_{k} t^2} \label{eq:expct3} \\
    g_{iik}(t) & = & \lambda_i \left(\frac{t+A_{i}}{t+B_{i}}\right)^{C_{k}} F(t)^2 \label{eq:expct4} .
\end{eqnarray}
The most effective of these parametrisations, and the one we use here, is given by equation~(\ref{eq:expct2}), where $A_{i}, B_{i}$ and $C_{k}$ are free parameters given by the fit to the data. The chosen parametrisation works perfectly at low energy but gives too high a cross section at $\sqrt{s}$ = 546/640 GeV. For this reason the triple Pomeron coupling term $g_{\mathbbm{PPP}}(t)$ is parametrized in a different way based on the possibility that it vanishes at $t=0$~\cite{luna2010} 
\begin{equation}
\left(A_{i} e^{B_{i}t} + C_{k}\right) \left(\frac{t}{t+t_0}\right),
\end{equation}
where $t_0 = -0.05$ GeV$^2$ is the optimum value.  

At small $t$, where the triple-Pomeron coupling term dominates, the vanishing term in the parametrisation improves the double differential cross section fit but it reduces the agreement of the differential cross section $d\sigma/dt$ with high-energy data. At the same time, at high energy and high $t$, the predicted differential cross section is lower than the data and this simple parametrisation fails. To improve the fit for both differential and double differential cross section the triple-Pomeron coupling parametrisation is divided into three different regions of $t$. For $-0.25 \le t < -0.0001$ we use
\begin{equation}
g_{\mathbbm{PPP}}(t)  = (0.4 + 0.5t),
\end{equation}
for $-1.15 \le  t < -0.25$ we use
\begin{equation}
g_{\mathbbm{PPP}}(t)  = (A_{\mathbbm{P}}e^{B_{\mathbbm{P}}t}+C_{\mathbbm{P}}) \left(\frac{t}{t-0.05}\right),
\end{equation}
and for $-4.00 \le  t < -1.155$ we use
\begin{eqnarray}
g_{\mathbbm{PPP}}(t)  =(A_{\mathbbm{P}}e^{B_{\mathbbm{P}}t}+C_{\mathbbm{P}})\left(\frac{t}{t-0.05}\right) \times \nonumber \\
 \left(1+0.4597(|t|-1.15) + 5.7575(|t|-1.15)^2\right).
\end{eqnarray}


We use a linear parameterisation at low $|t|$ to avoid unphysical behaviour and a modified form at high $|t|$ to increase the integrated cross section. 

One additional term is required: a ``Reggeized''  pion exchange term which is important at low $t$~\cite{DL_2003}, this term (\ref{pion_ex}) is kept fixed during the fitting procedure. In Regge theory the pion exchange term is given by
\begin{equation}
\frac{\partial^2 \sigma_{\pi}}{\partial t \partial \xi} = \frac{g^2_{\pi\pi p}}{16\pi^2} \frac{|t|}{(t-m_{\pi})^2}F^2(t)\xi^{1-2\alpha_{\pi}(t)}\sigma_{\pi^0 p}(s\xi)
\label{pion_ex}
\end{equation}
where $g^2_{\pi\pi p}/4\pi$ = 14.4~\cite{Field1974367} is the on-mass-shell coupling, $m_{\pi}$ is the pion mass, $\alpha_{\pi}(t)=0.93(t-m^2_{\pi})$ is the pion trajectory and $F^2(t)$ is the proton form factor.
In equation~(\ref{pion_ex}), $\sigma_{\pi^0 p}(s\xi)$ [mb], denotes the pion-proton cross section, modelled by~\cite{Field1974367}
\begin{equation}
\sigma_{\pi^0 p}(s\xi) = 13.63 (\xi s)^{0.0808} + 31.79 (\xi s)^{-0.4525},
\label{eq:sigma_pion_p}
\end{equation}
so the overall SD double differential cross for high missing mass can be written as:
\begin{eqnarray*}
\frac{\partial^2 \sigma^{HM}}{\partial t \partial \xi} &=& g_{\mathbbm{PPP}}(t)s^{\alpha_{\mathbbm{P}}(0)-1}\xi^{\alpha_{\mathbbm{P}}(0)-2\alpha_{\mathbbm{P}}(t)} + \nonumber\\
 & & g_{\mathbbm{PPR}}(t)s^{\alpha_{\mathbbm{R}}(0)-1}\xi^{\alpha_{\mathbbm{R}}(0)-2\alpha_{\mathbbm{P}}(t)} + \nonumber \\
 & & g_{\mathbbm{RRP}}(t)s^{\alpha_{\mathbbm{P}}(0)-1}\xi^{\alpha_{\mathbbm{P}}(0)-2\alpha_{\mathbbm{R}}(t)} + \nonumber \\
 & & g_{\mathbbm{RRR}}(t)s^{\alpha_{\mathbbm{R}}(0)-1}\xi^{\alpha_{\mathbbm{R}}(0)-2\alpha_{\mathbbm{R}}(t)} + \nonumber \\
 & & \frac{g^2_{\pi\pi p}}{16\pi^2} \frac{|t|}{(t-m_{\pi})^2}F^2(t)\xi^{1-2\alpha_{\pi}(t)}\sigma_{\pi^0 p}(s\xi).
\label{eqn:triple_regge_model}
\end{eqnarray*}
The fitting procedure described in section~\ref{sec:part4-4}.

\subsection{\label{sec:part4-2}Low mass: background and resonances}

The single-diffractive dissociation at low mass is a delicate issue in diffractive dissociation studies. A lot of $pp\rightarrow pX$ data in the resonance region is available at very low energy, much of which is not relevant, but some is used as a guide. Useful information to model the resonance region comes from data at $s$ = 565~GeV$^2$ for $t$ = -0.05~GeV$^2$~\cite{schamdiffractive}, where the $d^2\sigma/d\xi dt$ are averaged over $1.5 \le M_X^2 \le 2.5 $GeV$^2$. 

Both Pomeron and Reggeon exchange conserve helicity, so resonance excitation is primarily through incremental angular momentum with no change in the quark spin.
On the basis of the Gribov-Morrison rule~\cite{Gribov,Morrison} we expect the resonance to have spin-parity $(1/2)^+$,$(3/2)^-$,$(5/2)^+$,$(7/2)^-$ etc.  Also the dominant exchanges (Pomeron, $f_2$, $\omega$) are isoscalar, so we expect that the leading resonances produced are $P_{11}(1440)$, $D_{13}(1520)$, $F_{15}(1680)$ and $G_{17}(2190)$.
The background to these leading resonances comes from the low-mass continuation of the high-mass model of Pomeron and Reggeon exchange.

Hadron-hadron scattering at low energies can be described by the sum of a few amplitudes for direct $s$-channel production; as the energy increases these resonances increasingly overlap and more amplitudes need to be considered. At high energies it can be described by the sum of a few simple Reggeon exchange amplitudes in the $t$-channel; as the energy falls more of these are required. The principle of duality asserts that these are two descriptions of the same physics, valid at lower and higher energies.  

This can be extended to the principle of two component duality~\cite{Field1974367} in which the $s$-channel amplitudes comprise a smooth background which is dual to Pomeron exchange, and a set of resonances which is dual to 
Reggeon (non-Pomeron) exchanges.  

We determine the background term first. We assume it is quadratic, unlike a previous analysis~\cite{schamdiffractive} which used a
general polynomial, as this can match the triple-Regge form. The contribution vanishes at threshold,
$\xi_{th}=(M_p + m_\pi)^2/s$, and can therefore be written
\begin{equation}
B(\xi,t,s)=a(t,s)(\xi-\xi_{th})^2 + b(t,s)(\xi-\xi_{th}).
\end{equation}
$a$ and $b$ are then determined by requiring that the background matches smoothly onto the high mass region at some chosen value $\xi_c$ which represents the division between `low' and `high' mass.

Writing the triple-Regge function ${\partial^2 \sigma^{HM} \over \partial  t \partial \xi} (\xi,t,s)$ as $A(\xi,t,s)$
the boundary conditions can be written as
 \begin{eqnarray}
A(\xi_c,t,s) & = & B(\xi_c,t,s) \\
A'(\xi_c,t,s) & = & B'(\xi_c,t,s) 
\end{eqnarray}
and the resulting background coefficients are given by 
\begin{eqnarray}
a(t,s) & = & \frac{(\xi_c-\xi_{th}) A'(\xi_c,t,s) - A(\xi_c,t,s)}{(\xi_c-\xi_{th})^2}\\
b(t,s) & = & 2{A(\xi_c,t,s) \over \xi_c-\xi_{th}} - A'(\xi_c,t,s).
\end{eqnarray}

To complete the low mass model we now add the resonances contribution.
Each baryon resonance is parametrised by a Breit-Wigner function, with a mass $m_l$ and a width $\gamma_l$ .
So the total contributions from resonances to the SD cross section at low-mass is given by 
\begin{equation}
\frac{d \sigma_{Res}}{d M_{X}^{2}} = \sum_{l=1}^4\left[\frac{c_l}{M_X^2}\frac{m_l \Gamma_l}{(M_X^2-m_l^2)^2+(m_l \Gamma_l)^2}\right],\\
\label{eq:resonance}
\end{equation}

with
\begin{equation}
\Gamma_l  =  \gamma_l\left(\frac{q}{q_l}\right)^{2l+1} \left(\frac{1+5q_l}{1+5q}\right)^l, \\
\label{eq:Gamma_l}
\end{equation}

where $q$ and $q_l$ are respectively the 3-momenta at $M_X$ and $m_l$ in the resonance rest frame, assuming that $\pi p$ is the dominant final state. They are given by

 \begin{eqnarray}
q(M_X^2) & = &  \sqrt{\frac{(M_X^2 - (M_p + M_{\pi})^2)(M_X^2 - (M_p - M_{\pi})^2)}{4M_X^2}}\\
q_l & = & \sqrt{\frac{(m_l^2 - (M_p + M_{\pi})^2)(m_l^2 - (M_p - M_{\pi})^2)}{4m_l^2}}.
\end{eqnarray}
 The data from~\cite{schamdiffractive} for $t$ = -0.05 GeV$^2$ at $\sqrt{s} = 23.7$~GeV are fitted as a sum of the background and these four leading resonances.



The $t$ dependence in the resonance region comes from Schamberger~\cite{schamdiffractive}. For $1.5 \le M_X^2 \le 2.5 $ GeV$^2$ , $d\sigma/ dt \approx$ exp$((13.2\pm0.3) t)$.
 As there is some slight $t$-dependence in the background the double-differential cross section is obtained by multiplying equation~(\ref{eq:resonance}) by~exp$(13.5(t+0.05))$
In terms of the variable $\xi$, this is 
\begin{equation}
\resizebox{.995\hsize}{!}{$\frac{\partial^2 \sigma_{Res}}{\partial \xi \partial t}(\xi,t,s) = e^{13.5(t+0.05)}\sum_{l=1}^4\left[\frac{c_l}{\xi}\frac{m_l \Gamma_l}{(\xi s-m_l^2)^2+(m_l \Gamma_l)^2}\right].$}
\label{eq:resonance_tot}
\end{equation}

If the resonance contribution is simply added to the background there will be a small step between high and low mass regions at $\xi=\xi_c$. This is remedied by subtracting a small matching term $R_m$ linear in $\xi$, which is zero at threshold and 
equal to the magnitude of the resonance term at the matching point,
\begin{equation}
R_{m}(\xi,t,s) = -\frac{\partial^2 \sigma_{Res}}{\partial \xi \partial t}(\xi_c,t,s) \frac{\xi - \xi_{th}}{\xi_c - \xi_{th}}.
\label{eq:resonance_match}
\end{equation}
The total resonance contribution can then be written as
\begin{equation}
R(\xi,t,s) = \frac{\partial^2 \sigma_{Res}}{\partial \xi \partial t}(\xi,t,s) + R_{m}(\xi,t,s),
\label{eq:total_res}
\end{equation}
and the complete single-diffractive double differential cross section at low mass is given by
\begin{equation}
\frac{\partial^2 \sigma^{LM}}{\partial t \partial \xi}(\xi,t,s) =  R(\xi,t,s) + B(\xi,t,s).\\ 
\label{eq:BTot}
\end{equation}

\subsection{\label{sec:part4-4}Fitting procedures}
The large amount of data on soft diffraction dissociation is given in~\ref{sec:app_datalist2}. It covers the ranges $17.2 < \sqrt{s} < 546$~GeV and $0.015 < |t| < 4.15$~GeV$^2$, thus spanning the energies and the range of momentum transfer required. However there are clear inconsistencies of normalisation between different data sets and considerable variation in quality. In some data, for example from the ISR, the experimental resolution is insufficient to delineate clearly the resonance from the triple-Regge region, so fits to these data were restricted to $\xi > 0.01$. 
The twelve parameters of the parametrisation of equations~\ref{eq:expct1}-\ref{eq:expct4}, $g_{iik}$, are obtained from a global fit over all the available data using MINUIT within ROOT~\cite{ROOTcite}. 

Full systematic errors and correlations between experimental data sets are taken into account, and we consider two ways of doing this. 
The data are quoted with statistical and systematic errors. The former are due to Poisson statistics on the number of particles counted.  The latter are dominated by uncertainties in the acceptance, and are common to all measurements made by a given experiment. These errors are strictly multiplicative but they are small enough in practice to be taken as additive, greatly simplifying the analysis.

If the systematic errors are not considered, the quantity to be minimised is $$\chi^2=\sum_i \sum_{j=1}^{N_i} \frac{(f_{ij}-d_{ij})^2}{\sigma_{ij}^2}, $$ where $i$ is the number of the experiment, $j$ the measurement within that experiment's dataset, and $f_{ij}$, $d_{ij}$ and $\sigma_{ij}$ are respectively the fitted function, the measured cross section, and the quoted statistical error.

If the experiment also quotes a systematic uncertainty on the acceptance of $\alpha_i$, so that each measurement has an error $S_{ij}=\alpha_i d_{ij}$, 
it is not possible to merely replace $\sigma_{ij}^2$ in the denominator by $\sigma_{ij}^2+S_{ij}^2$, because although this correctly expresses the variance of an individual point, it does not take into account the correlation between points. One way to allow for the correlations is to include them in the $\chi^2$ using the formula $\chi^2 = (\vec f^T - \vec d^T) {\bf V}^{-1} (\vec f - \vec d)$ where $\bf V$ 
is the covariance matrix for the measurements. This is included in the fitting procedure and gives reasonable results, although it is not clear what the fit is doing to the normalisations of the individual experiments. 

An alternative approach is therefore taken, in which each experiment's results are adjusted by a factor $F_i$. These factors are included as parameters in the fit, with values initially set to 1, and a contribution of ${(F_i-1)^2 \over \alpha_i^2}$ is added to the $\chi^2$ for each such term. 
The results of this procedure are very similar (and, for test cases, identical) and the fitted value of $F$ yielded useful information as to what the fit was doing to the normalisation of the individual experiments.  
$F_i$ values initially set to 1 have been found to vary between 0.9 and 1.15 which is within the 20-25$\%$ of quoted systematic errors between different experiments. 

\begin{figure}[!htb]
\includegraphics[width=0.5\textwidth]{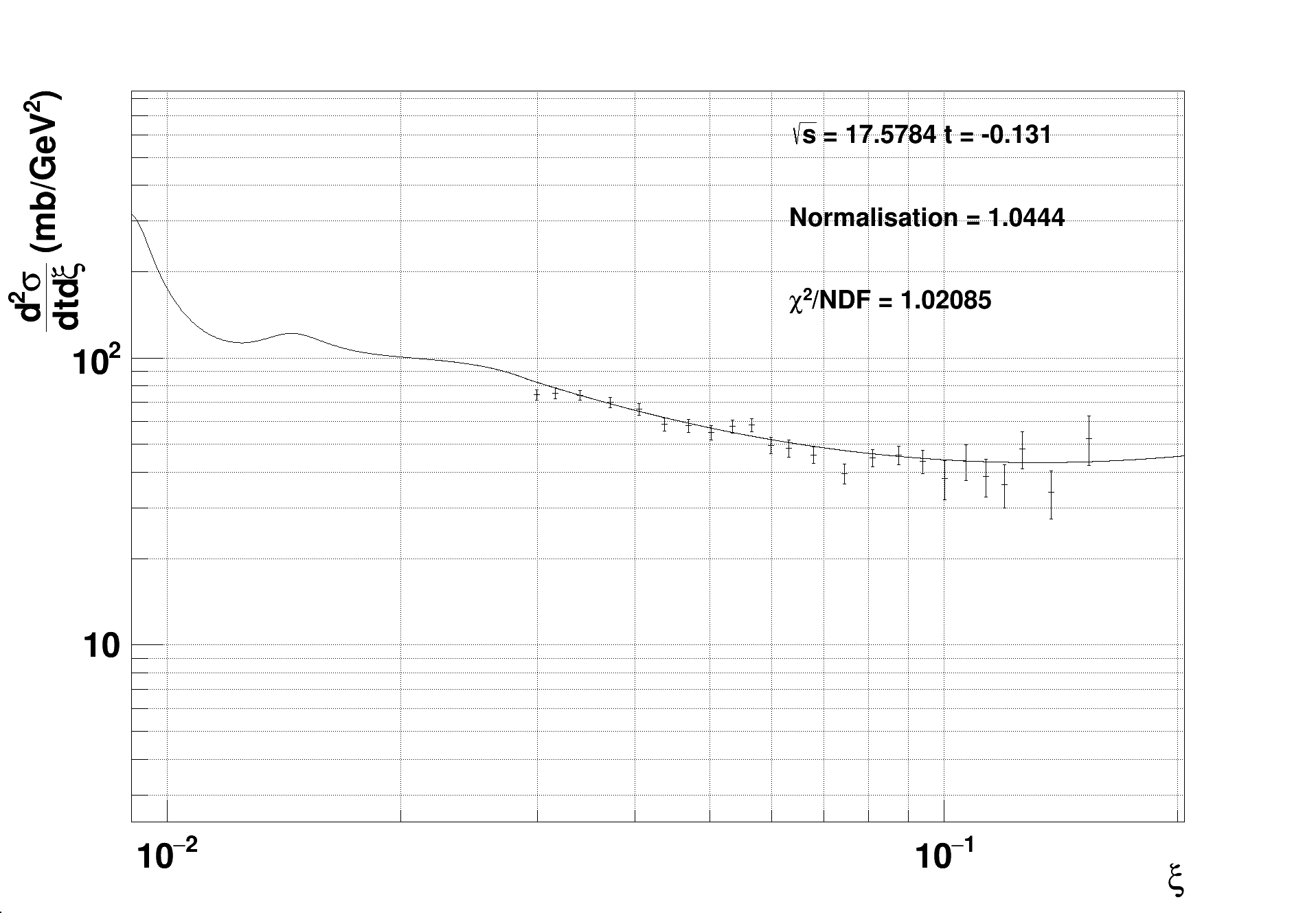}
\caption{\label{fig:309-1} The single diffraction DL model fit shown for $\sqrt{s}$ = 17.57 GeV and $t$ = - 0.131 GeV$^2$.}
\end{figure}

\begin{figure}[!htb]
\includegraphics[width=0.5\textwidth]{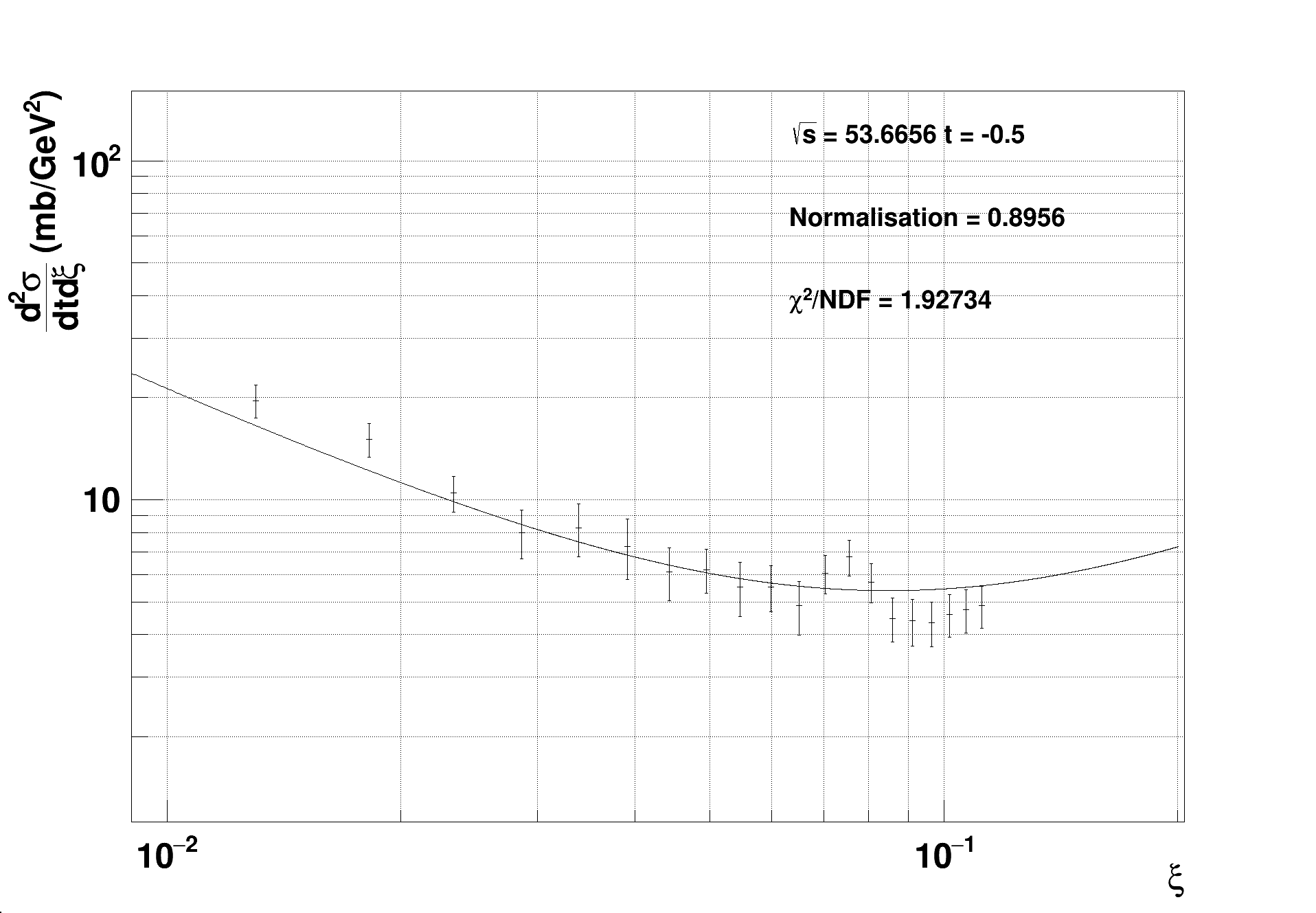}
\caption{\label{fig:3892-1} The single diffraction DL model fit shown for $\sqrt{s}$ = 53.66 GeV and $t$ = - 0.5 GeV$^2$}
\end{figure}

The minimisation is performed over 5562 data points yielding a $\chi^2/NDF = 8.61$.  The fit parameters are listed in table~\ref{tab:Regge_fit_terms} and the coefficients of the resonance contribution $c_l$ in the low mass region arising from the fit  (we do not float the resonance widths or locations) are given in table \ref{tab:Resonance_terms}). The normalisation for each experimental data set is presented in table~\ref{tab:SD_normalisation}.

\begin{table}[h]
\caption{Fit parameters for the triple-Regge model.}
\begin{center}
\begin{tabular}{cccc}
\hline
Term & $ A_i$ & $B_i$ & $C_k$\\ \hline
$\mathbbm{PPP}$ & 0.625 & 2.58 & 0\\
$\mathbbm{PPR}$ & 3.09 & 4.51 & 0.186 \\
$\mathbbm{RRP}$ &  4.00 & 3.03 & 10.0 \\
$\mathbbm{RRR}$ & 177.0 & 5.86 & 21.0 \\
\hline
\end{tabular}
\label{tab:Regge_fit_terms}
\end{center}
\end{table}

\begin{table}[htdp]
\begin{center}
\begin{tabular}{ccccc}
\hline
Resonance & $l$ & $m_l$ [GeV] & $\gamma_l$ & $c_l$\\ 
\hline
$P_{11}$ & 1 & 1.44 & 0.325 & 3.07\\
$D_{13}$ & 2 & 1.52 & 0.130   & 0.415 \\
$F_{15} $&  3 & 1.68 & 0.140   & 1.11 \\
$G_{17} $& 4 & 2.19 & 0.450  & 0.952 \\
\hline
\end{tabular}
\end{center}
\caption{Resonance parameters}
\label{tab:Resonance_terms}
\end{table}

\begin{table}[htdp]
\begin{center}
\begin{tabular}{cc}
\hline
Experiment  & Normalisation\\ 
\hline
 Albrow & 0.8698 \\
Armitage & 0.8956  \\
Schamberger &  1.0444  \\
Cool & 1.1111  \\
Akimov & 1.0629  \\
UA4 & 0.9775  \\
\hline
\end{tabular}
\end{center}
\caption{The high mass diffractive fit experimental normalisation used.}
\label{tab:SD_normalisation}
\end{table}

Typical fits are shown in figures~\ref{fig:309-1} and~\ref{fig:3892-1} for the double differential cross section; other results at different momentum transfer $t$ and energies are reported in~\ref{sec:app_datalist2}.  
  
\begin{figure}[htb]
\centering
  \includegraphics[width=0.5\textwidth]{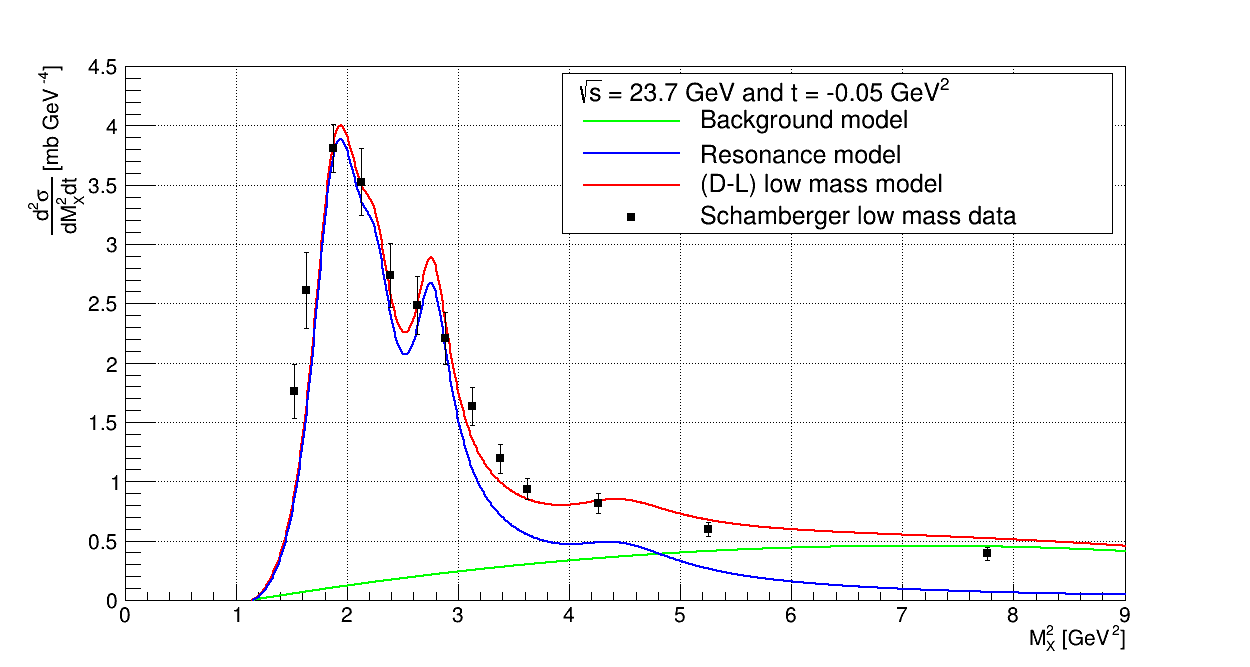}
   \caption{Contribution to $d^2\sigma/dM_X^2 dt$ from the background in blue, the resonance in red and the total in green. The black points represent data from Schamberger\cite{schamdiffractive} for $t$=-0.05~GeV$^2$ at $\sqrt{s} = 23.7$~GeV.}
   \label{fig:ResBckData}
\end{figure}

\begin{figure}[htb]
\centering
  \includegraphics[width=0.5\textwidth]{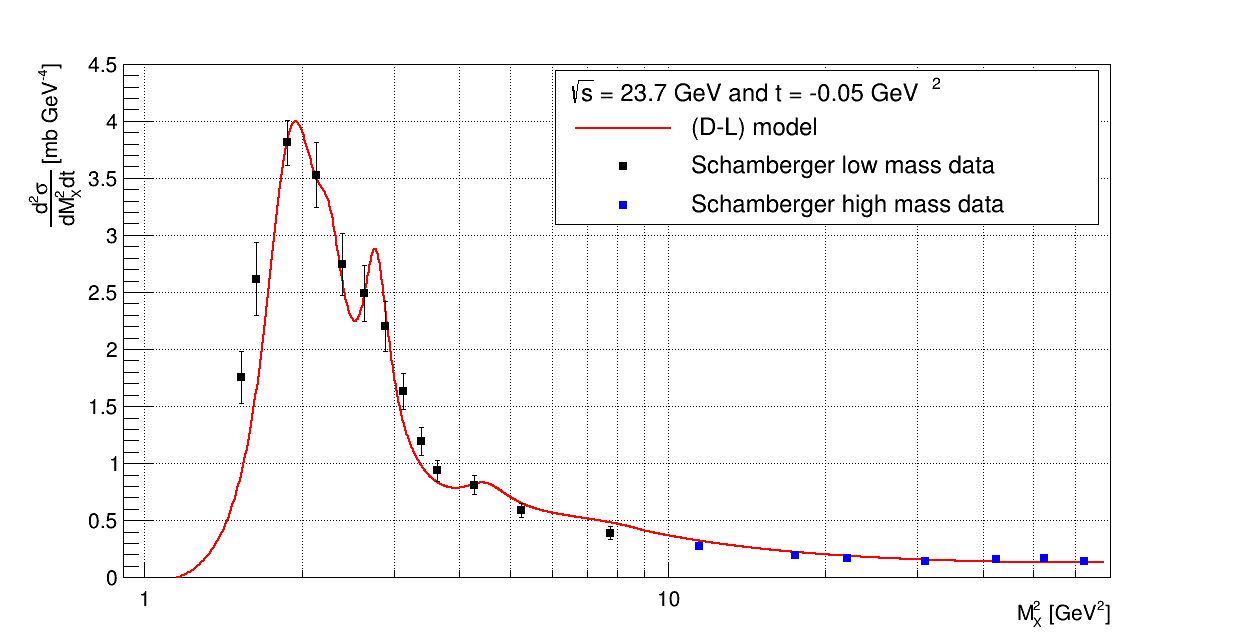}
   \caption{Double differential cross section at $\sqrt{s} = 23.7$~GeV and $t$ = -0.05~GeV$^2$. The red line is the DL model and, black and blue points represent, respectively, the data at low~\cite{schamdiffractive} and high $M_X$~\cite{PhysRevLett.34.1121}.}
   \label{fig:ResBckData2}
\end{figure}

 In figure~\ref{fig:ResBckData} we show the fit of $d^2\sigma^{LM}/ dM_X^2 dt$ to the data\cite{schamdiffractive} at low mass. The background is the green line, the resonance structure is blue and the total in red. 
 The resonance structure of the low mass region is reflected in the data with a strong peak at $P_{11}(1440)$ followed by a decreasing contribution from the remaining resonances. 
 A comparison for the full range of the missing mass $M_X$ is presented in figure~\ref{fig:ResBckData2}; the data at high mass are from~\cite{PhysRevLett.34.1121}.  

\subsection{\label{sec:part4-5} Single differential SD cross section}

\begin{figure}[!htb]
\includegraphics[width=0.5\textwidth]{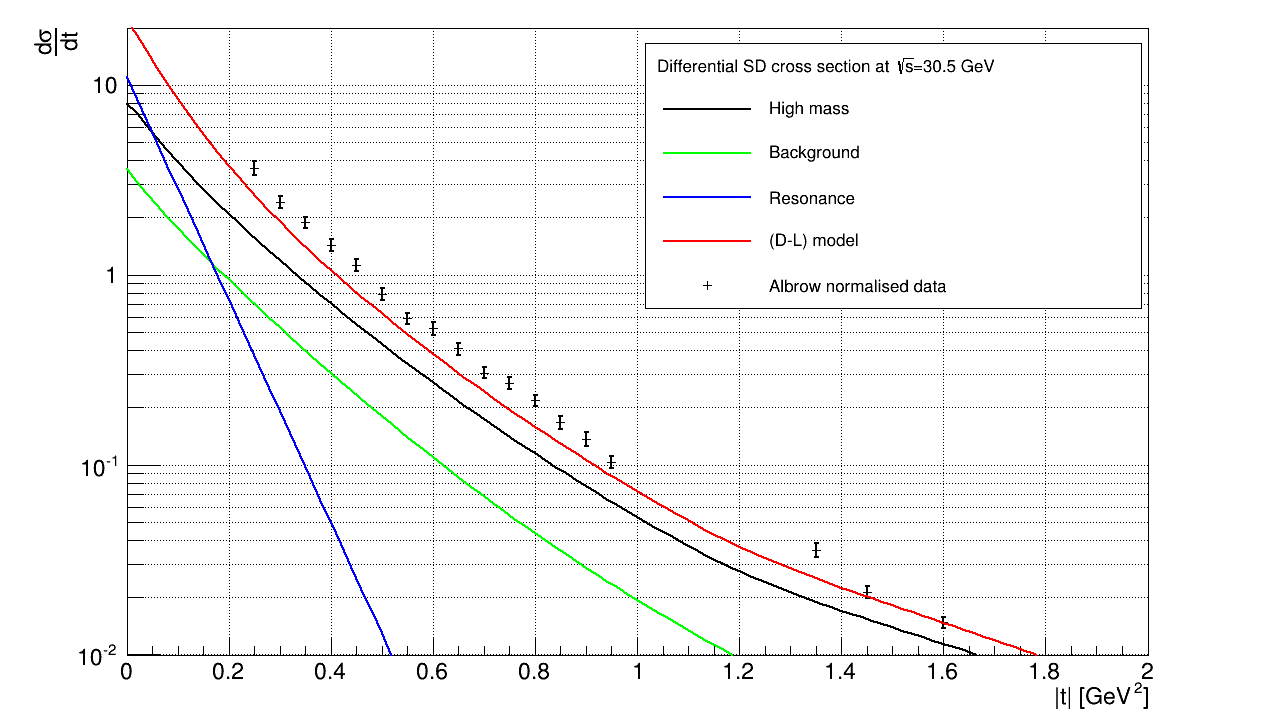}
\caption{\label{fig:934dcs} The single diffraction differential cross section at $\sqrt{s} = 30.5$ GeV. In blue the resonance, in green the background and in black the high mass contribution to the differential cross section. The total DL model is shown in red. The black points represent data from Albrow.}
\end{figure}

\begin{figure}[!htb]
\includegraphics[width=0.5\textwidth]{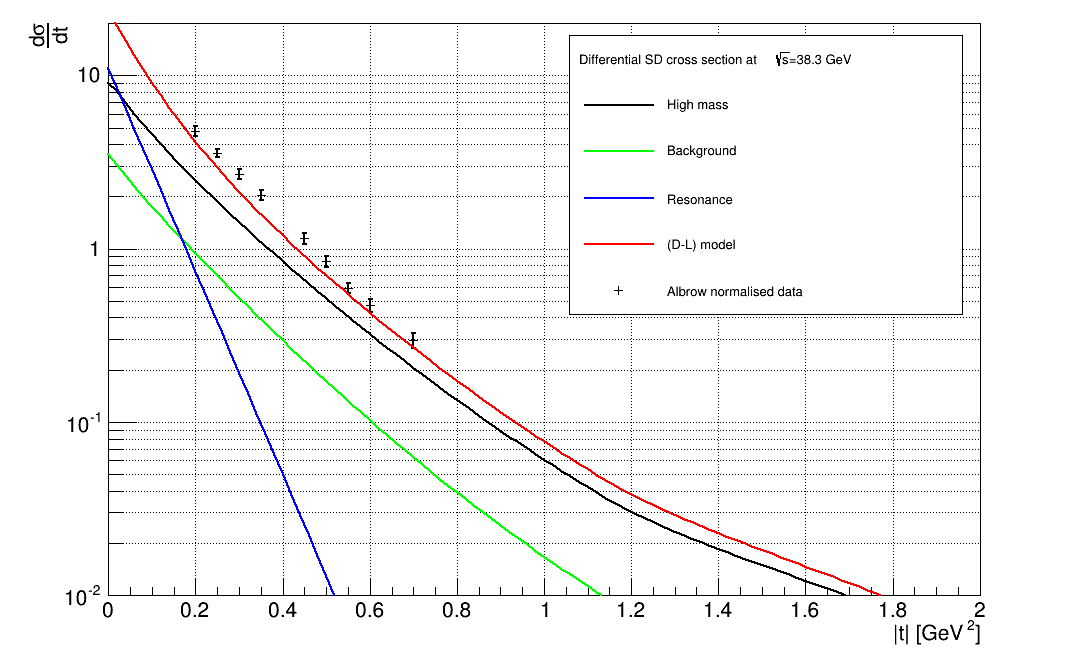}
\caption{\label{fig:1464dcs} The single diffraction differential cross section at $\sqrt{s} = 38.3$ GeV. In blue the resonance, in green the background and in black the high mass contribution to the differential cross section. The total DL model is shown in red. The black points represent data from Albrow.}
\end{figure}

\begin{figure}[!htb]
\includegraphics[width=0.5\textwidth]{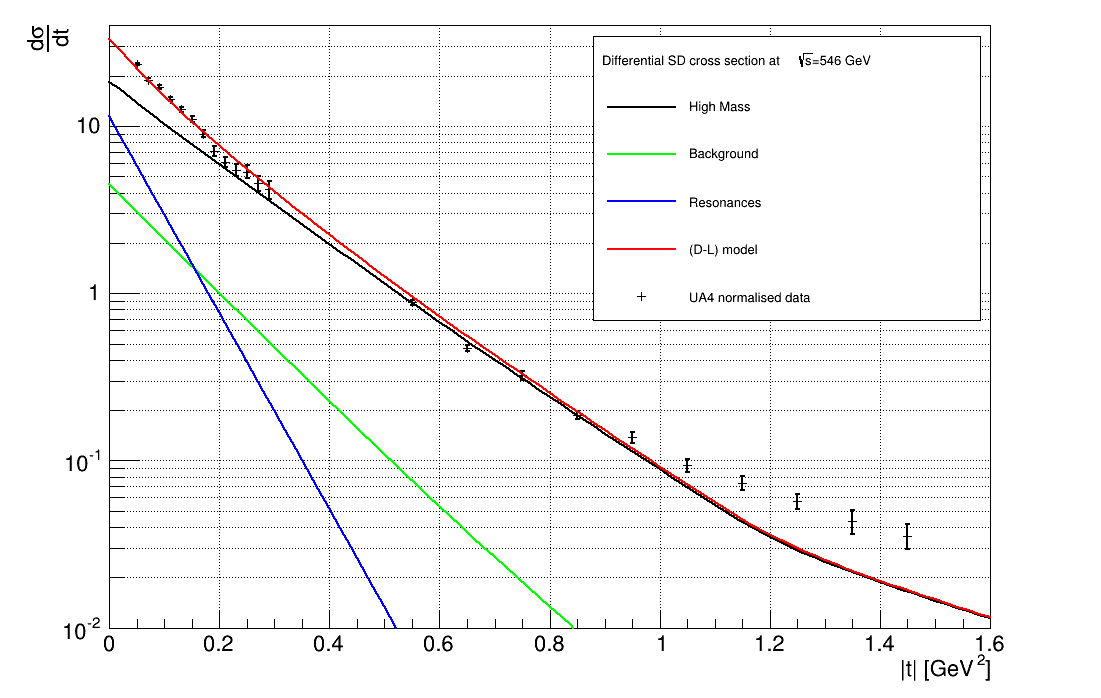}
\caption{\label{fig:UA4dcs} The single diffraction differential cross section at $\sqrt{s} = 546$ GeV. In blue the resonance, in green the background and in black the high mass contribution to the differential cross section. The total DL model is shown in red. The black points represent data from UA4 collaboration.}
\end{figure}

Data for the single differential SD cross section, $d\sigma/dt$ after integration over $\xi$, are available for a large range of energies and are used to compare the fit results and as a guide in the fitting procedure of the double differential cross section data.  Some examples are shown in figure~\ref{fig:934dcs} and \ref{fig:1464dcs} for low energies, $\sqrt{s} $ = 30.5 and 38.3 GeV, and figure~\ref{fig:UA4dcs} for high energy UA4 data~\cite{Bernard1987227}.  We show in blue the resonance contribution, in green the background and in black the high mass contribution to the differential cross section. The total DL model is shown in red. 

The DL model at low energies does not match the experimental data perfectly, due to an underestimated contribution from the resonances.  At low $t$ there is a strong contribution to the differential cross section from resonances and background, for higher $t$ the high mass term dominates. The effect of the correction at low $t$ from the triple-Pomeron term of the high mass contribution is visible, for example, in figure~\ref{fig:UA4dcs}  where the DL model matches the data very well. The modification at high $t$ is a reasonable compromise at low and high energies.

\subsection{\label{sec:part4-6} Integrated SD cross section $s$ dependence}

The energy dependence of the total SD cross section is a controversial topic. For energy ($\sqrt{s}$) below  25~GeV the standard Regge theory reproduces the SD cross section well, however it rises faster than the experimental observations at higher energy.  This behaviour was already expected theoretically due to problems related with the violation of the unitarity at high energy, i.e. $\sigma_{SD} > \sigma^{tot}$ and the Froissart bound~\cite{Froissart19611053}.  Some different theoretical approaches have been attempted to overcome this problem, including the renormalisation of standard pomeron flux to agree with the data~\cite{Goulianos1995379} or decoupling of the triple Pomeron vertex~\cite{Roy-Roberts}. 

Single-diffraction data at $\sqrt{s}$=7 TeV have been obtained by ALICE~\cite{Abelev:2012sea}, CMS~\cite{Chatrchyan:1699728} and TOTEM~\cite{Antchev:1954699}. The ALICE data are consistent with the integrated single-diffraction cross section increasing with energy. In contrast the CMS and TOTEM integrated cross sections, for 8 $< M_X <$ 350 GeV and 6.5 $< M_X <$ 1100 GeV respectively, appear to show a decrease with increasing energy when compared to extrapolations of conventional models. For $M_X <$ 3.4 GeV TOTEM~\cite{Antchev_2013} give an integrated cross section of (2.2 $\pm$ 2.17) mb and an upper limit of 6.31 mb at 95\% confidence level which is not inconsistent with the UA4 value of 3.0 $\pm$ 0.8 mb at $\sqrt{s}$=546 GeV and still allows, in principle, some increase with energy.  

A full discussion of the problem and possible modifications to the models is given in~\cite{KhozeMartin}. Including sophisticated modifications in our fitting of multiple data sets is impractical, so we adopted the simple alternative of modifying the chosen matching
point $M_c=\sqrt{s} \xi_c$. We do not include data above $\sqrt{s}$ = 546/630 GeV and ensure agreement with the UA4 integrated cross sections~\cite{Bernard1987227} and $d\sigma/dt$. This is achieved with a parametrisation given by
\[
 M_c(s)  =
  \begin{cases}
  3 & \text{for }  s < s_0 \\
  3 +  \alpha \ln(\frac{s}{s_0})  & \text{for } s > s_0, \\
  \end{cases}
\]  
where $\alpha$ = 0.6~GeV and $s_0$ = 4000~GeV$^2$. 


\begin{table}[htdp]
\begin{center}
\begin{tabular}{cccc}
\hline
Source & $\sigma_{SD}^{\mathrm{expt}}$ [mb] & $\sigma_{SD}^{\mathrm{expt}}$ [mb] & $\sigma_{SD}^{\mathrm{expt}}$ [mb]  \\
 & $M_X < 4$ GeV & $M_X > 4$ GeV &  $\xi <$ 0.05\\
\hline
 DL model & 2.89 & 6.59 & 9.485\\
 UA4 data & 3$\pm$0.8 & 6.4$\pm$0.4 & 9.4$\pm$0.4 \\
\hline
\end{tabular}
\end{center}
\caption{DL model prediction and UA4 data for low-high mass and total SD cross section at $\sqrt{s} = 546$~GeV. }
\label{tab:UA4_SD}
\end{table}

\begin{figure*}[!htb]
\includegraphics[scale=0.4]{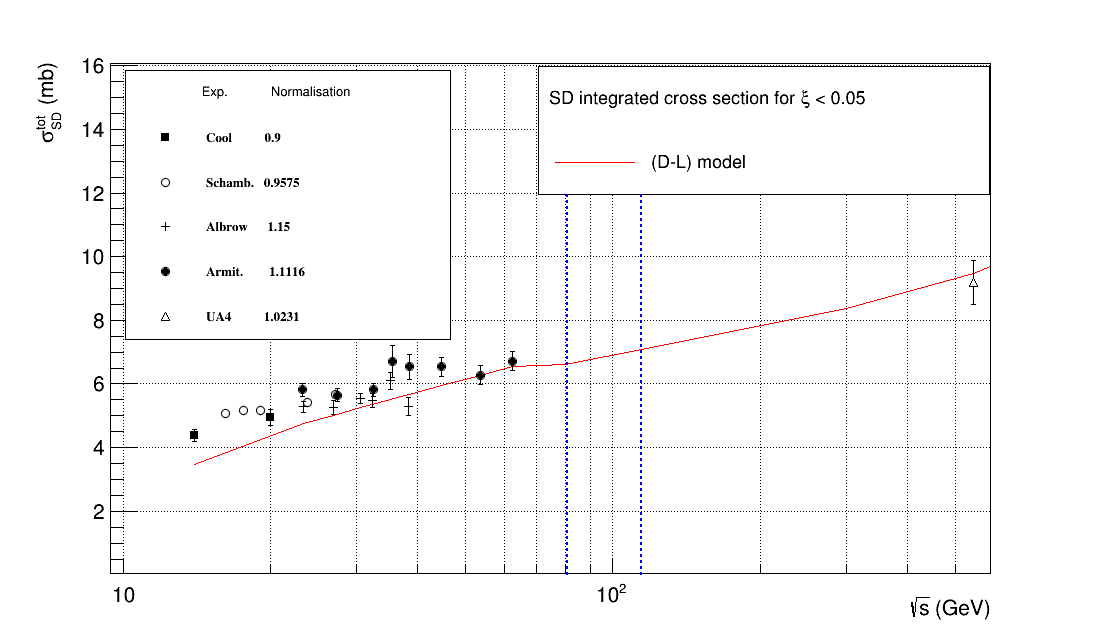}
\caption{\label{fig:totalSD} Total Integrated SD cross section $\sigma_{SD}^{tot}$ at different energies. The red line represents the DL model results and the points are the experimental data reported with their normalisation. The vertical dashed blue lines represent the energies range of interest for the LHC.}
\end{figure*}

A comparison between our model and the UA4 results is presented in table~\ref{tab:UA4_SD}, with the experimentally determined SD cross section $\sigma_{SD}^{\mathrm{expt}} = 2\sigma_{SD}$, where $\sigma_{SD}$ is the integrated cross section, to take into account both arms of the SD diagram as usually quoted by experimentalists. The agreement between the data and the integrated model is very good. The total integrated single diffractive cross section, integrated up to $\xi < 0.05$,  is shown in figure~\ref{fig:totalSD}. The red line is the DL model and the experimental points are indicated with their normalisations. The vertical dashed blue lines represents the energy range of the LHC.

The model works well in this region, and some distance beyond. At very low energy the Regge model does not accurately describe
the data, which is dominated by $s$-channel resonances, but this is well below our region of applicability.


\subsection{\label{sec:part4-7}Application of the model at the LHC energies}

For the LHC collimation studies the range of centre-of-mass energy $\sqrt{s}$ is between 29-115~GeV. We use our model to predict the total and double differential SD cross section at LHC energies.
The low mass, high mass and total cross sections are reported in table~\ref{tab:LHC_SD}, the values are reported for $\xi < 0.05$ and $\xi < 0.12$, the first is given as reference because it is the upper limit used in many publications on the subject and the second to present the values that we are using in our model to predict the LHC loss maps.
 \begin{table}[htdp]
\begin{center}
\begin{tabular}{cccc}
\hline
E[GeV] & $\sqrt{s}$[GeV] & $\sigma_{SD}$($\xi <$ 0.05) & $\sigma_{SD}$($\xi <$ 0.12)\\
\hline
3500 & 81 &  3.39  & 4.37\\
7000 & 115 & 3.55    & 4.53 \\
\hline
\end{tabular}
\end{center}
\caption{Low mass, high mass and total SD cross section at LHC energies. The cross section unit are mb.}
\label{tab:LHC_SD}
\end{table}



\begin{figure*}[!htb]
\includegraphics[width=1\textwidth]{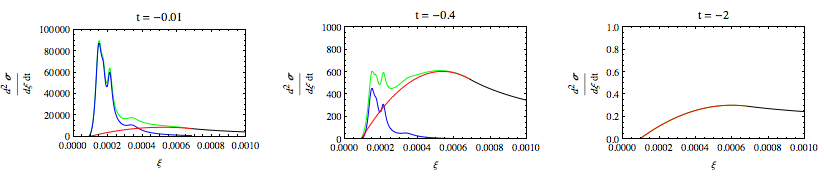}
\caption{\label{fig:SDbck} Individual contributions for the low mass SD double differential cross section for the LHC at top energy, $s$ = 114.6$^2$ GeV$^2$, from left to right $t$ = -0.01, $t$ = -0.4 and $t$ = -2 GeV$^2$. In blue the resonance function $R(\xi,t,s_{LHC})$, in red the background $B(\xi,t,s_{LHC})$ and in green the total SD double differential cross section at low mass. The black line represents the triple-Regge fit at high mass, note that it goes up to $\xi$ = 0.12}
\end{figure*}

Figure~\ref{fig:SDbck} shows the contribution of the background and resonances to the double differential cross section 
at $s=$114$^2$ GeV$^2$ for different values of the transfer momentum $t$ (from left to right $t$ =  0.01, $t$ = 0.4 and $t$ = 2 GeV$^2$). The blue line represents the resonance contribution, the red line the background contribution and the green line the total low mass fit. The black line represents the triple-Regge fit at high mass. The contributions of resonances and background are stronger at low $t$ and decrease at medium $t$; for high $t$ the contribution from resonances disappears and the cross section is dominated by the background. 

\begin{figure*}[!htb]
\includegraphics[width=1\textwidth]{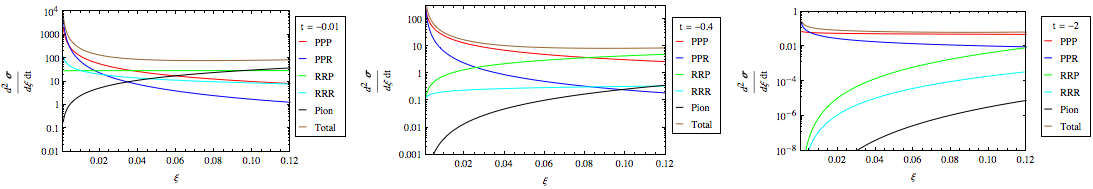}
\caption{\label{fig:SDxi} Individual contributions of triple-Reggeon exchanges and pion exchange for the SD double differential cross section  for the LHC at top energy,  $s$ = 114.6$^2$ GeV$^2$, from left to right $t$ = -0.01, $t$ = -0.4 and $t$ = -2 GeV$^2$.}
\end{figure*}

For the high mass region the contribution of the leading trajectories and the pion-exchange term are shown in figure~\ref{fig:SDxi}  for different value of the momentum transfer~$t$.  The plot on the left shows the contributions for low $t$, the triple pomeron term, $\mathbbm{PPP}$ in red, and the $\mathbbm{PPR}$  term in blue, dominate at low $\xi$, with some contribution from  $\mathbbm{RRP}$ in green and  $\mathbbm{RRR}$ in cyan. At higher $\xi$ the main contribution to the sum of all terms, in brown, is given by the $\mathbbm{RRP}$ and the pion-exchange term in black. At medium and high $t$ the SD double differential cross section is dominated by the triple-pomeron  $\mathbbm{PPP}$ term with some slight contribution from the remaining terms.  

\section{\label{sec:part5}Loss maps simulation using MERLIN code}

MERLIN~\cite{molsonicap12,serlucaIpac1,serlucaIpac2,MerlinWEB} is a C++ accelerator physics library, which has been extended to be used for large scale proton collimation simulations~\citep*{molsonicap12,serlucaIpac1,serlucaIpac2}, with the aim of providing an accurate simulation of the Large Hadron Collider (LHC) collimation system. MERLIN is used to simulate the nominal optics at 7~TeV of the LHC in order to generate loss maps, using a 
thick lens tracking model. These can be generated for different optics configurations, e.g. the $\beta$-function at the interaction points ($\beta$*) and beam crossing angles. The collimation process simulates all proton-collimator interactions and performs all aperture checking. If a proton undergoes an inelastic interaction inside the collimators or touches the beam pipe it is considered lost. If this takes place, the particle is removed from the bunch and the location at which this takes place is recorded. This can be done at any desired longitudinal accuracy, and by default a bin size of 10 cm is used. 

MERLIN has been benchmarked against MAD-X~\cite{MADX_ref} for the optical functions and also, using the same scattering physics models, with SixTrack+K2~\citep*{newSixtrack} for the loss map calculation~\citep*{serlucaIpac2}. SixTrack is the main tool used to calculate dynamical aperture in the LHC and, with the addition of the Monte Carlo scattering routine K2, to calculate the loss maps. MERLIN is part of the ongoing effort to improve the simulation tools for the HL-LHC project and future high energy colliders~\cite{bruceIpac14}. The modular nature of MERLIN allows one to easily switch between the Sixtrack+K2 scattering model and the DL model. 

The loss maps are characterised by the local inefficiency defined as
\begin{equation}\label{eq:inefficiency}
    \eta={N_{ABS}\over \Delta z \cdot N^{tot}_{coll}},
\end{equation}
where $\Delta z$ is the longitudinal resolution (10 cm), $N_{ABS}$ is the number of particles absorbed in $\Delta z$ and $N^{tot}_{coll}$ is the total loss in the collimators along the whole machine. For the collimator $\Delta z$ is set to the collimator length and $N_{ABS}$ are the total losses in the collimator.

The main parameters for the loss map calculation are summarised in table~\ref{tab:simulation_parameters} for a squeezed
beam with an IP beam separation and crossing angle applied. Beam is injected at the horizontal primary collimators (TCP.C6L7) in IR7, and tracked for 200 turns.  The transverse offset between the jaw surface and the impact point, called the impact parameter, is set to 1~$\mu$m.

\begin{table}[h]
\caption{A list of the relevant parameters required for the loss maps simulation. The LHC optics sequence is the version V6.503 for beam 1.}
\begin{center}
\begin{tabular}{cc}
\hline
\textrm{Parameter}&
\textrm{Value}\\ \hline
Energy & 7 TeV \\
Norm. Emittance $\epsilon_n$ & 3.75 mm mrad\\
$\beta^*$(IR1 \& IR5) & 0.55 m \\
$\beta^*$(IR2 \& IR8) & 10 m \\
Crossing angle(IR1) & -145 $\mu$rad\\
Crossing angle(IR5) & 145$\mu$rad\\
Crossing angle(IR2) & -90 $\mu$rad\\
Crossing angle(IR8) & -220$\mu$rad\\
Longitudinal Resolution & 10 cm\\
Turn number & 200\\
\hline
\end{tabular}
\end{center}
\label{tab:simulation_parameters}
\end{table}

\begin{figure*}
\includegraphics[scale=0.4]{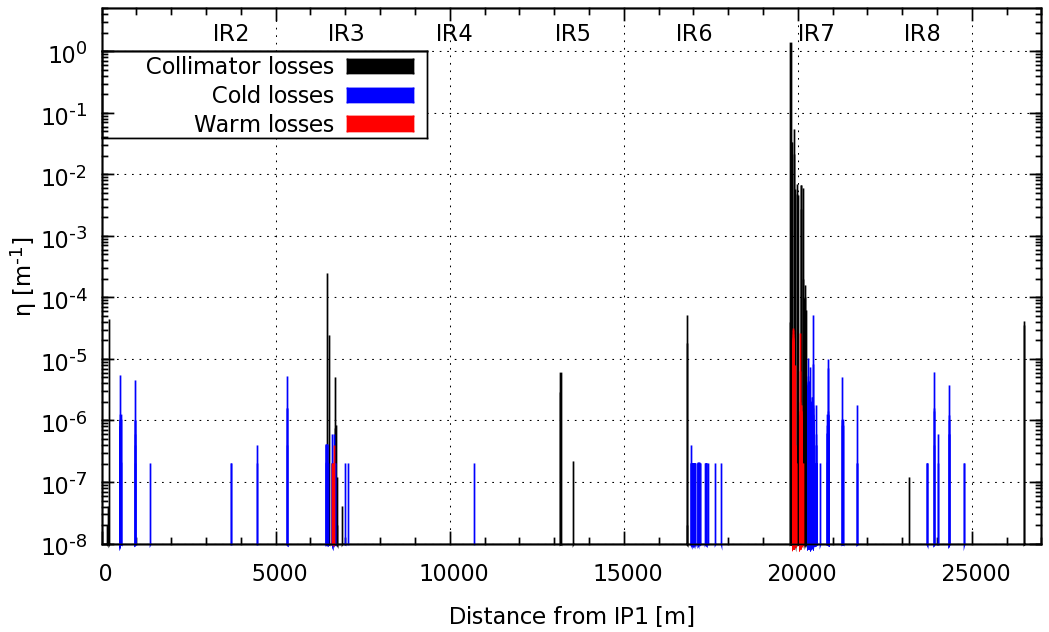}
\caption{\label{fig:LossMap_AllRing}Loss map result for beam 1 with DL scattering physics. The beam is injected in front of the collimator TCP.C6L7 in IR7 region, corresponding to the highest peak in the plot. The black peaks correspond to the losses in the collimator elements, the red ones in the warm elements and the blue one are the losses in the cold magnetic element of the machine}
\end{figure*}

The loss map calculated for this machine configuration and the DL scattering models is shown in figure~\ref{fig:LossMap_AllRing}. 
The plot is colour coded: black spikes represent losses in the collimator jaws, red spikes losses in warm elements of the accelerator, and most importantly blue spikes which indicate losses in the superconducting magnets. Using 64M simulated protons,
MERLIN calculates a total loss inefficiency of 77.65\%, with 0.010\% lost in cold regions and 0.011\% in warm elements. The remaining
protons are lost in the collimators. 

The highest black peak in the map corresponds to the horizontal primary collimator in IR7, which has an aperture of 6 $\sigma$. This is the tightest aperture in the machine. The loss map is consistent with those computed with Sixtrack+K2~\cite{hilumibook}, with most of the losses in the dedicated collimation regions IR7 and IR3, and black peaks in the tertiary collimators used to protect the high luminosity insertion in IR1 (ATLAS ) and IR5 (CMS). The cold peaks downstream of IR7 and in the arcs IR7-IR1 and IR1-IR2 are particularly important because they show where halo protons touch the SC dipoles. These predictions allow us to understand where possible quenching events may occur and how to optimise the collimation system.

\begin{figure}
\includegraphics[scale=0.23]{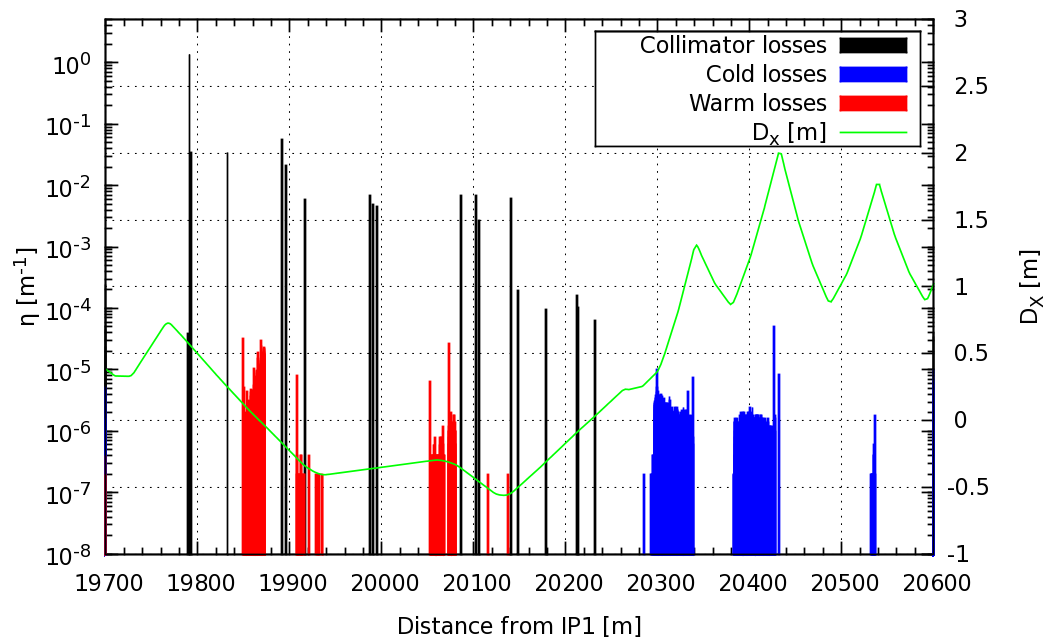}
\caption{\label{fig:LossMap_IR7}Loss map in the betatron collimation region IR7. In correspondence of the local maxima of the dispersion (green line) is visible the increment of the cold losses in the dispersion suppressor region}
\end{figure}

The losses in the dispersion suppressor region of IR7 are shown in figure~\ref{fig:LossMap_IR7} along with the horizontal dispersion in green. The highest collimator losses are in the primary collimators followed by lower losses in the secondary and tertiary collimators. 
The red spikes represent losses in the warm magnets among collimators and are mainly single diffracted protons with high momentum loss and scattered at high angles. There are more warm losses in this region compared to the results observed for the same machine configuration by K2 scattering routine~\cite{serlucaIpac2}. Losses downstream of IR7 in the dispersion suppressor, which are particularly sensitive areas, are shown in blue. Protons which experience single diffractive scattering in the bulk material of the collimator emerge with a transverse kick and a lower energy. Protons entering the dispersion suppressors, where the dispersion rises rapidly, can be lost in these cold areas. Most of the cold peaks in the arc between IR7/IR8 are located at the local maxima of the dispersion as shown for the first peak downstream of the dispersion suppressor in figure~\ref{fig:LossMap_IR7}.

\begin{figure}
\includegraphics[scale=0.125]{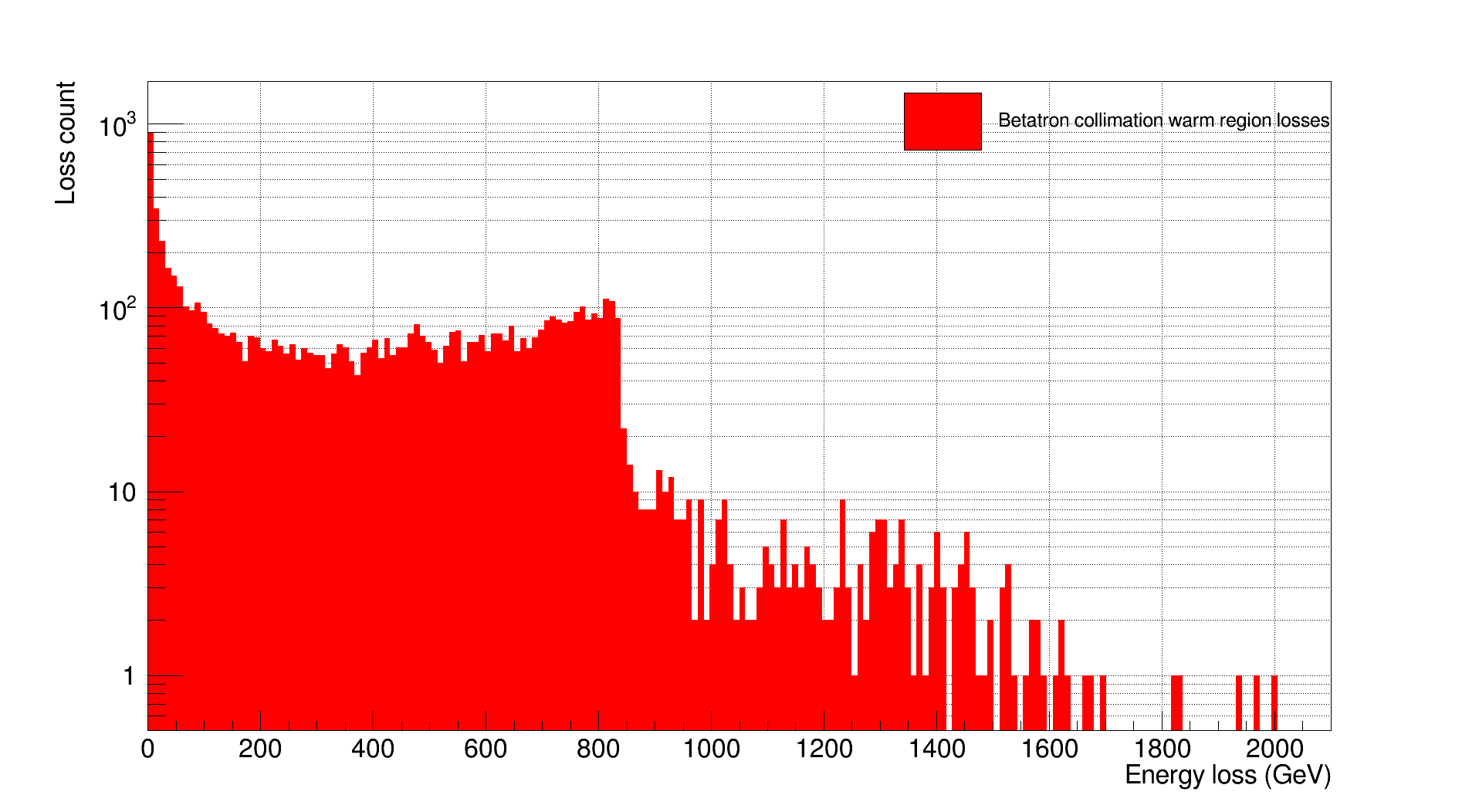}
\caption{\label{fig:warm_dp} Distributions of the $\delta p/p$ of the particles lost in the warm elements among collimators in IR7.}
\end{figure}

\begin{figure}
\includegraphics[scale=0.125]{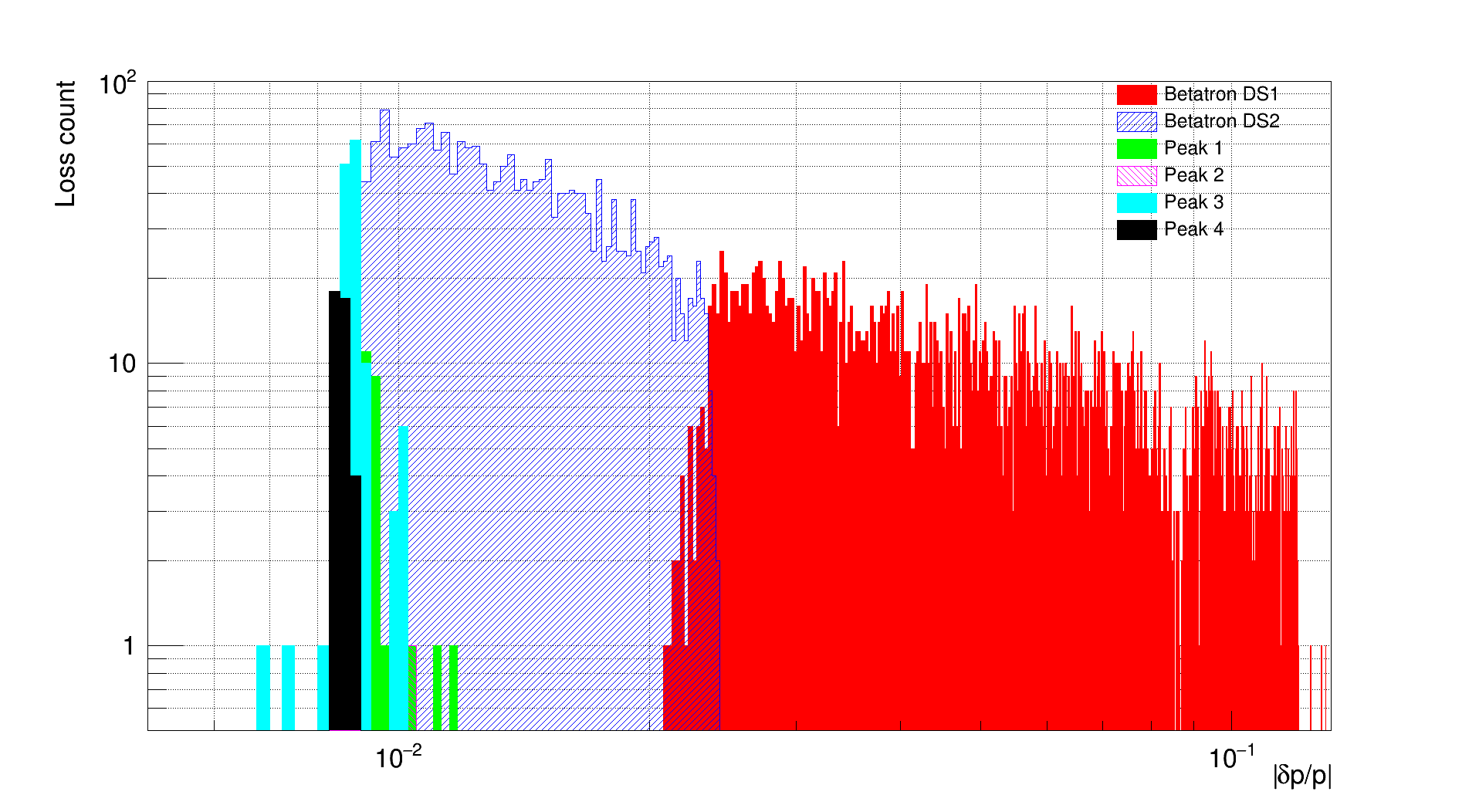}
\caption{\label{fig:cold_dp}Distributions of the $\delta p/p$ of the particles lost in the main peaks downstream IR7, grouped by dispersion suppressors and following cold peaks.}
\end{figure}

In figure~\ref{fig:warm_dp} we present the distribution of energy lost $|\delta p/p|$ by protons impacting on the warm elements among the collimators in IR7. The plot shows a peak at low lost energy followed by a plateau till 10\% and a rapid drop off. Figure~\ref{fig:cold_dp} shows the distribution of energy lost in the dispersion suppressors (DS1, DS2) and particles lost downstream of IR7. The range of energy lost is between 2-10\% for the first dispersion suppressor and 1-2\% for the second.

\section{CONCLUSION}

We have presented a development of the model of Donnachie and Landshoff for elastic and single-diffractive proton scattering for use in simulating collimation systems in high energy proton accelerators. The model includes a description of elastic scattering combining Coulomb with Regge exchange amplitudes, and a description of diffractive scattering that combines $s$-channel resonance formation with $t$-channel Regge exchange.  It is valid over a wide range in the centre of mass energy $\sqrt s$, the invariant 4-momentum transfer  $\sqrt t$ and the scaled missing mass $\xi$, covering the  relevant kinematical regions for the LHC (including the proposed high luminosity upgrade) and the Future Circular Collider. 

We have taken elastic and diffractive scattering data from a large number of previous experiments, with different systematic errors,  and fitted them with a small number of model parameters.  The results have been incorporated into the MERLIN tracking code and this has been used to predict loss maps for 7 TeV running at the LHC. This demonstrates that the model is a powerful tool to understand and improve the performance of collimation systems at present and future high energy proton accelerators.

\section{ACKNOWLEDGMENT}
The authors wish to thank Prof. Donnachie for his support, for assisting us during the development of the scattering models and for being always available to discuss and improve the work. In particular he was responsible for the form of the model in the low mass region of the diffractive scattering. We also thanks Stefano Redaelli and Roderick Bruce for providing support for calculations relating to the LHC collimation system. The HiLumi LHC Design Study is included in the High Luminosity LHC project and is partly funded by the European Commission within the Framework Programme 7 Capacities Specific Programme, Grant Agreement 284404.

\bibliographystyle{unsrt}
\bibliography{practical_pomeron}

\begin{thebibliography}{10}

\bibitem{LHCDesign}
O.~Br{\"{u}}ning, P.~Collier, P.~Lebrun, S.~Myers, R.~Ostojic, J.~Poole, and
  P.~Proudlock.
\newblock {\em {LHC Design Report}}.
\newblock CERN, Geneva, 2004.

\bibitem{hilumibook}
O.~Bruning and L.~Rossi.
\newblock {\em Advanced Series on Directions in High Energy Physics}, 24, 2015.

\bibitem{D-L1}
A.~Donnachie and P.V. Landshoff.
\newblock {\em Nuclear Physics B}, 244(2):322 -- 336, 1984.

\bibitem{QCDDonnachie}
A.~Donnachie, G.~Dosch, P.V. Landshoff, and O.~Nachtmann.
\newblock {\em {Pomeron Physics and QCD}}.
\newblock Cambridge University Press, 2002.

\bibitem{molsonicap12}
J.~Molson, R.B. Appleby, M.~Serluca, Adina Toader, and Roger Barlow.
\newblock Simulating the {LHC} collimation system with the accelerator physics
  library merlin, and loss map results.
\newblock In Dirk Hecht and Michaela Marx, editors, {\em Proceedings of the
  11th International Computational Accelerator Physics Conference}, ICAP 2012,
  pages 12--14, Rostock-Warnem{\"u}nde, Germany, October 2012.

\bibitem{serlucaIpac1}
M.~Serluca, R.B. Appleby, J.~Molson, R.~Barlow, A.~Toader, and H.~Rafique.
\newblock Hi-lumi lhc collimation studies with merlin code.
\newblock In {\em Proceedings of the 5th International Particle Accelerator
  Conference, MOPRI077}, IPAC 2014, pages 784--786, Dresden, Germany, June
  2014.

\bibitem{serlucaIpac2}
M.~Serluca, R.B. Appleby, J.~Molson, R.~Barlow, A.~Toader, H.~Rafique,
  R.~Bruce, A.~Marsili, S.~Redaelli, B.~Salvaucha, and C.~Tambasco.
\newblock Comparison of merlin/sixtrack for lhc collimation studies.
\newblock In {\em Proceedings of the 5th International Particle Accelerator
  Conference, MOPRO046}, IPAC 2014, pages 185--187, Dresden, Germany, June
  2014.

\bibitem{MerlinWEB}
R.B. Appleby, R.Barlow, J.Molson, H.~Rafique, M.Serluca, and A.Toader.
\newblock Merlin source code website.
\newblock \url{https://github.com/MERLIN-Collaboration/MERLIN}.

\bibitem{Shiltsev}
Vladimir Shiltsev, Kip Bishofberger, Vsevolod Kamerdzhiev, Sergei Kozub,
  Matthew Kufer, Gennady Kuznetsov, Alexander Martinez, Marvin Olson, Howard
  Pfeffer, Greg Saewert, Vic Scarpine, Andrey Seryi, Nikolai Solyak, Veniamin
  Sytnik, Mikhail Tiunov, Leonid Tkachenko, David Wildman, Daniel Wolff, and
  Xiao-Long Zhang.
\newblock {\em Phys. Rev. ST Accel. Beams}, 11:103501, Oct 2008.

\bibitem{Biryukov200523}
V.M. Biryukov, V.N. Chepegin, Yu.A. Chesnokov, V.~Guidi, and W.~Scandale.
\newblock {\em Nuclear Instruments and Methods in Physics Research Section B:
  Beam Interactions with Materials and Atoms}, 234(1--2):23 -- 30, 2005.
\newblock Relativistic Channeling and Related Coherent Phenomena in Strong
  Fields.

\bibitem{accphys}
A.~Chao and M.~Tigner.
\newblock {\em {Handbook of accelerator physics and engineering}}.
\newblock World Scientific, 1999.

\bibitem{Kwak1975233}
N.~Kwak, E.~Lohrmann, E.~Nagy, M.~Regler, W.~Schmidt-Parzefall, K.R. Schubert,
  K.~Winter, A.~Brandt, H.~Dibon, G.~Fl{\"u}gge, F.~Niebergall, P.E.
  Schumacher, J.J. Aubert, C.~Broll, G.~Coignet, J.~Favier, L.~Massonnet,
  M.~Vivargent, W.~Bartl, H.~Eichinger, Ch. Gottfried, and G.~Neuhofer.
\newblock {\em Physics Letters B}, 58(2):233 -- 236, 1975.

\bibitem{Bohm1974491}
A.~B{\"o}hm, M.~Bozzo, R.~Ellis, H.~Foeth, M.I. Ferrero, G.~Maderni,
  B.~Naroska, C.~Rubbia, G.~Sette, A.~Staude, P.~Strolin, and G.~de~Zorzi.
\newblock {\em Physics Letters B}, 49(5):491 -- 496, 1974.

\bibitem{Amaldi1978367}
U.~Amaldi, G.~Cocconi, A.N. Diddens, Z.~Dim{\v c}ovski, R.W. Dobinson,
  J.~Dorenbosch, P.~Duinker, G.~Matthiae, A.M. Thorndike, A.M. Wetherell,
  G.~Bellettini, P.L. Braccini, R.~Carrara, R.~Castaldi, V.~Cavasinni,
  F.~Cervelli, T.~Del Prete, P.~Laurelli, M.M. Massai, M.~Morganti,
  G.~Sanguinetti, M.~Valdata-Nappi, C.~Vannini, A.~Baroncelli, C.~Bosio,
  G.~Abshire, J.~Crouch, G.~Finocchiaro, P.~Grannis, H.~J{\"o}stlein,
  R.~Kephart, D.~Lloyd-Owen, and R.~Thun.
\newblock {\em Nuclear Physics B}, 145(2--3):367 -- 401, 1978.

\bibitem{D-L4}
A.~Donnachie and P.V. Landshoff.
\newblock {\em Physics Letters B}, 123(5):345 -- 348, 1983.

\bibitem{D-L2}
A.~Donnachie and P.V. Landshoff.
\newblock {\em Nuclear Physics B}, 267(3--4):690 -- 701, 1986.

\bibitem{D-L3}
A.~Donnachie and P.~V. Landshoff.
\newblock {\em Zeitschrift f{\"{u}}r Physik C Particles and Fields}, 2:55--62,
  1979.
\newblock 10.1007/BF01546237.

\bibitem{D-L5}
A.~Donnachie and P.V. Landshoff.
\newblock {\em Physics Letters B}, 387(3):637 -- 641, 1996.

\bibitem{D-L-arxiv}
A.~Donnachie and P~V. Landshoff.
\newblock Elastic scattering at the {LHC}, arxiv 1112.2485.
\newblock Dec 2011.

\bibitem{JamesThesis}
J.~Molson.
\newblock {\em "Proton scattering and collimation for the LHC and LHC
  luminosity upgrade"}.
\newblock PhD thesis, University of Manchester Press, UK, 2014.

\bibitem{RevModPhys.57.563}
M.~M. Block and R.~N. Cahn.
\newblock {\em Rev. Mod. Phys.}, 57:563--598, Apr 1985.

\bibitem{Cahn}
R.~Cahn.
\newblock {\em Zeitschrift f{\"{u}}r Physik C Particles and Fields},
  15:253--260, 1982.
\newblock 10.1007/BF01475009.

\bibitem{Amos1985689}
N.~Amos, M.M. Block, G.J. Bobbink, M.~Botje, D.~Favart, C.~Leroy, F.~Linde,
  P.~Lipnik, J-P. Matheys, D.~Miller, K.~Potter, S.~Shukla, C.~Vander
  Velde-Wilquet, and S.~Zucchelli.
\newblock {\em Nuclear Physics B}, 262(4):689 -- 714, 1985.

\bibitem{ROOTcite}
Rene Brun and Fons Rademakers.
\newblock Root - an object oriented data analysis framework.
\newblock In {\em AIHENP'96 Workshop, Lausane}, volume 389, pages 81--86, 1996.

\bibitem{sandyplb1}
A.~Donnachiea and P.V. Landshoff.
\newblock {\em Physics Letters B}, 727:500, 2013.

\bibitem{sandyplb2}
A.~Donnachiea and P.V. Landshoff.
\newblock {\em Physics Letters B}, 750:669, 2015.

\bibitem{Gribov}
J.V.N. Gribov.
\newblock {\em Soviet Journal Nuclear Physics}, 5(5):138, 1967.

\bibitem{DL_2003}
A~Donnachie and P.V. Landshoff.
\newblock {Soft diffraction dissociation}.
\newblock 2003.

\bibitem{LKMR09}
E.G.S. Luna, V.A. Khoze, A.D. Martin, and M.G. Ryskin.
\newblock {\em Eur.Phys.J.}, C59:1--12, 2009.

\bibitem{sandyup}
A~Donnachie.
\newblock unpublished note.

\bibitem{Kaidalov}
A.B. Kaidalov, V.A. Khoze, Yu.F. Pirogov, and N.L. Ter-Isaakyan.
\newblock {\em Physics Letters B}, 45(5):493 -- 496, 1973.

\bibitem{Ostapchenko}
S~Ostapchenko, H~J Drescher, F~M Liu, T~Pierog, and K~Werner.
\newblock {\em Journal of Physics G: Nuclear and Particle Physics},
  28(10):2597, 2002.

\bibitem{Gotsman}
E.~Gotsman, E.~Levin, U.~Maor, and J.S. Miller.
\newblock {\em Eur.Phys.J.}, C57:689--709, 2008.

\bibitem{Ryskin}
M.G. Ryskin, A.D. Martin, and V.A. Khoze.
\newblock {\em Eur.Phys.J.}, C54:199--217, 2008.

\bibitem{luna2010}
E.G.S. Luna, V.A. Khoze, A.D. Martin, and M.G. Ryskin.
\newblock {\em The European Physical Journal C}, 69(1-2):95--101, 2010.

\bibitem{Field1974367}
R.D. Field and G.C. Fox.
\newblock {\em Nuclear Physics B}, 80(3):367 -- 402, 1974.

\bibitem{schamdiffractive}
R.~D. Schamberger, J.~Lee-Franzini, R.~McCarthy, S.~Childress, and P.~Franzini.
\newblock {\em Phys. Rev. D}, 17:1268--1291, Mar 1978.

\bibitem{Morrison}
D.R.O. Morrison.
\newblock {\em Physical Review}, 165(5):1699--1702, 1968.

\bibitem{PhysRevLett.34.1121}
R.~D. Schamberger, J.~Lee-Franzini, R.~McCarthy, S.~Childress, and P.~Franzini.
\newblock {\em Phys. Rev. Lett.}, 34:1121--1124, Apr 1975.

\bibitem{Bernard1987227}
D.~Bernard, M.~Bozzo, P.L. Braccini, F.~Carbonara, R.~Castaldi, F.~Cervelli,
  G.~Chiefari, E.~Drago, M.~Haguenauer, V.~Innocente, P.~Kluit, B.~Koene,
  S.~Lanzano, G.~Matthiae, L.~Merola, M.~Napolitano, V.~Palladino,
  G.~Sanguinetti, P.~Scampoli, S.~Scapellato, G.~Sciacca, G.~Sette, R.~Van
  Swol, J.~Timmermans, C.~Vannini, J.~Velasco, P.G. Verdini, and F.~Visco.
\newblock {\em Physics Letters B}, 186(2):227 -- 232, 1987.

\bibitem{Froissart19611053}
M.~Froissart.
\newblock {\em Physical Review}, 123(3):1053--1057, 1961.

\bibitem{Goulianos1995379}
K.~Goulianos.
\newblock {\em Physics Letters B}, 358(3--4):379 -- 388, 1995.

\bibitem{Roy-Roberts}
D.P. Roy and R.G. Roberts.
\newblock {\em Nuclear Physics B}, 77(2):240 -- 268, 1974.

\bibitem{Abelev:2012sea}
Betty Abelev et~al.
\newblock {\em Eur.Phys.J.}, C73:2456, 2013.

\bibitem{Chatrchyan:1699728}
S~Chatrchyan.
\newblock ``measurement of pseudorapidity distributions of charged particles in
  proton-proton collisions at $\sqrt{s}$ = 8 tev by the cms and totem
  experiments''.
\newblock Technical Report arXiv:1405.0722. CMS-FSQ-12-026.
  CERN-PH-EP-TOTEM-2014-002. CERN-PH-EP-2014-063, CERN, Geneva, May 2014.
\newblock Comments: Submitted to the European Physical Journal C.

\bibitem{Antchev:1954699}
G~Antchev and others (TOTEM~Collaboration).
\newblock Technical Report CERN-PH-EP-2014-260, CERN, Geneva, Oct 2014.

\bibitem{Antchev_2013}
G~Antchev and others (TOTEM~Collaboration).
\newblock {\em EPL (Europhysics Letters)}, 101(2):21003, 2013.

\bibitem{KhozeMartin}
V.A. Khoze, A.D. Martin, and M.G. Ryskin.
\newblock {\em The European Physical Journal C}, 73(7), 2013.

\bibitem{MADX_ref}
\url{http://mad.web.cern.ch/mad/}.

\bibitem{newSixtrack}
G.~Robert-Demolaize, R.~Assmann, S.~Redaelli, and F.~Schmidt.
\newblock A new version of sixtrack with collimation and aperture interface.
\newblock In {\em Proceedings of the Particle Accelerator Conference, 2005. PAC
  2005.}, pages 4084--4086, May 2005.

\bibitem{bruceIpac14}
R.~Bruce, C.~Bracco, F.~Cerruti, A.~Ferrari, A.~Lechner, D.~Mirarchi, P.G.
  Ortega, A.~Rossi, D.P. Sinuela, V.~Vlachoudis, A.~Mereghetti, A.~Assmann,
  L.~Lari, S.M. Gibson, L.J. Nevay, R.B. Appleby, J.~Molson, R.~Barlow,
  A.~Toader, H.~Rafique, R.~Bruce, A.~Marsili, S.~Redaelli, M.~Serluca,
  B.~Salvaucha, and C.~Tambasco.
\newblock Integrated simulation tools for collimation cleaning in hl-lhc.
\newblock In {\em Proceedings of the 5th International Particle Accelerator
  Conference, MOPRO039}, IPAC 2014, pages 160--162, Dresden, Germany, June
  2014.

\bibitem{Nagy1979221}
E.~Nagy, R.S. Orr, W.~Schmidt-Parzefall, K.~Winter, A.~Brandt, F.W. B{\"u}sser,
  G.~Fl{\"u}gge, F.~Niebergall, P.E. Schumacher, H.~Eichinger, K.R. Schubert,
  J.J. Aubert, C.~Broll, G.~Coignet, H.~De Kerret, J.~Favier, L.~Massonnet,
  M.~Vivargent, W.~Bartl, H.~Dibon, Ch. Gottfried, G.~Neuhofer, and M.~Regler.
\newblock {\em Nuclear Physics B}, 150(0):221 -- 267, 1979.

\bibitem{Albrow19761}
M.G. Albrow, A.~Bagchus, D.P. Barber, P.~Benz, A.~Bogaerts, B.~B{\v
  o}snjakovi{\'c}, J.R. Brooks, C.Y. Chang, A.B. Clegg, F.C. Ern{\'e}, C.N.P.
  Gee, P.~Kooijman, D.H. Locke, F.K. Loebinger, N.A. McCubbin, P.G. Murphy,
  D.~Radoji{\v c}i{\'c}, A.~Rudge, J.C. Sens, A.L. Sessoms, J.~Singh, D.~Stork,
  and J.~Timmer.
\newblock {\em Nuclear Physics B}, 108(1):1 -- 29, 1976.

\bibitem{PhysRevD.23.33}
W.~Faissler, M.~Gettner, J.~R. Johnson, T.~Kephart, E.~Pothier, D.~Potter,
  M.~Tautz, S.~Conetti, C.~Hojvat, D.~G. Ryan, K.~Shahbazian, D.~G. Stairs,
  J.~Trischuk, P.~Baranov, J.~L. Hartmann, J.~Orear, S.~Rusakov, and
  J.~Vrieslander.
\newblock {\em Phys. Rev. D}, 23:33--42, Jan 1981.

\bibitem{Breakstone1984253}
A.~Breakstone, R.~Campanini, H.B. Crawley, G.M. Dallavalle, M.M. Deninno,
  K.~Doroba, D.~Drijard, F.~Fabbri, A.~Firestone, H.G. Fischer, H.~Frehse,
  W.~Geist, G.~Giacomelli, R.~Gokieli, M.~Gorbics, P.~Hanke, M.~Heiden,
  W.~Herr, P.G. Innocenti, E.E. Kluge, J.W. Lamsa, T.~Lohse, W.T. Meyer,
  G.~Mornacchi, T.~Nakada, M.~Panter, A.~Putzer, K.~Rauschnabel, B.~Rensch,
  F.~Rimondi, R.~Sosnowski, M.~Szczekowski, O.~Ullaland, D.~Wegener, and
  M.~Wunsch.
\newblock {\em Nuclear Physics B}, 248(2):253 -- 260, 1984.

\bibitem{Amaldi1980301}
U.~Amaldi and K.R. Schubert.
\newblock {\em Nuclear Physics B}, 166(2):301 -- 320, 1980.

\bibitem{PhysRevLett.54.2180}
A.~Breakstone, H.~B. Crawley, G.~M. Dallavalle, K.~Doroba, D.~Drijard,
  F.~Fabbri, A.~Firestone, H.~G. Fischer, H.~Frehse, W.~Geist, G.~Giacomelli,
  R.~Gokieli, M.~Gorbics, P.~Hanke, M.~Heiden, W.~Herr, E.~E. Kluge, J.~W.
  Lamsa, T.~Lohse, W.~T. Meyer, G.~Mornacchi, T.~Nakada, M.~Panter, A.~Putzer,
  K.~Rauschnabel, F.~Rimondi, G.~P. Siroli, R.~Sosnowski, M.~Szczekowski,
  O.~Ullaland, and D.~Wegener.
\newblock {\em Phys. Rev. Lett.}, 54:2180--2183, May 1985.

\bibitem{Ambrosio1982495}
M.~Ambrosio, G.~Anzivino, G.~Barbarino, G.~Carboni, V.~Cavasinni, T.~Del Prete,
  P.D. Grannis, D.~Lloyd Owen, M.~Morganti, G.~Paternoster, S.~Patricelli, and
  M.~Valdata-Nappi.
\newblock {\em Physics Letters B}, 115(6):495 -- 502, 1982.

\bibitem{LHC_TOTEM1}
G.~Antchev and others (TOTEM~Collaboration).
\newblock {\em EPL (Europhysics Letters)}, 95(4):41001, 2011.

\bibitem{LHC_TOTEM2}
G.~Antchev and others (TOTEM~Collaboration).
\newblock {\em EPL (Europhysics Letters)}, 96(2):21002, 2011.

\bibitem{Augier1993448}
C.~Augier, D.~Bernard, J.~Bourotte, M.~Bozzo, A.~Bueno, R.~Cases, F.~Djama,
  M.~Haguenauer, V.~Kundr{\'a}t, M.~Lokaj{\'\i}{\v c}ek, G.~Matthiae,
  A.~Morelli, F.~Natali, S.~N{\v e}me{\v c}ek, M.~Nov{\'a}k, E.~Sanchis,
  G.~Sette, M.~Smi{\v z}ansk{\'a}, and J.~Velasco.
\newblock {\em Physics Letters B}, 316(2--3):448 -- 454, 1993.

\bibitem{Battiston1983472}
R.~Battiston, M.~Bozzo, P.L. Braccini, F.~Carbonara, R.~Carrara, R.~Castaldi,
  F.~Cervelli, G.~Chiefari, E.~Drago, M.~Haguenauer, B.~Koene, G.~Matthiae,
  L.~Merola, M.~Napolitano, V.~Palladino, G.~Sanguinetti, G.~Sciacca, G.~Sette,
  R.~van Swol, J.~Timmermans, C.~Vannini, J.~Velasco, and F.~Visco.
\newblock {\em Physics Letters B}, 127(6):472 -- 475, 1983.

\bibitem{Bernard1987583}
D.~Bernard, M.~Bozzo, P.L. Braccini, F.~Carbonara, R.~Castaldi, F.~Cervelli,
  G.~Chiefari, E.~Drago, M.~Haguenauer, V.~Innocente, P.~Kluit, S.~Lanzano,
  G.~Matthiae, L.~Merola, M.~Napolitano, V.~Palladino, G.~Sanguinetti,
  P.~Scampoli, S.~Scapellato, G.~Sciacca, G.~Sette, J.~Timmermans, C.~Vannini,
  J.~Velasco, P.G. Verdini, and F.~Visco.
\newblock {\em Physics Letters B}, 198(4):583 -- 589, 1987.

\bibitem{Bozzo1984385}
M.~Bozzo, P.L. Braccini, F.~Carbonara, R.~Castaldi, F.~Cervelli, G.~Chiefari,
  E.~Drago, M.~Haguenauer, V.~Innocente, B.~Koene, S.~Lanzano, G.~Matthiae,
  L.~Merola, M.~Napolitano, V.~Palladino, G.~Sanguinetti, S.~Scapellato,
  G.~Sciacca, G.~Sette, R.~van Swol, J.~Timmermans, C.~Vannini, J.~Velasco,
  P.G. Verdini, and F.~Visco.
\newblock {\em Physics Letters B}, 147(4--5):385 -- 391, 1984.

\bibitem{Bozzo1985197}
M.~Bozzo, P.L. Braccini, F.~Carbonara, R.~Castaldi, F.~Cervelli, G.~Chiefari,
  E.~Drago, M.~Haguenauer, V.~Innocente, B.~Koene, S.~Lanzano, G.~Matthiae,
  L.~Merola, M.~Napolitano, V.~Palladino, G.~Sanguinetti, S.~Scapellato,
  G.~Sciacca, G.~Sette, R.~Van Swol, J.~Timmermans, C.~Vannini, J.~Velasco,
  P.G. Verdini, and F.~Visco.
\newblock {\em Physics Letters B}, 155(3):197 -- 202, 1985.

\bibitem{PhysRevD.50.5518}
F.~Abe and others (CDF~collaboration).
\newblock {\em Phys. Rev. D}, 50:5518--5534, Nov 1994.

\bibitem{Bernard1986142}
D.~Bernard, M.~Bozzo, P.L. Braccini, F.~Carbonara, R.~Castaldi, F.~Cervelli,
  G.~Chiefari, E.~Drago, M.~Haguenauer, V.~Innocente, P.~Kluit, B.~Koene,
  S.~Lanzano, G.~Matthiae, L.~Merola, M.~Napolitano, V.~Palladino,
  G.~Sanguinetti, P.~Scampoli, S.~Scapellato, G.~Sciacca, G.~Sette, R.~van
  Swol, J.~Timmermans, C.~Vannini, J.~Velasco, P.G. Verdini, and F.~Visco.
\newblock {\em Physics Letters B}, 171(1):142 -- 144, 1986.

\bibitem{Amos1990127}
N.A. Amos, C.~Avila, W.F. Baker, M.~Bertani, M.M. Block, D.A. Dimitroyannis,
  D.P. Eartly, R.W. Ellsworth, G.~Giacomelli, B.~Gomez, J.A. Goodman, C.M.
  Guss, A.J. Lennox, M.R. Mondardini, J.P. Negret, J.~Orear, S.M. Pruss,
  R.~Rubinstein, S.~Sadr, S.~Shukla, I.~Veronesi, and S.~Zucchelli.
\newblock {\em Physics Letters B}, 247(1):127 -- 130, 1990.

\bibitem{D0elastic}
{{D0 Collaboration}}.
\newblock Measurement of the differential cross section d$\sigma$/d$|t|$ in
  elastic $p\bar{p}$ scattering at $\sqrt{s} = 1.96$~{T}e{V}.
\newblock D0 Note 6056-CONF, 2010.

\bibitem{Albrow1974376}
M.G. Albrow et~al.
\newblock Missing mass spectra in pp inelastic scattering at total energies of
  23 {G}e{V} and 31 {G}e{V}.
\newblock {\em Nuclear Physics B}, 72(3):376 -- 392, 1974.

\bibitem{Armitage1982365}
J.C.M. Armitage, P.~Benz, G.J. Bobbink, F.C. Erne, P.~Kooijman, F.K. Loebinger,
  A.A. Macbeth, H.E. Montgomery, P.G. Murphy, A.~Rudge, J.C. Sens, D.~Stork,
  and J.~Timmer.
\newblock {\em Nuclear Physics B}, 194(3):365 -- 372, 1982.

\bibitem{kooijman_thesis}
P.M. Kooijman.
\newblock {\em Investigation of Diffraction Dissociation in Proton-Proton
  Collisions at high energies}.
\newblock PhD thesis, University of Utrecht, April 1979.

\bibitem{PhysRevLett.47.701}
R.~L. Cool et~al.
\newblock Diffraction dissociation of $\pi^{\pm}$, ${K}^{\pm}$, and ${p}^{\pm}$
  at 100 and 200 {G}e{V}/c.
\newblock {\em Phys. Rev. Lett.}, 47:701--704, Sep 1981.

\bibitem{PhysRevLett.39.1432}
Y.~Akimov, V.~Bartenev, R.~Cool, K.~Goulianos, D.~A. Gross, E.~Jenkins,
  E.~Malamud, P.~Markov, S.~Mukhin, D.~Nitz, S.~L. Olsen, A.~Sandacz, S.~L.
  Segler, H.~Sticker, and R.~Yamada.
\newblock {\em Phys. Rev. Lett.}, 39:1432--1435, Dec 1977.

\bibitem{PhysRevLett.32.389}
S.~Childress, P.~Franzini, J.~Lee-Franzini, R.~McCarthy, and R.~D. Schamberger.
\newblock Small-momentum-transfer $p-p$ inelastic scattering at 300 gev/c.
\newblock {\em Phys. Rev. Lett.}, 32:389--392, Feb 1974.

\bibitem{PhysRevD.17.1268}
R.~D. Schamberger, J.~Lee-Franzini, R.~McCarthy, S.~Childress, and P.~Franzini.
\newblock Mass spectrum of proton-proton inelastic interactions from 55 to 400
  gev/c at small momentum transfer.
\newblock {\em Phys. Rev. D}, 17:1268--1291, Mar 1978.

\bibitem{Bozzo1984217}
M.~Bozzo, P.L. Braccini, F.~Carbonara, R.~Carrara, R.~Castaldi, F.~Cervelli,
  G.~Chiefari, E.~Drago, M.~Haguenauer, B.~Koene, G.~Matthiae, L.~Merola,
  M.~Napolitano, V.~Palladino, G.~Sanguinetti, G.~Sciacca, G.~Sette, R.~van
  Swol, J.~Timmermans, C.~Vannini, J.~Velasco, and F.~Visco.
\newblock {\em Physics Letters B}, 136(3):217 -- 220, 1984.

\bibitem{Brandt19983}
A.~Brandt et~al.
\newblock Measurements of single diffraction at $\sqrt{s}$ = 630 gev; evidence
  for a non-linear $\alpha(t)$ of the pomeron.
\newblock {\em Nuclear Physics B}, 514(1-2):3 -- 44, 1998.

\bibitem{PhysRevD.59.114017}
K.~Goulianos and J.~Montanha.
\newblock Factorization and scaling in hadronic diffraction.
\newblock {\em Phys. Rev. D}, 59:114017, May 1999.

\end{thebibliography}

\appendix
\section{\label{sec:app_datalist}Elastic data sources and model fit}
\begin{figure*}
\begin{minipage}[b]{1.0\linewidth}
\includegraphics[width=0.5\textwidth]{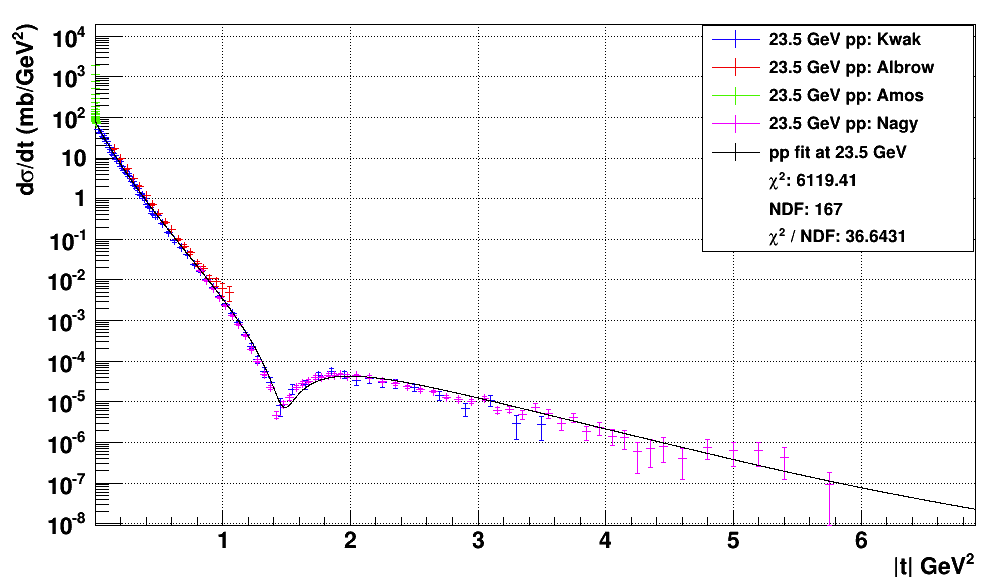}
\includegraphics[width=0.5\textwidth]{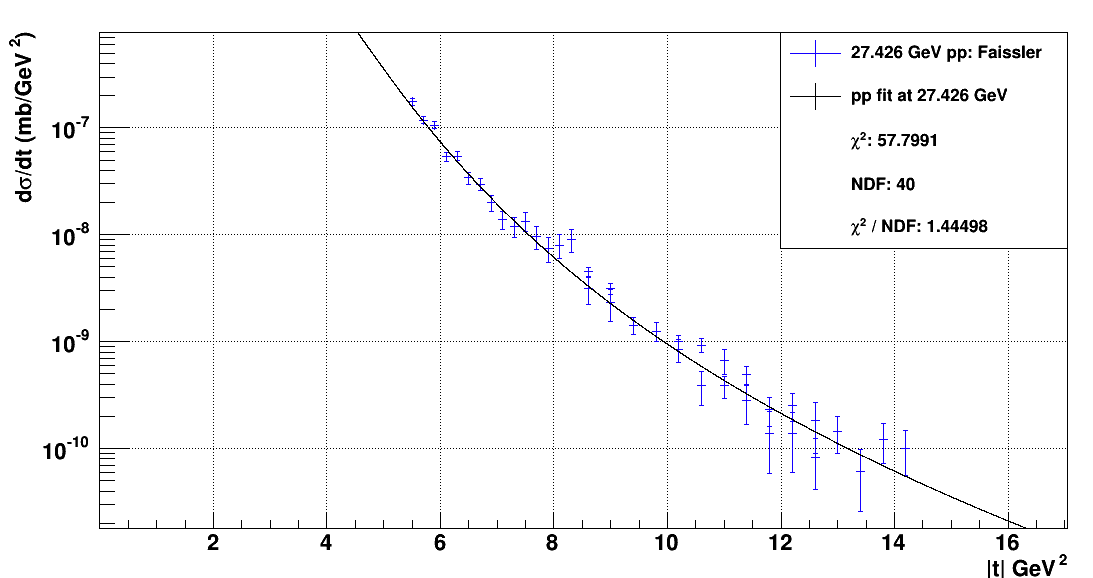}
\end{minipage}
\begin{minipage}[b]{1.0\linewidth}
\includegraphics[width=0.5\textwidth]{plots/2.png}
\includegraphics[width=0.5\textwidth]{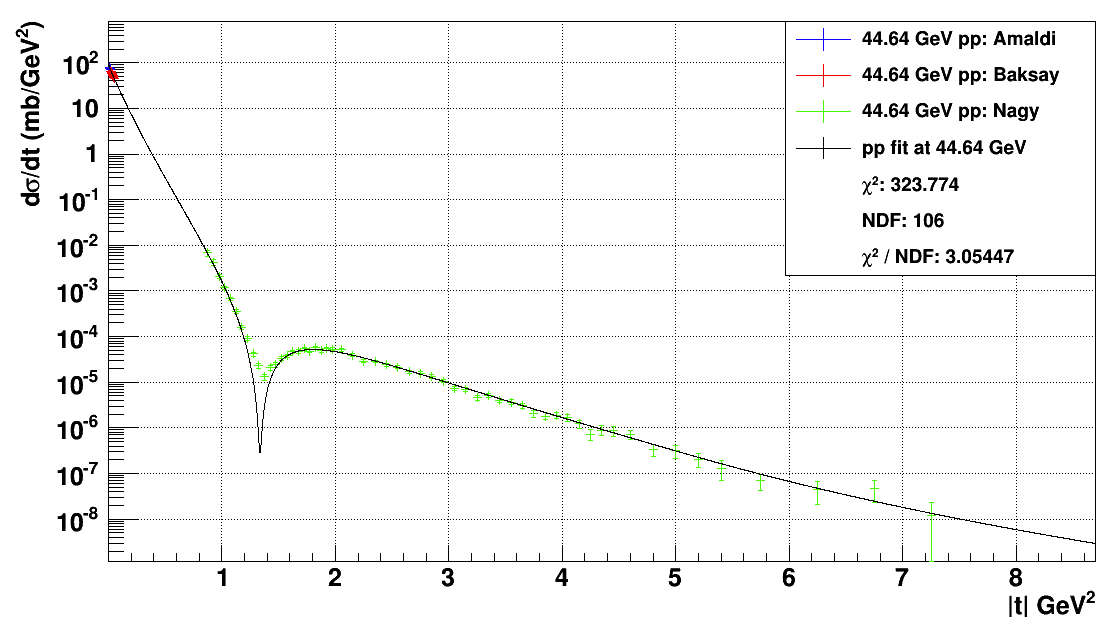}
\end{minipage}
\begin{minipage}[b]{1.0\linewidth}
\includegraphics[width=0.5\textwidth]{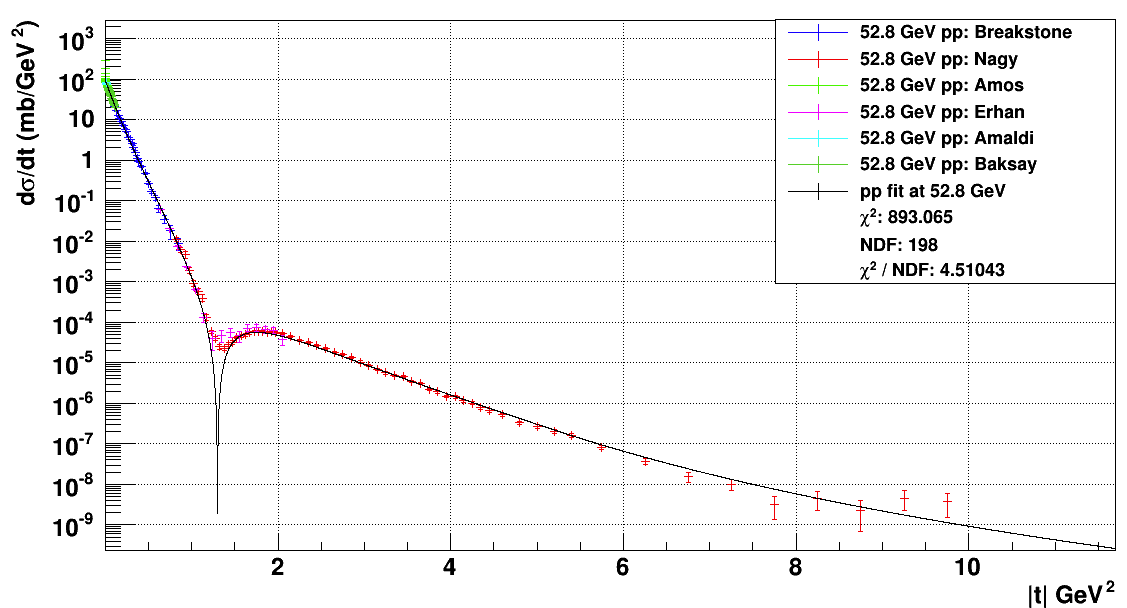}
\includegraphics[width=0.5\textwidth]{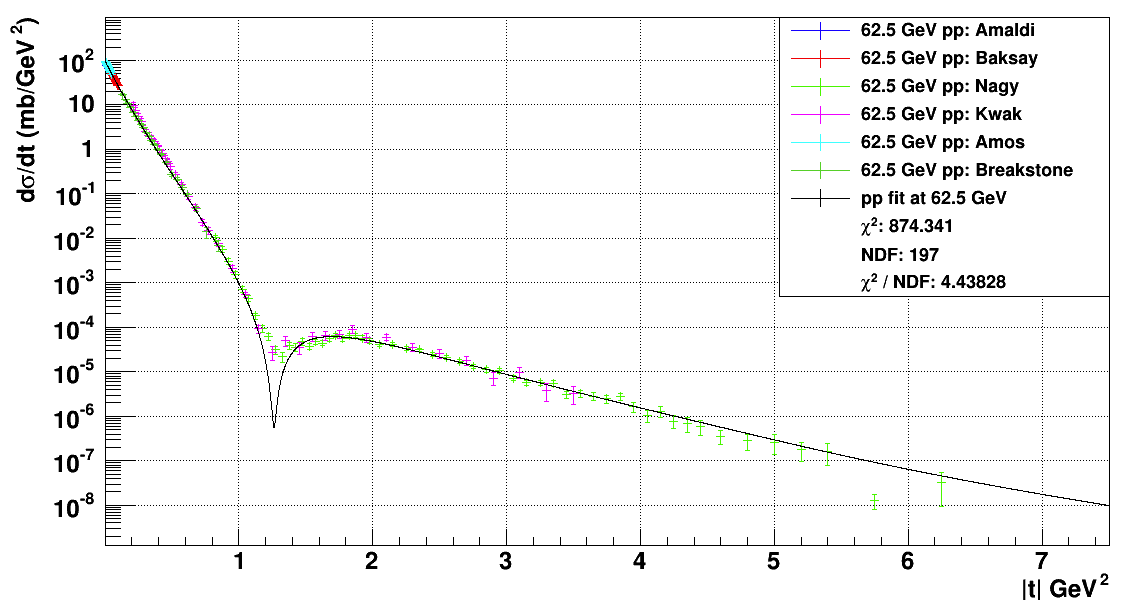}
\end{minipage}
\caption{\label{fig:elastic_full_pp}The $pp$ elastic scattering model fit (fitted over all data) shown
for a range of $\sqrt{s}$,  over the full $t$ range, including the coulomb peak and down to the lower cut on $t$.}
\end{figure*}

The elastic data sources with their energies and references are reported in table~\ref{tab:elastic_data_sources}. 
The normalisations chosen by the fitter can be found in~\cite{JamesThesis}.
In figures~\ref{fig:elastic_full_pp} we present the $pp$ elastic scattering model fit (fitted over all data) shown for a range of $\sqrt{s}$,  over the full $t$ range, including the Coulomb peak and down to the lower cut on $t$. 

\begin{table}[h]
\caption{\label{tab:elastic_data_sources}%
A list of elastic data used in the fit and its source.
}
\begin{tabular}{cccc}
\hline
\textrm{Accelerator (Experiments)}&
\textrm{Particles}&
\textrm{$\sqrt{s}$ (GeV)}&
\textrm{Sources}\\ \hline
ISR & $pp$ & 23.46 & \cite{Amos1985689,Nagy1979221,Kwak1975233,Albrow19761}\\
Fermilab (E177A) & pp & 27.426 & \cite{PhysRevD.23.33} \\
ISR & $pp$ & 30.54 & \cite{Amos1985689,Breakstone1984253,Nagy1979221,Albrow19761} \\
ISR & $pp$ & 44.64 & \cite{Nagy1979221,Amaldi1980301} \\
ISR (R211,SFM)& pp & 52.81 & \cite{PhysRevLett.54.2180,Amos1985689,Breakstone1984253,Ambrosio1982495,Nagy1979221} \\
ISR & $pp$ & 62.5 & \cite{Amos1985689,Breakstone1984253,Nagy1979221,Kwak1975233} \\
LHC (TOTEM) & pp & 7000 & \cite{LHC_TOTEM1,LHC_TOTEM2}\\
\hline
ISR & $p\bar{p}$ & 30.4 & \cite{Amos1985689,Breakstone1984253} \\
ISR & $p\bar{p}$ & 52.6 & \cite{PhysRevLett.54.2180,Amos1985689,Breakstone1984253,Ambrosio1982495} \\
ISR & $p\bar{p}$ & 62.3 & \cite{Amos1985689,Breakstone1984253} \\
SP$\bar{\text{P}}$S (UA4/2) & $p\bar{p}$ & 541 & \cite{Augier1993448} \\
SP$\bar{\text{P}}$S (UA4),Tevatron (CDF) & $p\bar{p}$ & 546 & \cite{Battiston1983472,Bernard1987583,Bozzo1984385,Bozzo1985197,PhysRevD.50.5518} \\
SP$\bar{\text{P}}$S (UA4) & $p\bar{p}$ & 630 & \cite{Bernard1986142} \\
Tevatron (CDF,E710) & $p\bar{p}$ & 1800 & \cite{PhysRevD.50.5518,Amos1990127} \\
Tevatron (D\O) & $p\bar{p}$ & 1980 & \cite{D0elastic} \\
\hline
\end{tabular}
\end{table}

\begin{figure*}
\begin{minipage}[b]{1.0\linewidth}
\includegraphics[width=0.5\textwidth]{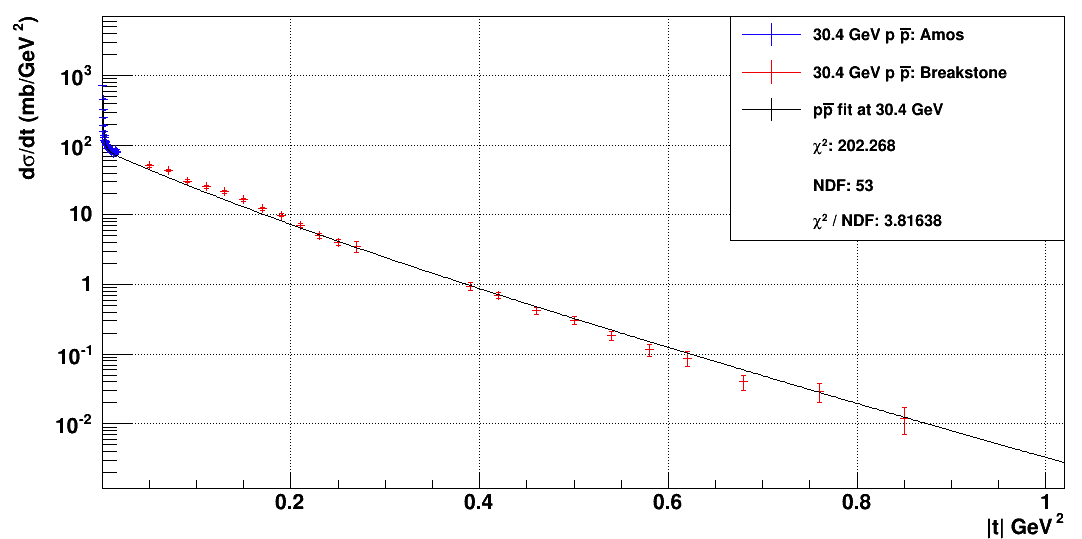}
\includegraphics[width=0.5\textwidth]{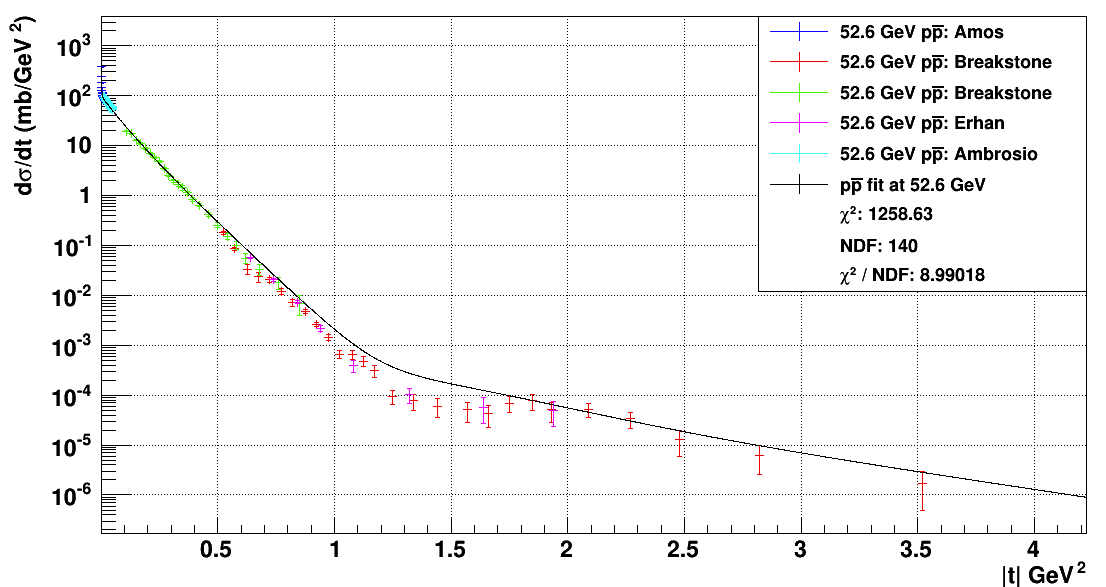}
\end{minipage}
\begin{minipage}[b]{1.0\linewidth}
\includegraphics[width=0.5\textwidth]{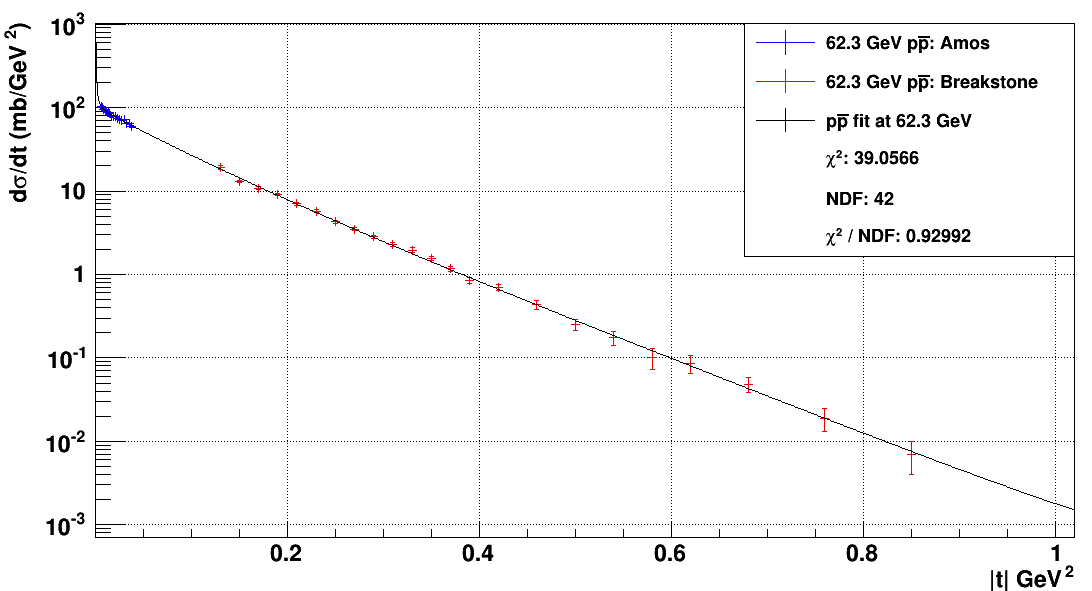}
\includegraphics[width=0.5\textwidth]{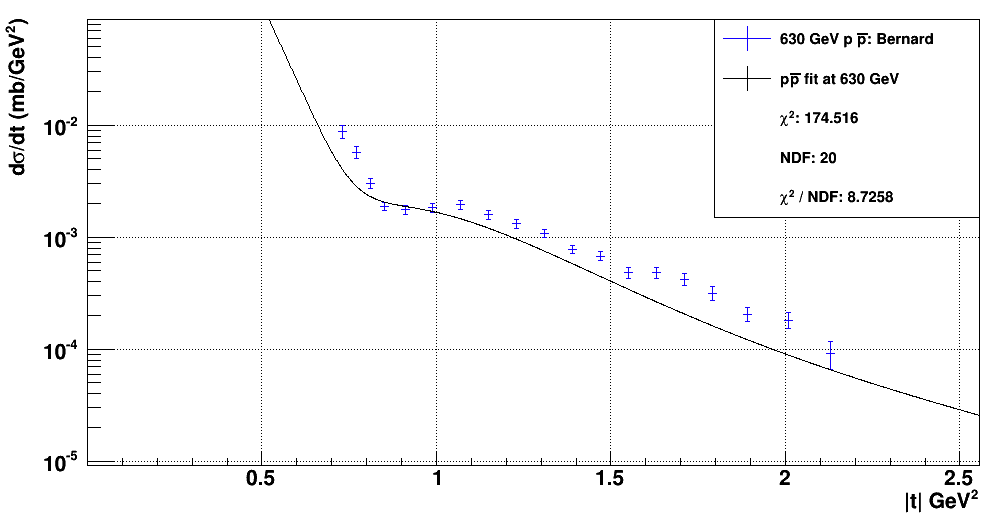}
\end{minipage}
\begin{minipage}[b]{1.0\linewidth}
\includegraphics[width=0.5\textwidth]{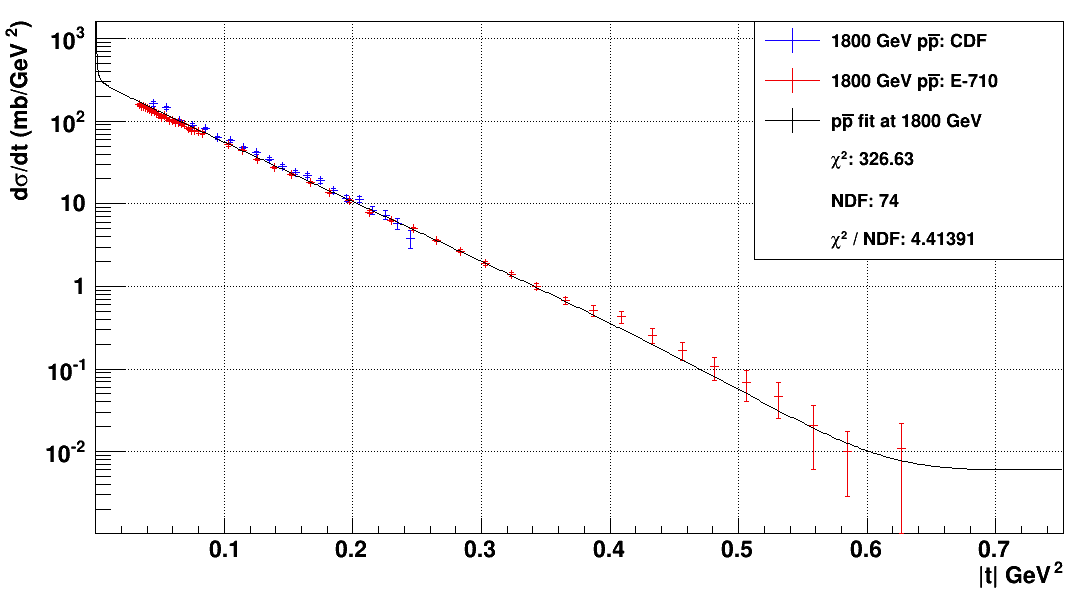}
\includegraphics[width=0.5\textwidth]{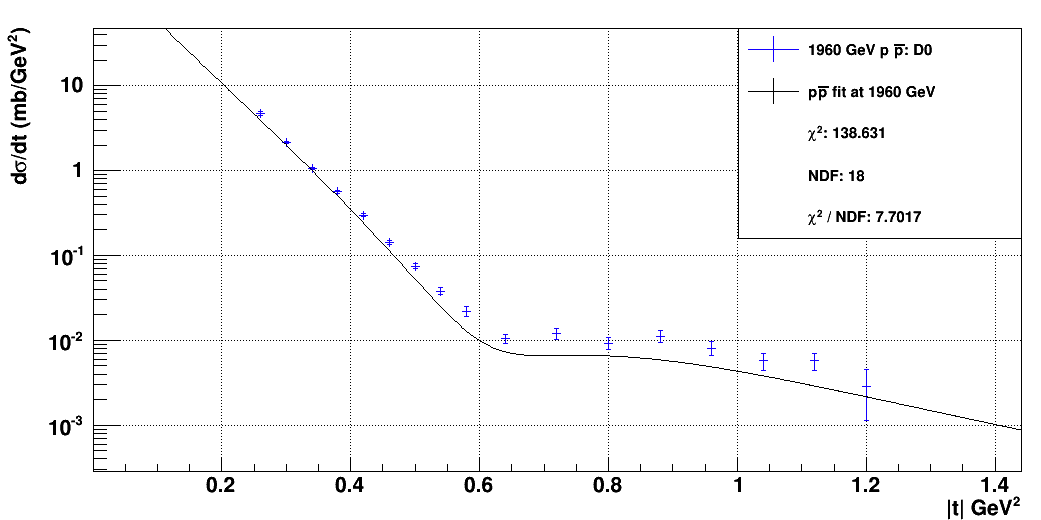}
\end{minipage}
\caption{\label{fig:elastic_full_ppbar}The $p\bar{p}$ elastic scattering model fit (fitted over all data) shown
for a range of $\sqrt{s}$,  over the full $t$ range, including the coulomb peak and down to the lower cut on $t$.}
\end{figure*}


\section{\label{sec:app_datalist2}Single diffraction dissociation data sources and model fit}
\begin{figure*}
\begin{minipage}[b]{1.0\linewidth}
\includegraphics[width=0.5\textwidth]{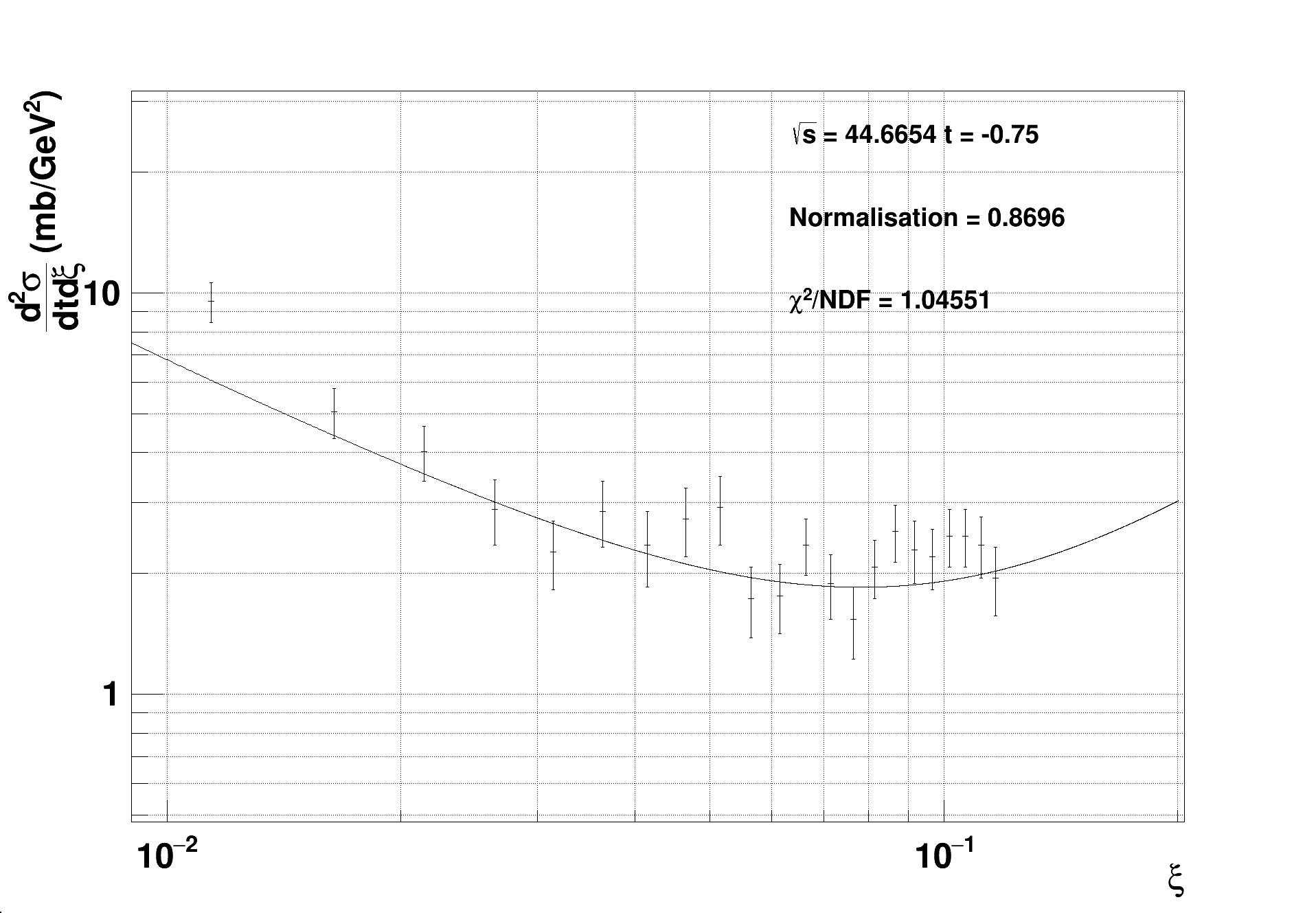}
\includegraphics[width=0.5\textwidth]{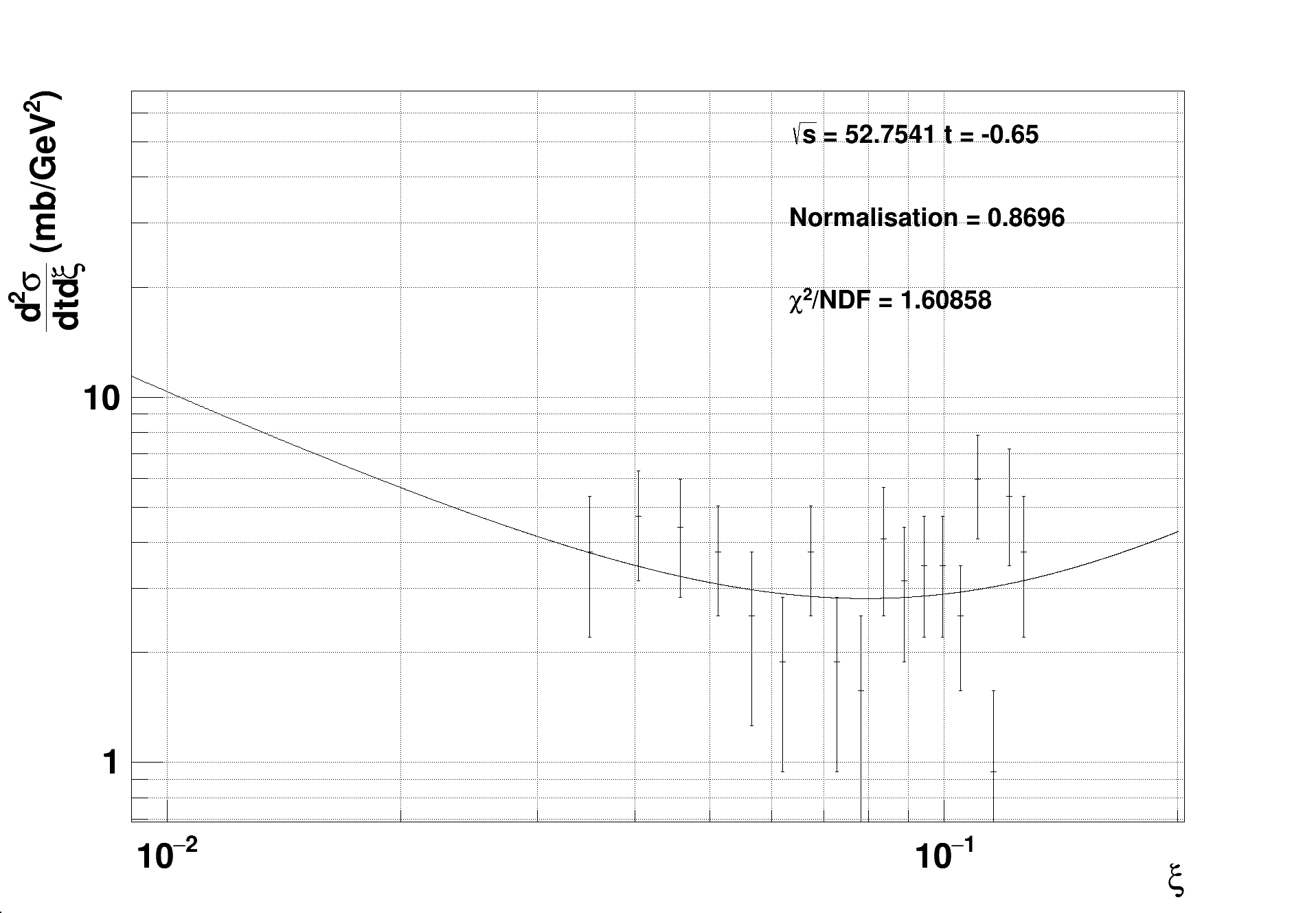}
\end{minipage}
\begin{minipage}[b]{1.0\linewidth}
\includegraphics[width=0.5\textwidth]{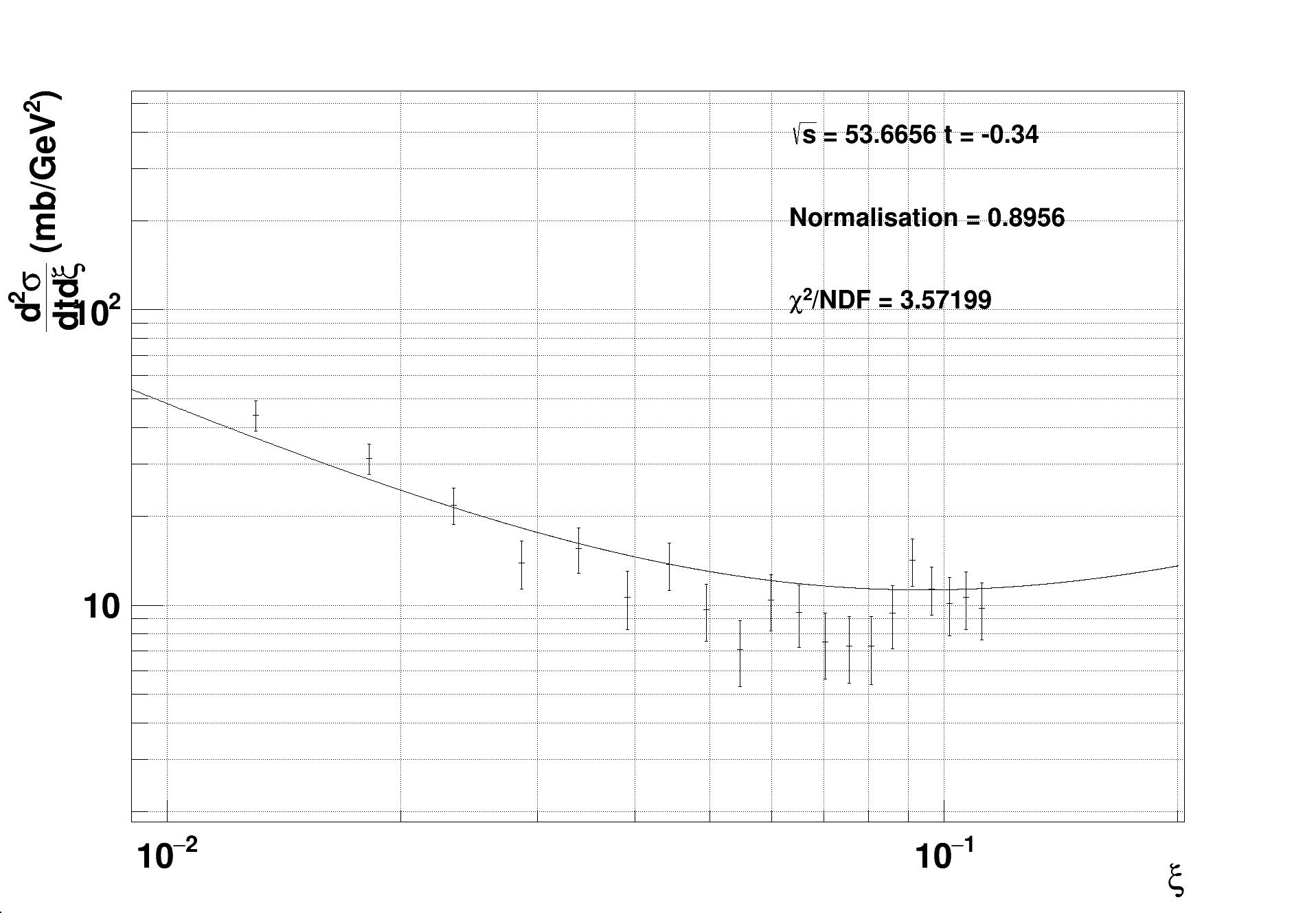}
\includegraphics[width=0.5\textwidth]{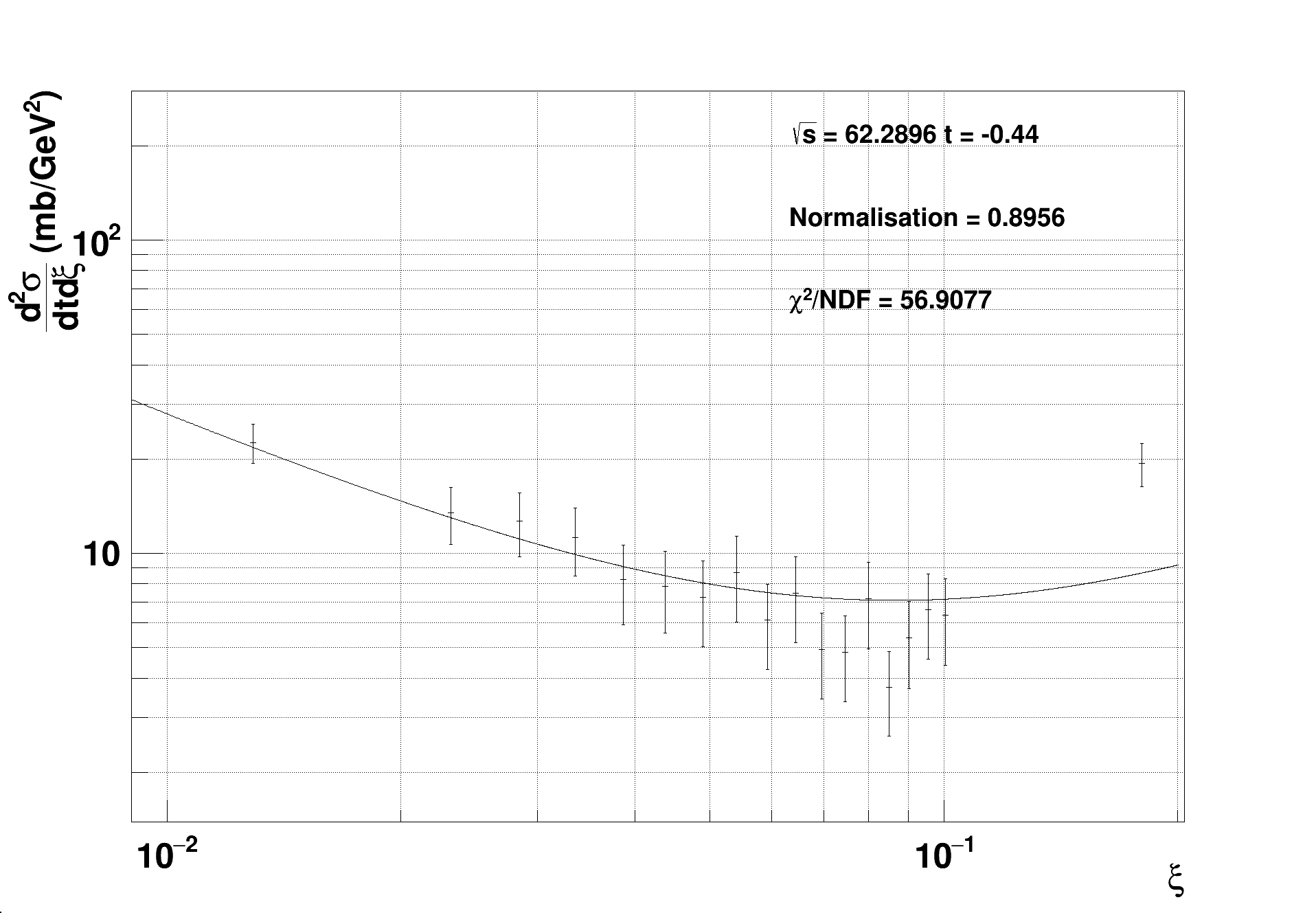}
\end{minipage}
\begin{minipage}[b]{1.0\linewidth}
\includegraphics[width=0.5\textwidth]{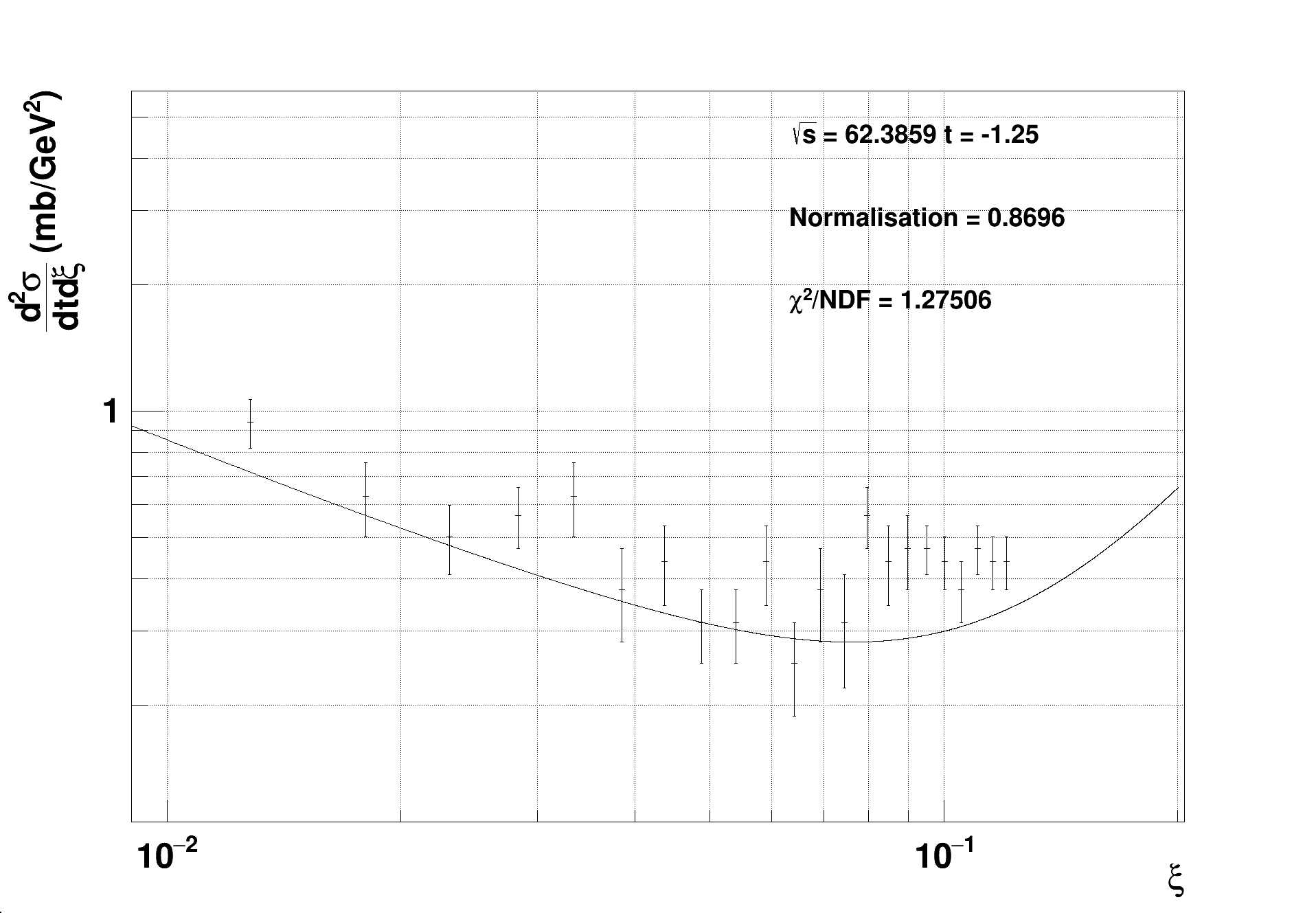}
\includegraphics[width=0.5\textwidth]{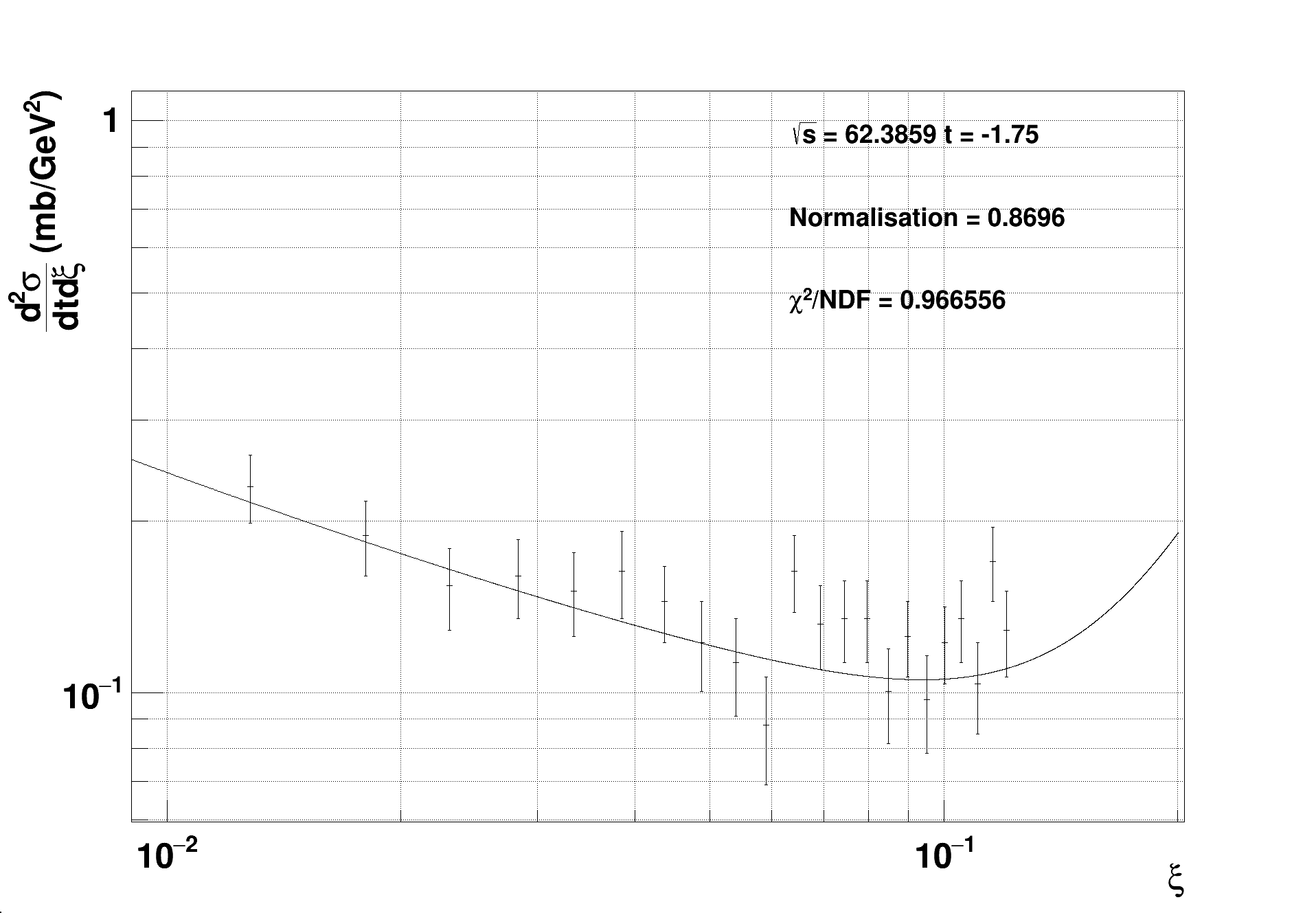}
\end{minipage}
\caption{\label{fig:SD_full_06}The SD scattering model fit (fitted over all data) shown
for a range of $\sqrt{s}$,  over the full $\xi$ range.}
\end{figure*}

\begin{figure*}
\begin{minipage}[b]{1.0\linewidth}
\includegraphics[width=0.5\textwidth]{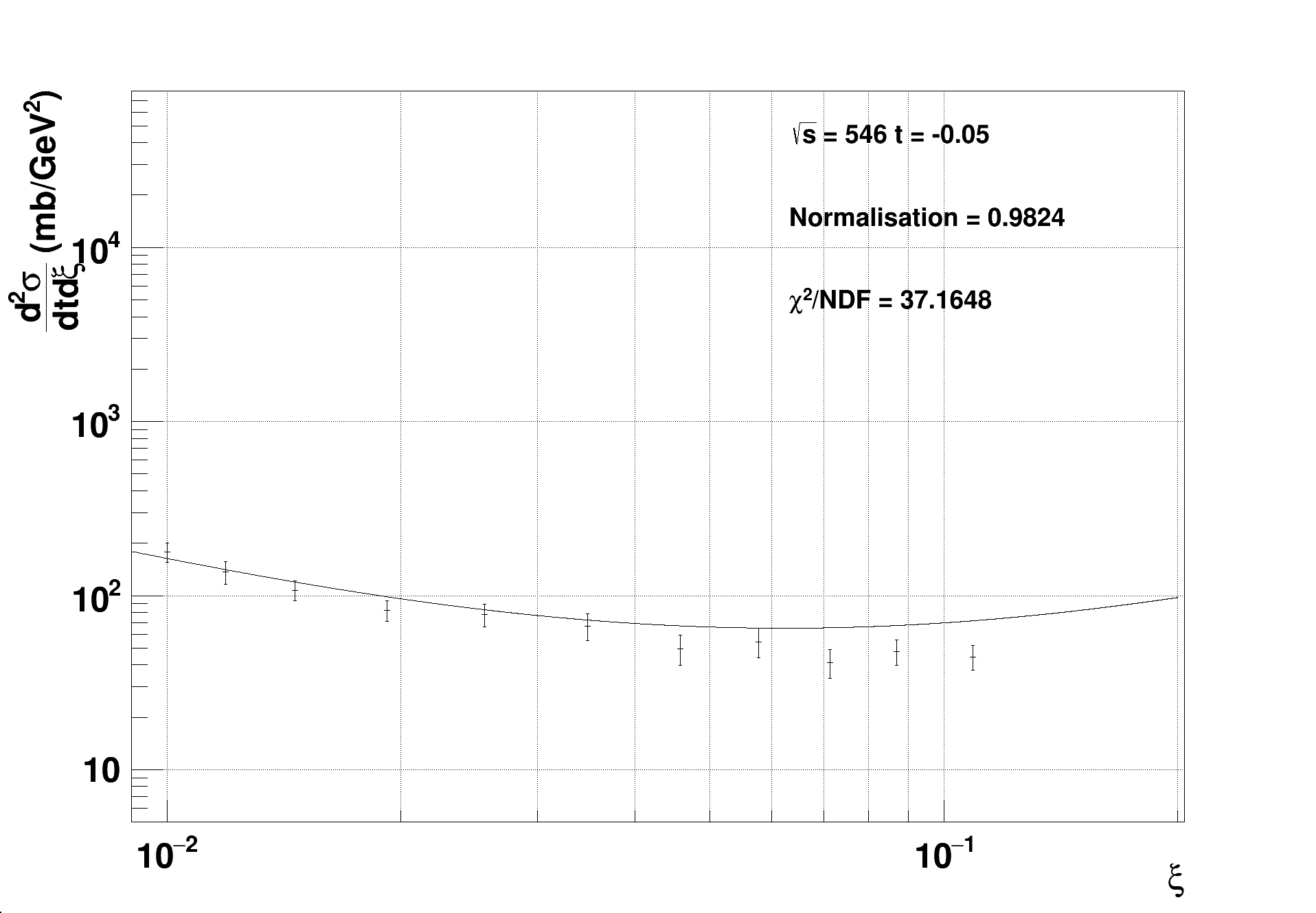}
\includegraphics[width=0.5\textwidth]{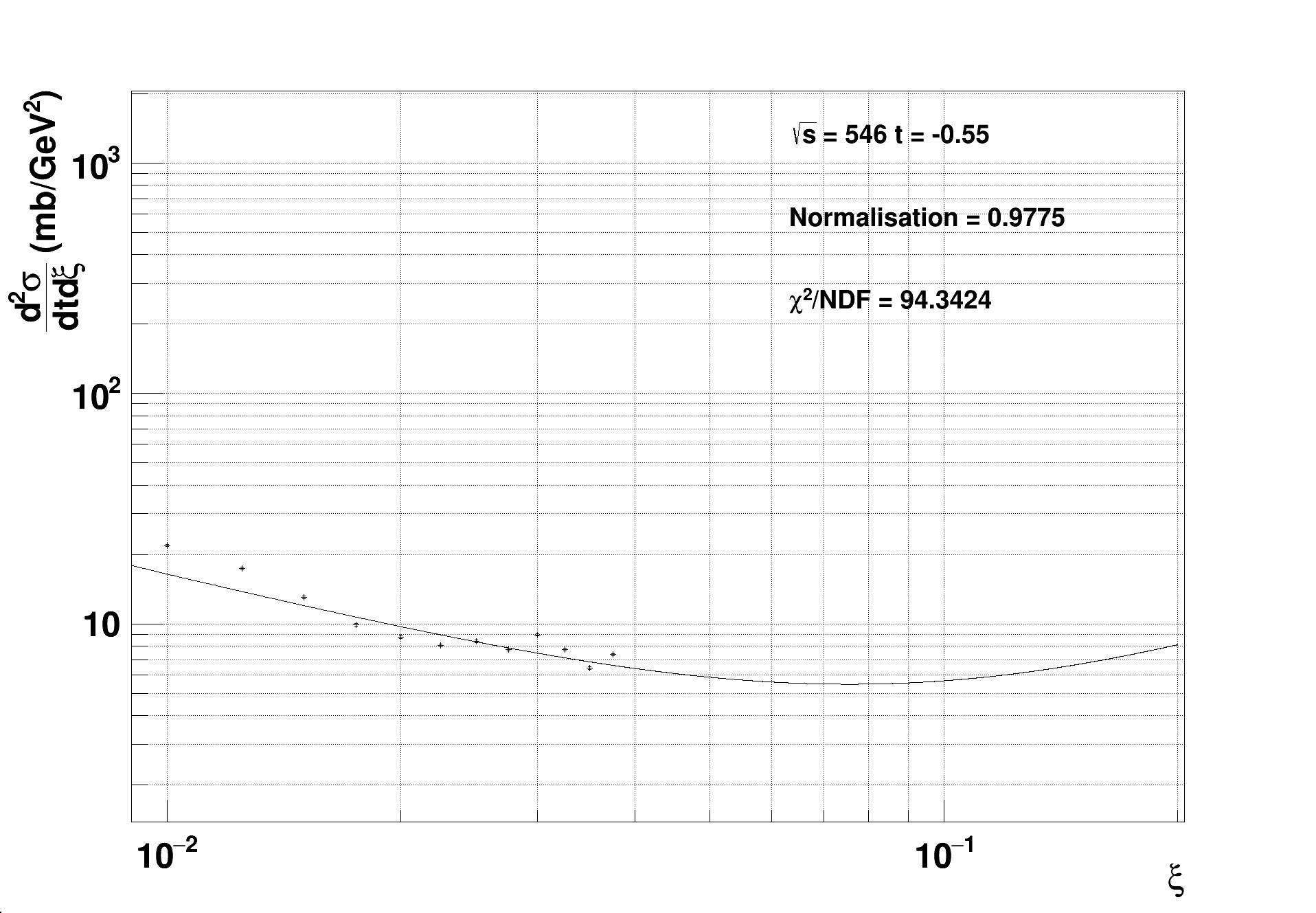}
\end{minipage}
\begin{minipage}[b]{1.0\linewidth}
\includegraphics[width=0.5\textwidth]{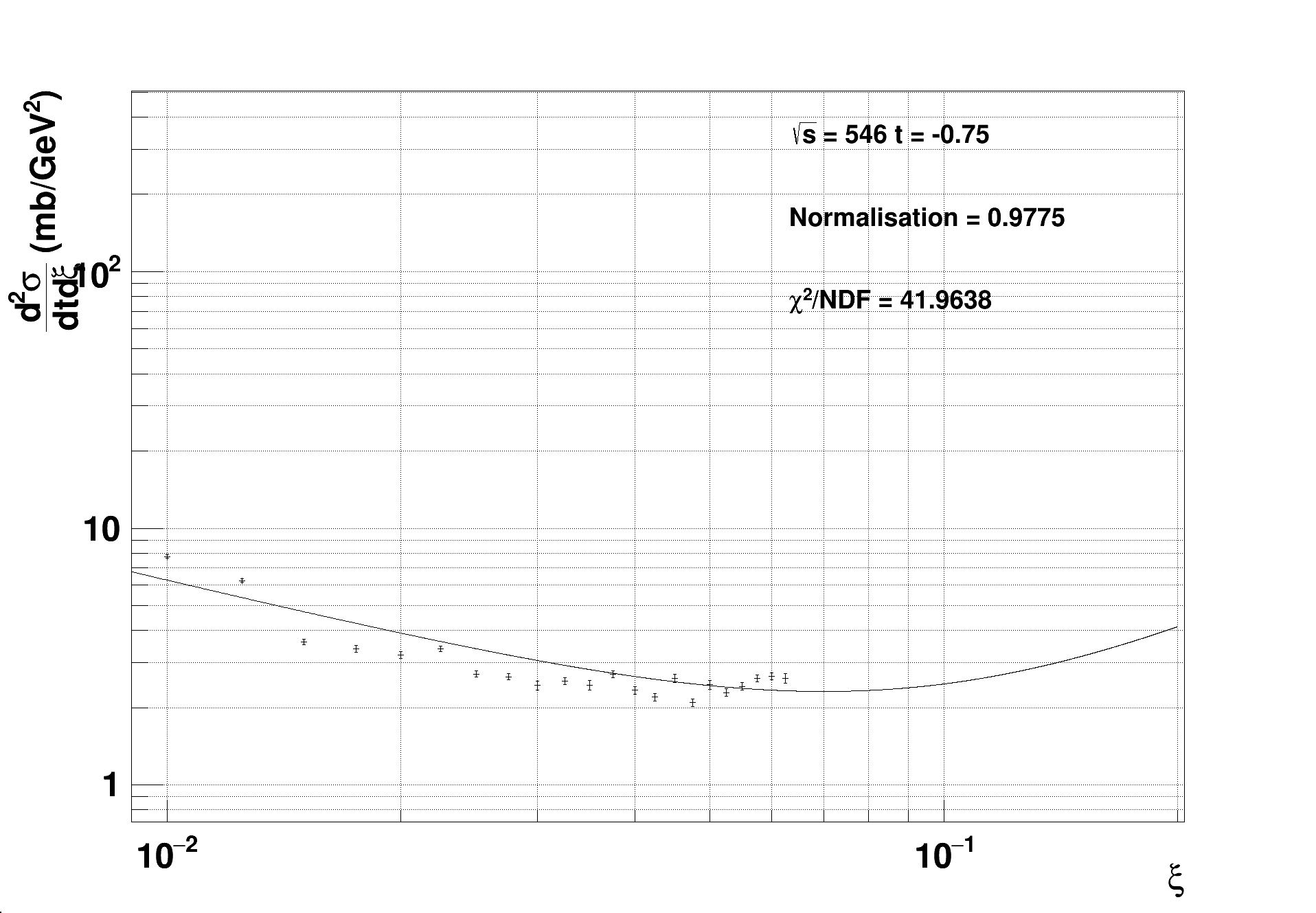}
\includegraphics[width=0.5\textwidth]{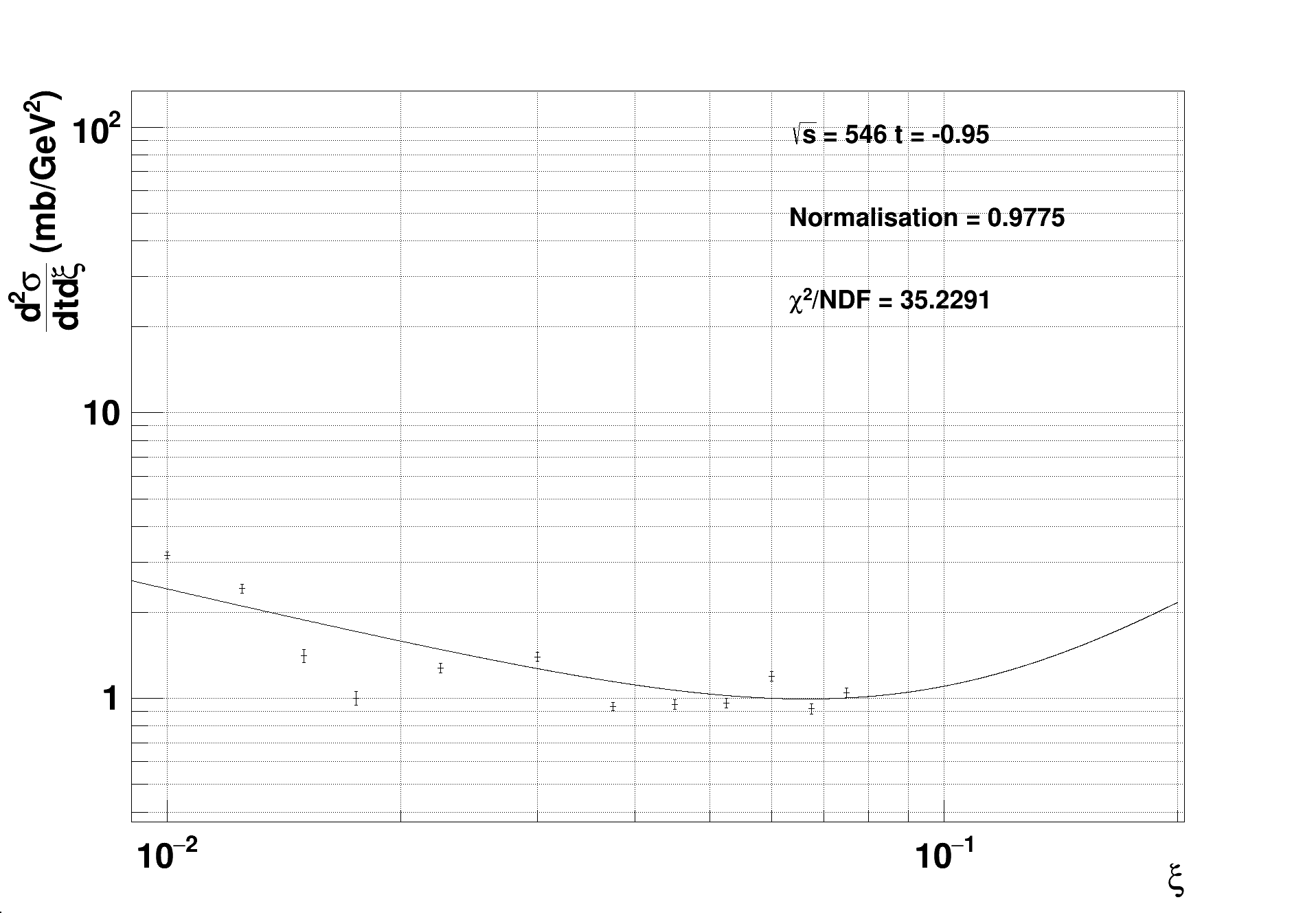}
\end{minipage}
\caption{\label{fig:SD_full_07}The SD scattering model fit (fitted over all data) shown
for a range of $\sqrt{s}$,  over the full $\xi$ range, for $p\bar{p}$ scattering.}
\end{figure*}

The single diffraction dissociation data sources with their energies and references are reported in table~\ref{tab:diffractive_data_sources}. The complete set of model fit comprise more than 300 plots. In figures~\ref{fig:SD_full_06} and~\ref{fig:SD_full_07} we present a selection of these plots to cover the full range of energies ($\sqrt{s}$) and $\xi$ and for low and high momentum transfer $t$.   

\begin{center}
\begin{table*}[h]
\caption{\label{tab:diffractive_data_sources}A list of available single diffraction data and its sources.}
\begin{tabular}{cccc}
\hline
\textrm{Experiments}&
\textrm{Particles}&
\textrm{$\sqrt{s}$ ($\text{GeV}$)}&
\textrm{Diffractive data sources}\\ \hline
CHLM & $pp$ & 23.4-62.3 & \cite{Albrow19761,Albrow1974376,Armitage1982365,kooijman_thesis}\\ 
Cool & $pp$ & 13.7-19.4& \cite{PhysRevLett.47.701} \\ 
Akimov & $pp$,$pd$ & 8.1,12.4,19.3 & \cite{PhysRevLett.39.1432} \\ 
Schamberger & $pp$ & 16.2-30.7 & \cite{PhysRevLett.32.389,PhysRevLett.34.1121,PhysRevD.17.1268} \\ \hline
UA4 &$p\bar{p}$ &546 & \cite{Bozzo1984217,Bernard1987227}\\ 
UA8 & $p\bar{p}$ & 630 & \cite{Brandt19983}\\ 
CDF & $p\bar{p}$ & 546,1800 & \cite{PhysRevD.59.114017}\\ \hline 
\end{tabular}
\end{table*}
\end{center}

\end{document}